\documentclass[final,authoryear,3p,times,twocolumn]{elsarticle}
\pdfoutput=1

\usepackage{graphicx}

\usepackage{amssymb}

\newcommand\araa{{ARA\&A}}%
\newcommand\apjs{{ApJS}}%
\newcommand\aap{{A\&A}}%
\newcommand\aapr{{A\&A~Rev.}}%
\def\chandra{{\it Chandra\/}}
\def\xray{\hbox{X-ray}}

\def\lum{erg~s$^{-1}$}
\def\flux{erg~cm$^{-2}$~s$^{-1}$}
\def\etal{{et\,al.\,}}
\def\lsim{\mathrel{\rlap{\lower4pt\hbox{\hskip1pt$\sim$}}
    \raise1pt\hbox{$<$}}}                
\def\gsim{\mathrel{\rlap{\lower4pt\hbox{\hskip1pt$\sim$}}
    \raise1pt\hbox{$>$}}}                
\def\AA{\buildrel _{\circ} \over {\mathrm{A}}}
\def\cdfs{\hbox{CDF-S}}
\def\cdfn{\hbox{CDF-N}}
\def\ecdfs{\hbox{E-CDF-S}}

\def\zs{$z_{\rm spec}$}
\def\zp{$z_{\rm phot}$}

\def\nh{$N_{\rm H}$}
\def\lx{$L_{\rm X}$}
\def\edd{$\lambda_{\rm Edd}$}

\journal{New Astronomy Reviews}

\newcommand{\msun}{$M_{\odot}$}

\begin{document}

\begin{frontmatter}



\title{The \chandra\ Deep Fields: Lifting the Veil on Distant Active Galactic Nuclei and X-Ray Emitting Galaxies}


\author[ustc1,ustc2]{Y.~Q.~Xue}
\ead{xuey@ustc.edu.cn}

\address[ustc1]{CAS Key Laboratory for Research in Galaxies and Cosmology, Department of Astronomy, University of Science and Technology of China, Hefei 230026, China}
\address[ustc2]{School of Astronomy and Space Science, University of Science and Technology of China, Hefei 230026, China}

\begin{abstract}

The \chandra\ Deep Fields (CDFs), being a major thrust among extragalactic X-ray surveys 
and complemented effectively by multiwavelength observations, have critically contributed 
to our dramatically improved characterization of the 0.5--8 keV cosmic X-ray background sources, 
the vast majority of which are distant active galactic nuclei (AGNs) and starburst and normal galaxies.
In this review, I highlight some recent key observational results, mostly from the CDFs, on the AGN demography, the interactions between AGNs and their host galaxies, 
the evolution of non-active galaxy X-ray emission, and the census of X-ray galaxy groups and clusters through cosmic time, after providing the necessary background information. 
I then conclude by summarizing some significant 
open questions and discussing future prospects for moving forward.

\end{abstract}

\begin{keyword}

black holes \sep galaxies \sep active galactic nuclei \sep quasars \sep accretion


\end{keyword}

\end{frontmatter}



\section{Introduction}

\subsection{Effectiveness of extragalactic \xray\ surveys}\label{sec:utility}

\begin{figure*}[!t]
\centering
\includegraphics[width=165mm]{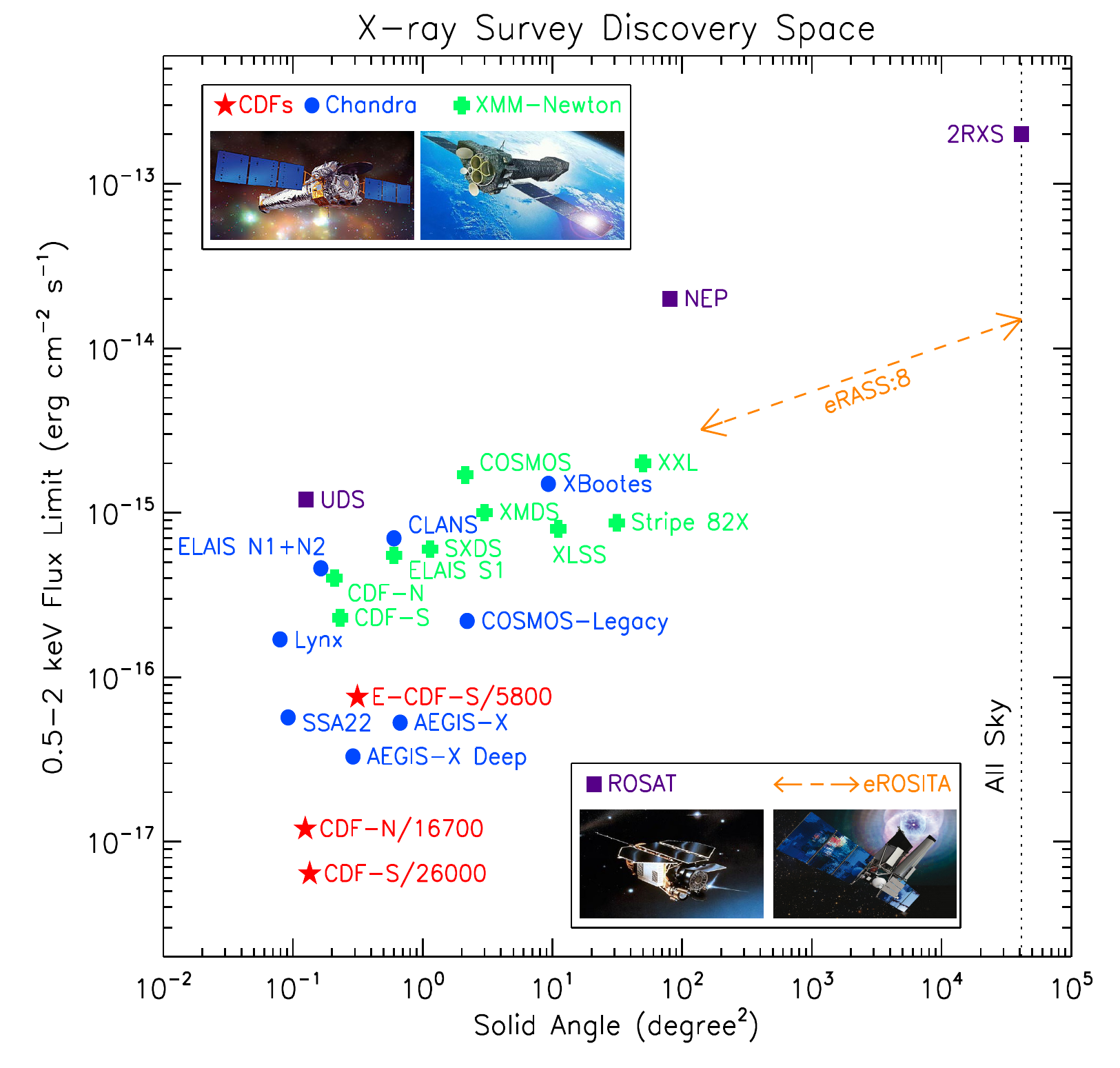}
\caption{Flux limits achieved versus solid angles covered by some selected \xray\ surveys in the 
0.5--2~keV band from \chandra, {\it XMM-Newton}, 
{\it ROSAT}, and {\it eROSITA}. 
The vertical dotted line indicates the solid angle for the whole sky.
The surveys plotted are listed below, with their corresponding references shown in the parentheses:
(1) for the \chandra\ Deep Fields (CDFs; red stars):
the 7~Ms \cdfs\ survey (Luo et~al. 2017),
the 2~Ms \cdfn\ survey (Xue et~al. 2016),
and the 250~ks \ecdfs\ survey (Xue et~al. 2016), with the numbers annotated after the survey names
indicating the observed \xray\ source densities in their respective central 3-arcmin areas
(these numbers have not been corrected for detection incompleteness or Eddington bias; see Table~\ref{tab:cdf}; cf.~Section~\ref{sec:numcts}); 
(2) for the other \chandra\ surveys (blue bullets):
the \chandra\ Deep Survey of the Extended Groth Strip (AEGIS-X; Laird et~al. 2009),
the AEGIS-X Deep survey (Nandra et~al. 2015),
the SSA22 protocluster survey (Lehmer et~al. 2009),
the Lynx survey (Stern et~al. 2002),
the \chandra\ COSMOS Legacy survey (Civano et~al. 2016),
the ELAIS N1+N2 deep \xray\ survey (Manners et~al. 2003),
the \chandra\ Lockman Area North Survey (CLANS; Trouille et~al. 2008), and
the \xray\ survey of the NDWFS Bootes field (XBootes; Murray et~al. 2005);
(3) for the {\it XMM-Newton} surveys (green crosses):
the \cdfs\ survey (Ranalli et~al. 2013),
the \cdfn\ survey (Miyaji et~al. 2003),
the ELAIS-S1 field survey (Puccetti et~al. 2006),
the Subaru/{\it XMM-Newton} Deep Survey (SXDS; Ueda et~al. 2008),
the XMM-Large Scale Structure survey (XLSS; Chiappetti et~al. 2013),
the Stripe~82X survey (LaMassa et~al. 2016),
the XMM Medium Deep Survey (XMDS; Chiappetti et~al. 2005), 
the COSMOS survey (Cappelluti et~al. 2009), and
the XXL survey (Pierre et~al. 2017);
(4) for the {\it ROSAT} surveys (purple squares):
the {\it ROSAT} Ultra Deep Survey (UDS; Lehmann et~al. 2001); 
the {\it ROSAT} North Ecliptic Pole survey (NEP; Henry et~al. 2006), and
the Second {\it ROSAT} all-sky survey source catalog (2RXS; Boller et~al. 2016);
and (5) 
the proposed final (4~years) {\it eROSITA} All-Sky Survey (eRASS:8; the orange dashed line with two arrow heads; Merloni et~al. 2012).
\label{fig:xspace}}
\end{figure*}

Since the discovery of the cosmic \xray\ background (CXRB; e.g., Giacconi \etal\ 1962),
various major \xray\ observatories have joined the efforts of resolving it into discrete cosmic 
sources as well as characterizing such sources, by carrying out different tiers of
extragalactic \xray\ surveys that range from shallow all-sky surveys to ultradeep pencil-beam
surveys. 
Together, these surveys, being highly complementary to each other, 
effectively occupy the practically-accessible half of the so-called \xray\ survey
discovery space (i.e., \xray\ flux limits achieved vs. solid angles covered; 
see Fig.~\ref{fig:xspace}),\footnote{Also see Fig.~3 and Table~1 of Brandt \& Alexander (2015) for a recent demonstration and the information of additional \xray\ surveys, respectively.}
powerfully providing 
a comprehensive understanding of \xray\ source populations in the universe, and  
essentially reveal that the large portion ($\approx 80\%$) of the CXRB up to $\approx 10$~keV 
can be accounted for by discrete \xray\ point sources detected, among which the vast majority are accreting 
supermassive black holes (SMBHs; with millions to billions of solar masses), i.e., active galactic nuclei (AGNs), in addition to
normal and starburst galaxies (see, e.g., Brandt \& Hasinger 2005 for a review).

X-ray AGN surveys are arguably the most effective method of identifying highly reliable     
and fairly complete samples of AGNs, due to several reasons (see, e.g., Section~1.1 of Brandt
\& Alexander 2015 for detailed reasoning and caveats):
(1) Observationally, \xray\ emission is nearly a universal feature of
optically, infrared (IR), radio-selected AGNs that are neither
highly Compton-thick (CT, i.e., with neutral hydrogen column densities of 
$N_{\rm H}\gsim 1.5\times10^{24}$~cm$^{-2}$; highly CT: 
$N_{\rm H}\gg 1.5\times10^{24}$~cm$^{-2}$; e.g., Lanzuisi \etal\ 2015b) nor intrinsically \xray\ weak
(such sources are very rare; see, e.g., Wu \etal\ 2011; Luo \etal\ 2014a). 
Theoretically,
\xray\ emission can be produced in various accretion disk models for AGNs that are applicable 
for a wide range of mass accretion rates (from sub-Eddington to super-Eddington accretion), 
disk temperatures (cold vs. hot accretion flows), gas opacities (optically thick vs. thin), and geometric structures (thin vs. thick); these models invoke a corona or corona-like component to
Compton up-scatter soft photons into hard \hbox{X-rays} when necessary (see, e.g., 
Yuan \& Narayan 2014 for a review).
(2) \hbox{X-rays} can penetrate through non-highly CT columns that
are common among the majority AGN populations, 
and become even more penetrating at high redshifts due to positive \hbox{$K$-correction}, 
thereby reducing significantly absorption biases, probing immediate vicinities of SMBHs,
and allowing for reliable \nh\ measurements to uncover intrinsic (i.e., 
absorption-corrected) AGN luminosities.
(3) \xray\ emission is subject to low dilution by host-galaxy stellar emission.
An \xray\ point source sitting right at the center of a galaxy is very likely to be
an AGN; this serves as an effective way to identify distant AGNs when it is typically unfeasible
to resolve spatially AGN light from host starlight.
(4) An AGN \xray\ spectrum is produced through numerous line and continuum emission processes
subject to obscuration, and can therefore be utilized to infer physical conditions close to
the SMBH, provided that the spectrum is of sufficient signal-to-noise ratio and energy resolution.

\subsection{The \chandra\ Deep Fields}\label{sec:cdf}

\begin{figure*}[!t]
\centering
\includegraphics[width=165mm]{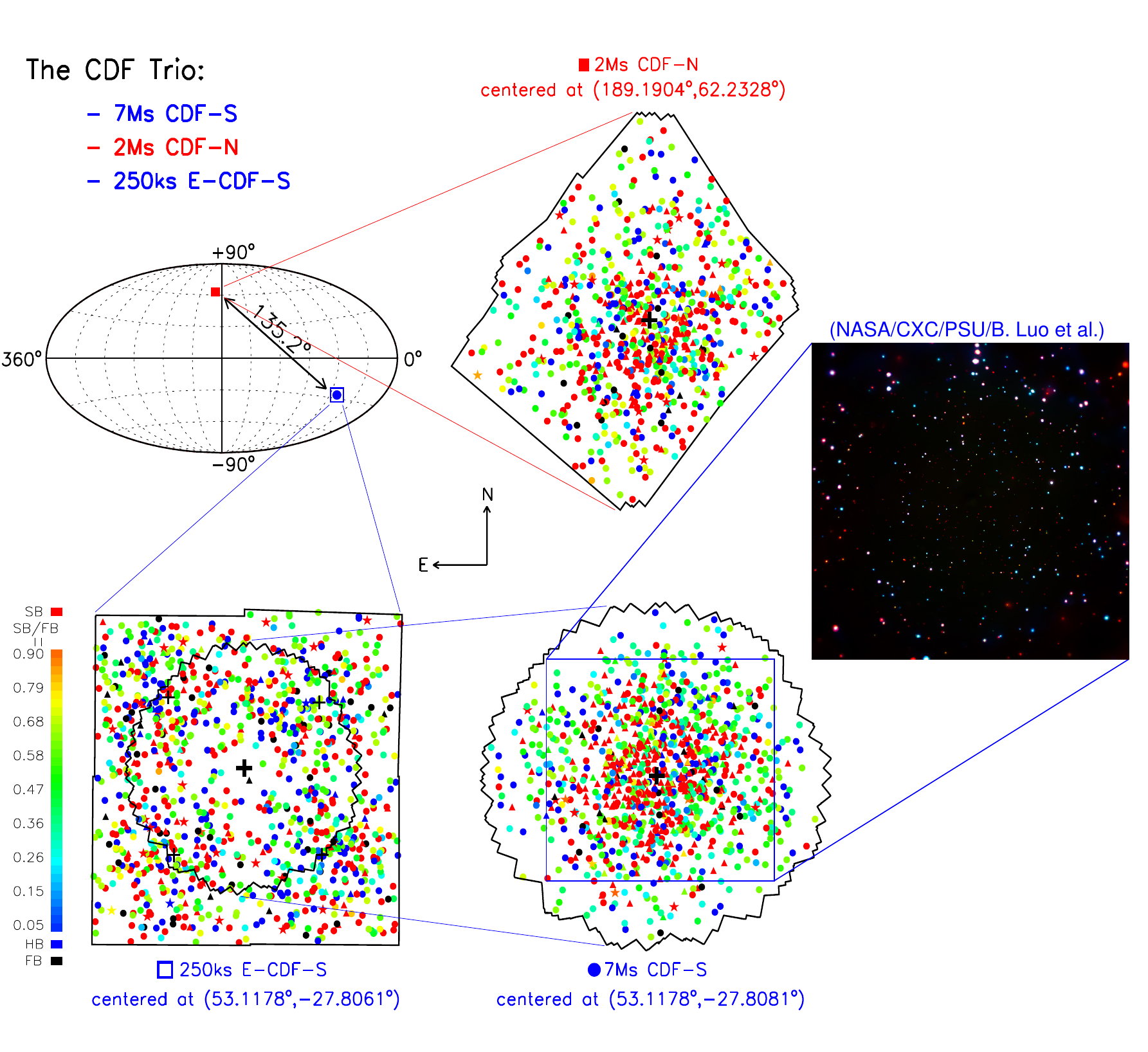}
\caption{The CDF trio --- the 7~Ms \cdfs, 2~Ms \cdfn, and 250~ks \ecdfs\ (see Table~\ref{tab:cdf} for more information). 
(Top left) The locations of the CDFs in the sky, with the separation between the \cdfn\ and 
\cdfs/\ecdfs\ annotated.
(Others) Spatial distributions of all point sources detected in the 2~Ms \cdfn\ (Xue \etal\ 2016),
250~ks \ecdfs\ (Xue \etal\ 2016), and 7~Ms \cdfs\ (Luo \etal\ 2017), respectively, as 
well as the false-color image of the central 7~Ms \cdfs\ (16~arcmin across; courtesy of B.~Luo).
The filled circles, triangles, and five-pointed stars represent AGNs, starburst and normal galaxies, and Galactic stars, respectively.
Sources detected neither in the 0.5--2~keV band (soft band; SB) nor in the 2--7~keV band (hard band; HB) are coded in black (i.e., detected only in the 0.5--7~keV band; full band; FB);
sources detected in the SB but not in the HB are coded in red;
sources detected in the HB but not in the SB are coded in blue; and
sources detected both in the SB and HB are coded in between red and blue based on ratios between the SB and SB+HB=FB net counts (see the color bar at the bottom-left corner).
Bold crosses roughly indicate exposure-weighted average aim points.\label{fig:cdf}}
\end{figure*}

The \chandra\ Deep Fields (CDFs; see Fig.~\ref{fig:cdf} and Table~\ref{tab:cdf}) 
consist of the \chandra\ Deep Field-South (\cdfs),
the \chandra\ Deep Field-North (\cdfn), and the Extended-\chandra\ Deep Field-South (\ecdfs).
The \cdfs\ survey was originally led by R.~Giacconi during 1999--2000 (1~Ms \cdfs; Giacconi \etal\ 2002),
extended to 2~Ms through the Director's Discretionary Time (DDT) by the CXC director H.~Tananbaum 
in 2007 (2~Ms \cdfs; Luo \etal\ 2008), 
awarded an additional 2~Ms DDT by H.~Tananbaum in 2010 (4~Ms \cdfs; 
Xue \etal\ 2011), and eventually pushed to 7~Ms by W.~N.~Brandt during 2014--2016 (7~Ms \cdfs;
Luo \etal\ 2017).\footnote{In addition to the 7~Ms \chandra\ $\approx$0.3--8~keV \cdfs\ coverage,
there are $\approx$3~Ms of \cdfs\ coverage with {\it XMM-Newton} at \hbox{$\approx$0.2--12~keV} (Comastri et~al. 2011; Ranalli et~al. 2013; see Fig.~\ref{fig:xspace}) and 200~ks of \ecdfs\ coverage with
{\it NuSTAR} at 3--24~keV (Mullaney et~al. 2015a).\label{ft:cdfs}}
The \cdfs\ patch of sky, lying in the Fornax constellation, was chosen because of
very low foreground Galactic \nh\ ($8.8\times10^{19}$~cm$^{-2}$; e.g., Stark \etal\ 1992),
no bright ($m_{\rm V}\le 14$) Galactic stars, and optimal visibility from large 
ground-based telescopes in Chile.
The \cdfn\ project was initiated by G.~Garmire (the first \hbox{$\approx 0.5$~Ms})
and W.~N.~Brandt (the second \hbox{$\approx 0.5$~Ms}) during 1999--2001 (1~Ms \cdfn; 
Brandt \etal\ 2001),
and subsequently enlarged by W.~N.~Brandt during 2001--2002 (2~Ms \cdfn; Alexander \etal\ 2003;
Xue \etal\ 2016).\footnote{In addition to the 2~Ms \chandra\ \cdfn\ coverage,
there are 180~ks of \cdfn\ coverage with {\it XMM-Newton} (Miyaji \etal\ 2003; see Fig.~\ref{fig:xspace}) and 200~ks of \cdfn\ coverage with
{\it NuSTAR} (completed in February 2016).\label{ft:cdfn}}
The \cdfn\ lies in the Ursa Major constellation, and was chosen for largely similar
considerations as the \cdfs\ (e.g., $N_{\rm H}=1.6\times10^{20}$~cm$^{-2}$; Stark \etal\ 1992),
in addition to the desire to cover the \hbox{HDF-N} that was unique back then. 
Furthermore, as a parallel field to the \cdfs, the \cdfn\ doubles the number of such deep
surveys, thus controlling for influence of cosmic variance and allowing for direct
comparative studies between fields.     
The \ecdfs\ survey was carried out (PI: W.~N.~Brandt) in 2004 (250~ks \ecdfs; Lehmer \etal\ 2005;
Xue \etal\ 2016), which significantly expands the sky coverage of the \cdfs\ proper with
four distinct, contiguous, and flanking pointings (thus totaling $\approx 1$~Ms exposure).

\begin{table*}[!t]
\caption{Properties of the CDFs$^a$}
\centering
\resizebox{1.00\textwidth}{!}{%
\begin{tabular}{lccc}\hline\hline
 & \cdfs & \cdfn & \ecdfs \\\hline
Galactic \nh\ (cm$^{-2}$) & $8.8\times10^{19}$ & $1.6\times10^{20}$ & $8.8\times10^{19}$ \\
Observational timespan & 1999/10 -- 2016/03 (16.4 yrs) & 1999/11 -- 2002/02 (2.3 yrs) & 2004/02 -- 2004/11 (0.8 yrs) \\
Total number of observations & 102 & 20 & 9 \\
Effective exposure (ks) & 6727 & 1896 & 235/209/240/241$^b$ \\
Solid angle covered (arcmin$^2$) & 484.2 & 447.5 & 1128.6 \\
Source detection criteria & {\sc WAVDETECT} at $10^{-5}$ & {\sc WAVDETECT} at $10^{-5}$ & {\sc WAVDETECT} at $10^{-5}$ \\
 & and $P<0.007$$^c$ & and $P<0.004$ & and $P<0.002$ \\
Number of sources detected$^d$ & 1008 & 683 & 1003 \\
FB (0.5--7 keV) detected counts & (11.2, 98.9, 56916.2)$^e$ & (8.1, 66.2, 19748.4) & (3.3, 27.1, 4010.6) \\
SB (0.5--2 keV) detected counts & (\ 6.1, 47.4, 38817.0) & (5.4, 35.0, 14227.3) & (2.2, 18.9, 2802.6) \\
HB (2--7 keV) detected counts & (\ 9.2, 94.6, 18137.8) & (7.7, 57.5, \ 5540.6) & (3.4, 20.4, 1210.8) \\
$1\sigma$ \xray\ positional uncertainty ('') & (0.11, 0.47, 1.28) & (0.10, 0.47, 2.02) & (0.10, 0.63, 1.30) \\
Logarithm of FB flux (\flux) & ($-16.76$, $-15.50$, $-12.96$) & ($-16.35$, $-15.09$, $-12.70$) & ($-15.73$, $-14.79$, $-12.88$) \\
Logarithm of SB flux (\flux) & ($-17.11$, $-16.19$, $-13.29$) & ($-16.83$, $-15.79$, $-13.07$) & ($-16.13$, $-15.27$, $-13.26$) \\
Logarithm of HB flux (\flux) & ($-16.46$, $-15.25$, $-13.13$) & ($-16.15$, $-14.95$, $-12.95$) & ($-15.73$, $-14.70$, $-13.02$) \\
Faintest sources detected & 1 count per $\approx10$ days & 1 count per $\approx4$ days & 1 count per $\approx1$ day \\
Logarithm of $L_{\rm 0.5-7\ keV}$ (\lum)$^f$ & (39.01, 42.48, 45.05) & ($39.28$, $42.94$, $45.07$) & ($39.89$, $43.34$, $45.50$) \\
\% of multiwavelength identifications & 98.4\% & 98.1\% & 95.5\% \\
\% of \zs\ ($z_{\rm adopted}$)$^g$ & 67.2\% (97.8\%) & 51.4\% (93.4\%) & 47.5\% (80.8\%) \\
$z_{\rm adopted}$$^h$ & (0.000, 1.156, 5.776) & (0.000, 1.130, 5.365) & (0.000, 1.193, 7.203) \\
\% of AGNs/galaxies/stars & 70.5\%/28.3\%/1.2\% & 86.5\%/11.0\%/2.5\% & 90.6\%/6.7\%/2.7\% \\
AGN/galaxy/star density (deg$^{-2})$$^i$ & 13600/12100/250 & 12400/4200/100 & 5200/500/100 \\\hline
\end{tabular}}
\begin{flushleft}
$^a$ For source properties, here I refer only to the sources from the main catalogs of the 7~Ms 
\cdfs\ (Luo et~al. 2017), 2~Ms \cdfn\ (Xue et~al. 2016), and 250~ks \ecdfs\ (Xue et~al. 2016). These three main catalogs were produced using essentially
the same approach (critically aided by the use of the ACIS Extract package; AE; Broos et~al. 2010), which incorporates a number of recent improvements in \chandra\ source-cataloging methodology, and therefore maximizes the number of reliable sources detected
and allows for best possible \xray\ characterization of source properties (see,
e.g., Table~1 of Xue et~al. 2016 for details). \\
$^b$ The \ecdfs\ consists of four distinct, contiguous pointings that flank the \cdfs\ proper (see Fig.~\ref{fig:cdf}). \\
$^c$ $P$ indicates the probability of a source not being real (i.e., due to background fluctuations). \\
$^d$ Among the 1008+683+1003=2694 CDF sources, 298 were detected both in the \cdfs\ and \ecdfs, which results in a total of 2396 unique CDF sources.\\
$^e$ The three numbers in parentheses denote the minimum, median, and maximum values. \\
$^f$ This is the absorption-corrected rest-frame 0.5--7~keV luminosity ($L_{\rm X}$ hereafter). \\
$^g$ $z_{\rm adopted}$ denotes the adopted redshifts, with secure spectroscopic redshifts (\zs)  preferred over photometric redshifts (\zp). \\   
$^h$ The maximum secure \zs's are 4.762 and 5.186 for the \cdfs/\ecdfs\ and \cdfn, respectively. The \zp\ estimates above these values are subject to large uncertainties. \\ 
$^i$ These are observed source densities calculated within the respective central $r\le3$~arcmin areas.
\end{flushleft}     
\label{tab:cdf}
\end{table*}

All the CDF observations were performed with the Advanced CCD Imaging Spectrometer
(ACIS; Garmire \etal\ 2003) onboard the \chandra\ \xray\ observatory (Weisskopf \etal\ 2000),
whose sharp point-spread function (PSF) and low background make higher sensitivities
achievable with longer exposures in an efficient manner. 
Indeed, the 7~Ms \cdfs\ and 2~Ms \cdfn\ images are the two deepest \xray\ images ever taken,  
thus being able to explore parameter space that has never been probed by
any other X-ray surveys and fulfill one of \chandra's central design goals --- revealing and characterizing the sources that constitute the CXRB.
Amazingly, the faintest sources detected in the 7~Ms \cdfs\ have only $\approx 1$ count per 10 days;
and the 7~Ms \cdfs\ will serve as a multi-decade \chandra\ legacy for advancing deep-survey science projects, owing to its unique combination of great depth and high angular resolution. 
The motivation of going ultradeep is at least two-fold:
on one hand, 
ultradeep exposures accumulate sufficient photons for known (faint) sources
that allow for better characterization of the AGNs producing most of cosmic accretion power
via \xray\ spectral and variability analyses;
on the other hand,
ultra-sensitivities reveal a substantial amount of new sources (typically fainter
and more obscured), thereby facilitating better understanding of obscured growing SMBHs
through the \hbox{$z\approx1$--4} era of massive galaxy assembly, majority AGN
populations in the 
first galaxies, and \xray\ binary (XRB) populations in cosmologically distant starburst and normal galaxies.

\subsection{Importance of multiwavelength observations}\label{sec:multi}

\begin{figure*}[!t]
\centering
\includegraphics[width=120mm]{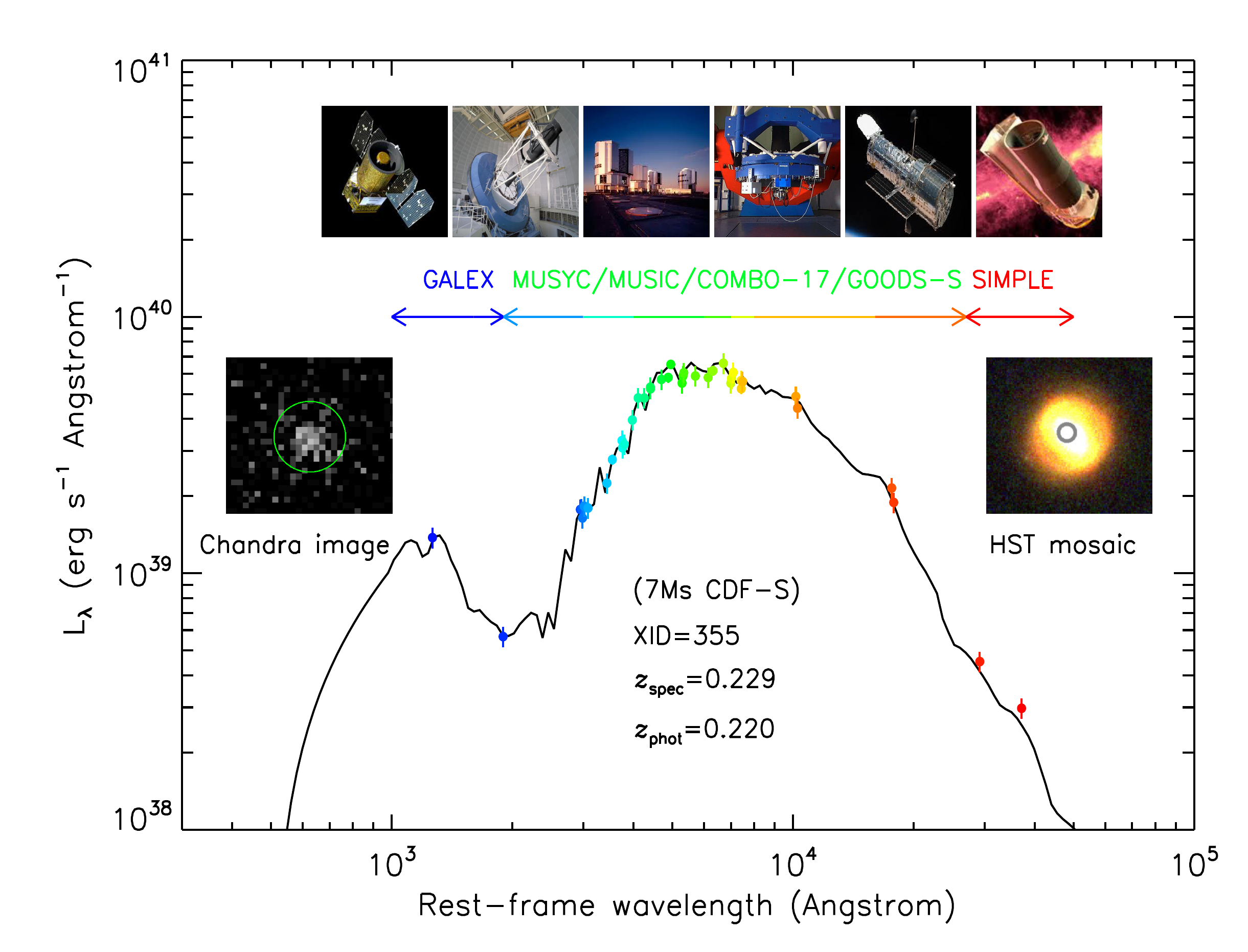}
\caption{Reliable \zp\ calculation for the source with XID=355 in the 7~Ms \cdfs\ main catalog, which is enabled by using the rich and high-quality broadband photometric data ($\approx 40$~filters; see Table~\ref{tab:data} for the detail of photometric data used) and 
the spectral energy distribution (SED) template fitting approach.\label{fig:zphot}}
\end{figure*}

The ultradeep multiwavelength coverage, well matched with the CDF \xray\ data, is critically important for many aspects, including   
source identification, source classification, measurements of host-galaxy physical properties (e.g., redshift, stellar mass --- $M_\star$, star formation rate --- SFR, absolute rest-frame magnitudes, morphology), and providing useful comparison samples of non-active galaxies.
The CDFs are among the most intensively observed sky patches with enormous observational investments at many wavelengths from both ground and space.
For instance, the \cdfs\ area has been covered by $\approx 50$ different photometric bands that span from extreme ultraviolet (UV) to far-IR (FIR) at great depths,
in conjunction with numerous deep spectroscopic and imaging observations,
thereby resulting in a very-high redshift success rate
(97.8\% of the 7~Ms \cdfs\ main-catalog sources have redshift measurements
with 67.2\% being spectroscopic ones;\footnote{For \zs\ references, see Section~4.3 of Luo et~al. 2017 and Section~3.3.4 of Xue et~al. 2016 for the \cdfs/\ecdfs, and 
Section~2.3.4 of Xue et~al. 2016 for the \cdfn.} Luo et~al. 2017; see Table~\ref{tab:cdf}).   
Table~\ref{tab:data} lists some of the most notable multiwavelength surveys and photometric-redshift catalogs in the CDF areas that were completed; and
Figure~\ref{fig:zphot} presents an example of calculating accurate \zp\ using rich multiwavelength photometric data.
By combining the \xray\ and multiwavelength coverage, the CDF datasets provide a unique opportunity to study both statistical and source-by-source properties of detected sources over a large range of $L_{\rm X}-z$ parameter space (see Table~\ref{tab:cdf}).
Furthermore, the CDF regions will be a testbed for future large observatories such as {\it JWST} and ELT, and the ultradeep multiwavelength coverage will continue to improve persistently, thereby keeping the science exciting.

\begin{table*}[!t]
\caption{Some Notable CDF Multiwavelength Surveys and Photometric-Redshift Catalogs}
\centering
\resizebox{1.00\textwidth}{!}{%
\begin{tabular}{lll}\hline\hline
 & Surveys & \zp \\\hline
\cdfs/ & VLA 1.4~GHz \ecdfs\ (Bonzini et~al. 2012; Miller et~al. 2013) & MUSYC \ecdfs\ (Cardamone et~al. 2010) \\
\ecdfs\ & ALMA LABOCA \ecdfs\ (ALESS; Hodge \etal\ 2013) & 2~Ms CDF-S (Luo et~al. 2010) \\
 & LABOCA \ecdfs\ (LESS; Wei\ss\ \etal\ 2009) & \ecdfs\ (Rafferty et~al. 2011) \\
 & {\it Herschel}/HerMEs (Oliver \etal\ 2012) & LABOCA/\ecdfs\ (Wardlow et~al. 2011) \\  
 & {\it Herschel} PEP (Lutz \etal\ 2011) & CANDELS/GOODS-S, CDF-S, E-CDF-S (Hsu et~al. 2014) \\
 & GOODS-{\it Herschel} (Elbaz \etal\ 2011) & CANDELS/3D-{\it HST} (Skelton et~al. 2014) \\
 & SEDS IRAC (Ashby et~al. 2013) & CANDELS/GOODS-S (Santini et~al. 2015) \\
 & SIMPLE IRAC (Damen et~al. 2011) & ZFOURGE/CDF-S (Straatman et~al. 2016) \\
 & TENIS WIRCam $K_s$-band (Hsieh et~al. 2012) & \\
 & GOODS-S MUSIC (Grazian et~al. 2006; Santini et~al. 2009) &  \\
 & MUSYC \ecdfs\ (Taylor et~al. 2009) &  \\
 & CANDELS GOODS-S (Grogin et~al. 2011; Koekemoer et~al. 2011) & \\
 & CANDELS/3D-{\it HST} (Skelton et~al. 2014) & \\
 & GEMS {\it HST} (Caldwell et~al. 2008) &  \\
 & COMBO-17 (Wolf et~al. 2004, 2008) &  \\
 & GOODS-S (Giavalisco et~al. 2004) &  \\
 & ESO WFI $R$-band (Giacconi et~al. 2002; Giavalisco et~al. 2004) &  \\
 & HUDF (Beckwith et~al. 2006) &  \\
 & {\it GALEX} DIS (Morrissey \etal\ 2007) &  \\\hline
\cdfn\ & VLA 1.4~GHz GOODS-N (Morrison et~al. 2010) & CANDELS/3D-{\it HST} (Skelton et~al. 2014) \\
 & SCUBA-2 CLS (S2CLS) GOODS-N (Geach \etal\ 2017) & H-HDF-N (Yang et~al. 2014) \\
 & SCUBA HDF-N (Wang, Cowie, \& Barger 2004) & \\
 & {\it Herschel}/HerMEs (Oliver \etal\ 2012) & \\  
 & {\it Herschel} PEP (Lutz \etal\ 2011) & \\
 & GOODS-{\it Herschel} (Elbaz \etal\ 2011) & \\
 & SEDS IRAC (Ashby et~al. 2013) & \\
 & GOODS-N WIRCam $K_s$-band (Wang et~al. 2010) & \\
 & H-HDF-N (Capak et~al. 2004; Yang et~al. 2014) &  \\
 & CANDELS GOODS-N (Grogin et~al. 2011; Koekemoer et~al. 2011) & \\
 & GOODS-N (Giavalisco et~al. 2004) & \\
 & HDF-N (Williams et~al. 1996) &  \\
 & {\it GALEX} DIS (Morrissey \etal\ 2007) & \\\hline
\hline
\end{tabular}}
\label{tab:data}
\end{table*}

\subsection{Identification of \xray\ AGNs in CDFs}\label{sec:agnid}

Over past few decades, a series of observational and mostly empirical methods have been developed and refined to identify AGN candidates from the general \xray\ source population
detected in extragalactic \xray\ deep surveys, 
including AGNs, starburst and normal galaxies, galaxy groups and clusters, and Galactic stars.
Table~\ref{tab:agnid} outlines the AGN identification criteria that have been routinely used in
the CDF catalog papers and some relevant works (see, e.g., Bauer et~al. 2004,
Alexander et~al. 2005a, Lehmer et~al. 2008 and 2012, Xue et al. 2010, 2011, and 2016,
and Luo et~al. 2017 for details and caveats), which rely upon either direct use of the \xray\ data
or combined use of both the \xray\ and multiwavelength data.\footnote{For additional methods to select AGN candidates that are undetected by \xray\ observations, I refer readers to, e.g.,  Section~3.3 of Brandt \& Alexander (2015), for details.}
These methods are effective in revealing distinct AGN signatures owing to three major factors: 
accurate \xray\ source positions (thanks to {\it Chandra}'s sharp PSF and thus superb sub-arcsecond angular resolution; see Table~\ref{tab:cdf}) that allow for reliable multiwavelength identifications with a sophisticated likelihood-ratio matching procedure (e.g., Brusa et~al. 2005, 2007; 
Luo et~al. 2010, 2017; Xue et~al. 2016),
robust \xray\ photometric measurements that well describe source \xray\ properties (e.g., Xue et~al. 2011, 2016; Luo et~al. 2017), and
rich multiwavelength data that critically complement the \xray\ observations
(see Section~\ref{sec:multi} and Table~\ref{tab:data}). 
It is worthing noting that the vast majority of the CDF AGNs satisfy multiple criteria listed in
Table~\ref{tab:agnid}; such cross-validations of AGN candidates therefore result in highly complete, reliable, and pure samples of distant AGNs for statistically meaningful investigations.
In addition to AGN candidates, Galactic stars are also identified based on spectroscopic and/or imaging observations (e.g., Feigelson \etal\ 2004; Xue et~al. 2011, 2016), and the remaining non-AGN and non-star CDF sources are regarded as galaxies, 
whose measured \xray\ emission can be produced by a mixture of XRBs and low-rate SMBH accretion.

\begin{table*}[!t]
\caption{Frequently-Used Criteria for \xray\ AGN Identification in CDFs}
\centering
\resizebox{1.00\textwidth}{!}{%
\begin{tabular}{ll}\hline\hline
AGN criterion & Basic reasoning (targeted AGN subpopulation) \\\hline
$L_{\rm X,intrinsic}\ge 3\times 10^{42}$ \lum & Typical starburst and normal galaxies: $L_{\rm X,intrinsic}\lsim 10^{42}$ \lum\ (\xray\ luminous AGNs) \\  
$\Gamma_{\rm effective}\le 1$ & Obscured ($N_{\rm H}\gsim10^{22}$~cm$^{-2}$) AGNs: hard \xray\ spectra observed (\xray\ obscured AGNs) \\
log$(f_{\rm X}/f_R)>-1$ & Elevated \xray\ emission compared to host optical emission (majority AGN populations) \\
log$(f_{\rm X}/f_{Ks})>-1.2$ & Elevated \xray\ emission compared to host near-IR (NIR) emission (majority AGN populations) \\
log$(f_{\rm X}/f_{3.6\ \mu\rm m})>-1.2$ & Elevated \xray\ emission compared to host NIR emission (majority AGN populations) \\
$L_{\rm X,intrinsic}/L_{\rm 1.4\ GHz}\ge 2.4\times 10^{18}$ & Excess \xray\ emission over that expected from pure star formation (AGNs with radio data) \\
Spectroscopic AGN features & Broad emission/absorption lines; high-excitation emission lines (AGNs with quality spectra) \\
Significant \xray\ variability & Large-amplitude \xray\ variability commonly seen in AGNs (AGNs with $L_{\rm X,intrinsic}\gsim 10^{41}$ \lum) \\\hline
\end{tabular}}
\label{tab:agnid}
\end{table*}

\subsection{Scope of this review}

This paper is meant to be a relatively focused and compact one, which highlights briefly some of the recent key observational results (my apology in advance for unavoidably embedding personal flavor as well as not possibly managing to cover all relevant works), mostly from the CDFs, mainly on distant AGNs, \xray\ emitting galaxies, as well as galaxy groups and clusters, thus demonstrating the beauty and power of the CDF treasure trove in these relevant fields of study and hopefully arousing 
interest of the even broader community, in light of the recent release of the most up-to-date CDF catalogs and products (Xue \etal\ 2016; Luo \etal\ 2017) yet to be fully exploited scientifically.
For additional relevant in-depth reviews of results mainly from extragalactic \xray\ surveys in general,
I refer readers to, e.g.,  
Comastri (2004), Gilli (2004, 2013),  
Brandt and Hasinger (2005), Shankar (2009), Brandt and Alexander (2010, 2015), 
Alexander and Hickox (2012), Fabian (2012), Treister \& Urry (2012), Kormendy \& Ho (2013),
Heckman \& Best (2014), Vignali (2014), Reines \& Comastri (2016), Padovani et al. (2017), 
and the references therein.

The remainder of this paper is structured as follows:
Section~\ref{sec:demo} describes the AGN demography results
(including AGN number counts and CXRB, high-redshift AGNs and AGN \xray\ luminosity function --- XLF, highly obscured and
CT AGNs, low-mass black holes/AGNs, and significantly variable AGNs);
Section~\ref{sec:inter} highlights results on the interactions between AGNs and their host galaxies
(including AGN \xray\ luminosity versus galaxy SFR, conducive host galaxy properties
for AGN activity, Eddington ratio distribution and 
correlation between intrinsic \xray\ photon index and Eddington ratio, and
coeval growth of SMBHs and their hosts);
Section~\ref{sec:gal} introduces results on the evolution of starburst and normal galaxy \xray\ emission;
Section~\ref{sec:group} presents the census of \xray\ galaxy groups and clusters;
Section~\ref{sec:addsci} gleans some additional results not formally fit into the above sections; and finally,
Section~\ref{sec:sum} summarizes this paper and discusses future prospects.
Apparently, many of the above subtopics are intertwined by nature and should be understood collectively.

\section{AGN demography}\label{sec:demo}

\subsection{Measuring AGN number counts and resolving the CXRB}\label{sec:numcts} 

The cumulative \xray\ number counts (i.e., log$N$--log$S$), quantifying the increase of the cumulative number of \xray\ sources per unit area ($N$) as a function of 
decreasing flux ($S$), has been routinely used to characterize the extragalactic \xray\
source populations (e.g., Brandt \etal\ 2001; Rosati \etal\ 2002; 
Bauer \etal\ 2004; Kim \etal\ 2007; Georgakakis \etal\ 2008; 
Luo \etal\ 2008; Lehmer \etal\ 2012; Ehlert \etal\ 2013; Ranalli \etal\ 2013).
In particular, Lehmer \etal\ (2012) have pushed such number-counts studies to a new level
(also see, e.g., Georgakakis \etal\ 2008) 
by presenting a state-of-the-art procedure for deriving 
number counts in the 4~Ms \cdfs, which 
performs simulations to obtain source recovery functions that account for detection
incompleteness, and
implements a new Bayesian approach to obtain flux-probability distributions that
account for Eddington bias,  
thereby enabling reliable computation of number counts down to flux limits that
are typically a factor of $\approx1.5$ lower than nominal survey sensitivities.

\begin{figure*}[!t]
\centering
\includegraphics[width=160mm]{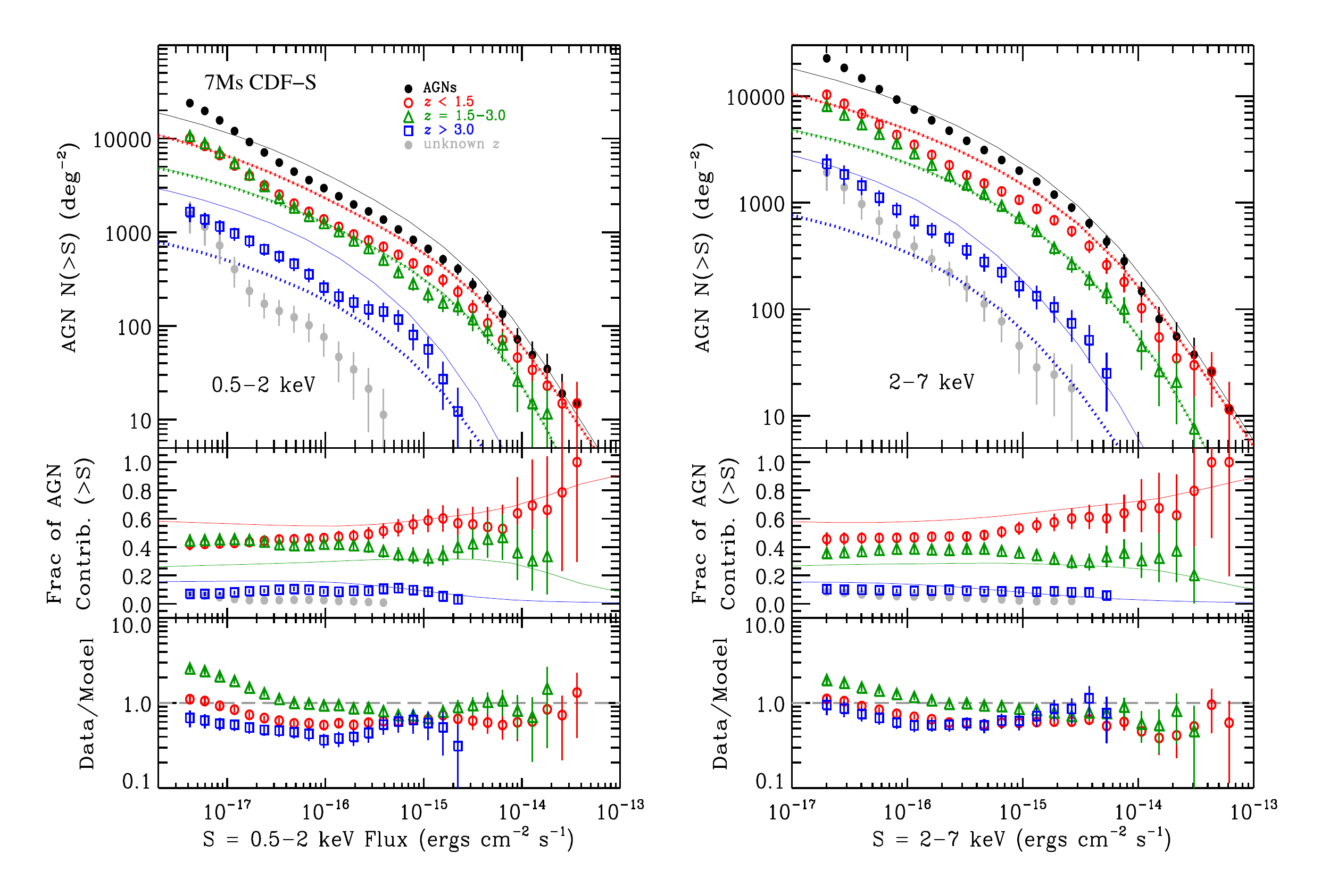}
\caption{Top panels: cumulative AGN number counts (filled black circles) for the 7~Ms \cdfs\ in the 0.5--2~keV (left) and 2--7~keV (right) bands, apportioned by AGN subsamples with $z<1.5$ (open red circles), $1.5\le z\le 3.0$ (open green triangles), $z>3.0$ (open blue squares), and unknown redshifts (filled gray circles).
Also shown are the Gilli et~al. (2007) AGN population-synthesis model predictions (in corresponding colors) either with (dotted curves) or without (solid curves) an exponential decline in the XLF at $z>2.7$. 
Middle panels: fractional contribution of each AGN subsample to the overall AGN number counts.
Bottom panels: the ratio between the AGN number-count data and the Gilli et~al. (2007) model without a declining XLF at $z>2.7$ 
(courtesy of B.~D.~Lehmer; cf. Fig.~9 of Lehmer et~al. 2012 for the 4~Ms \cdfs\ case).
\label{fig:ncts-z}}
\end{figure*}

\begin{figure*}[!t]
\centering
\includegraphics[width=160mm]{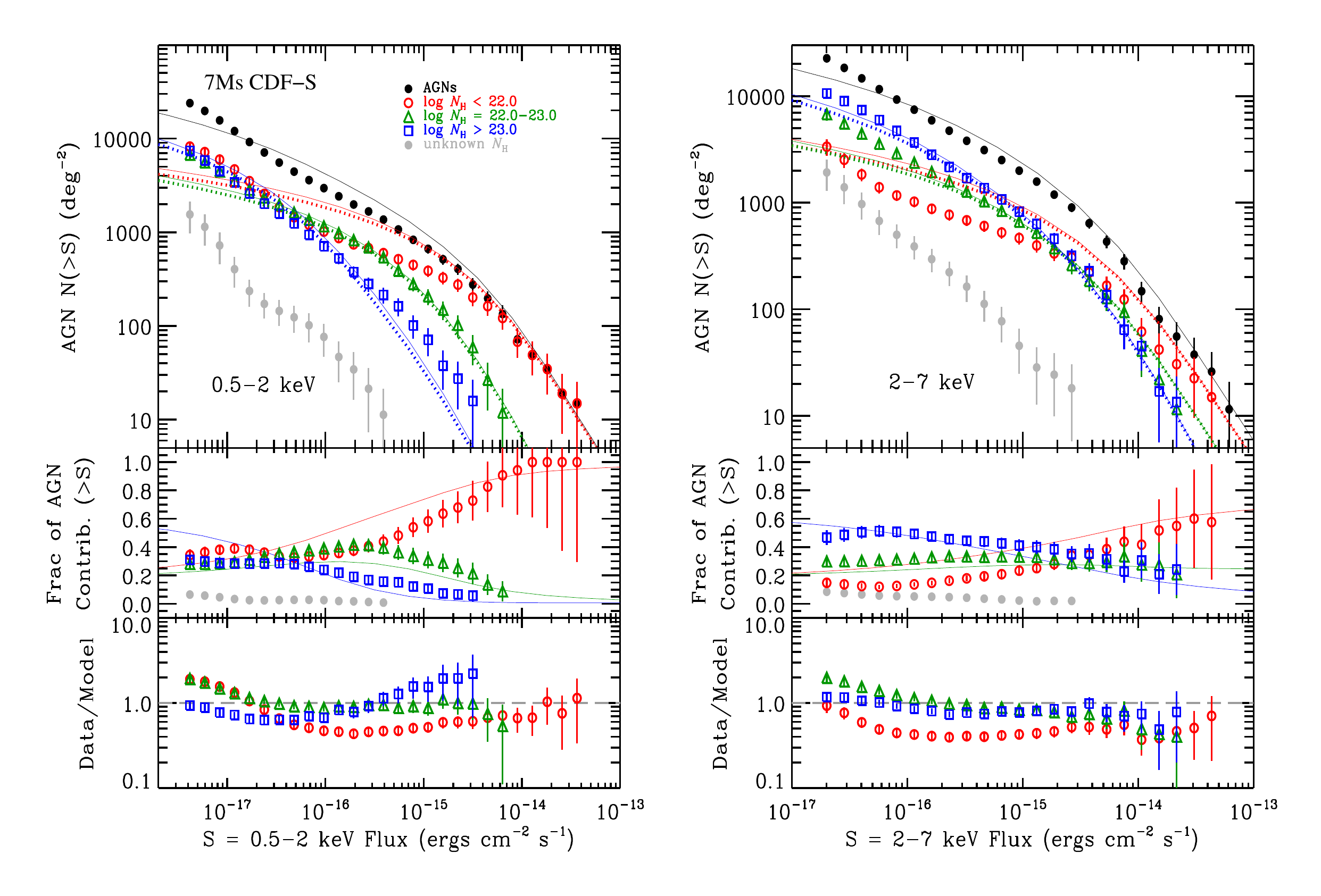}
\caption{Same as Fig.~\ref{fig:ncts-z} (i.e., cumulative AGN number counts for the 7~Ms \cdfs), but apportioned by AGN subsamples with different \nh\ values (courtesy of B.~D.~Lehmer; cf. Fig.~11 of Lehmer et~al. 2012 for the 4~Ms \cdfs\ case).
\label{fig:ncts-nh}}
\end{figure*}

\begin{figure*}[!t]
\centering
\includegraphics[width=160mm]{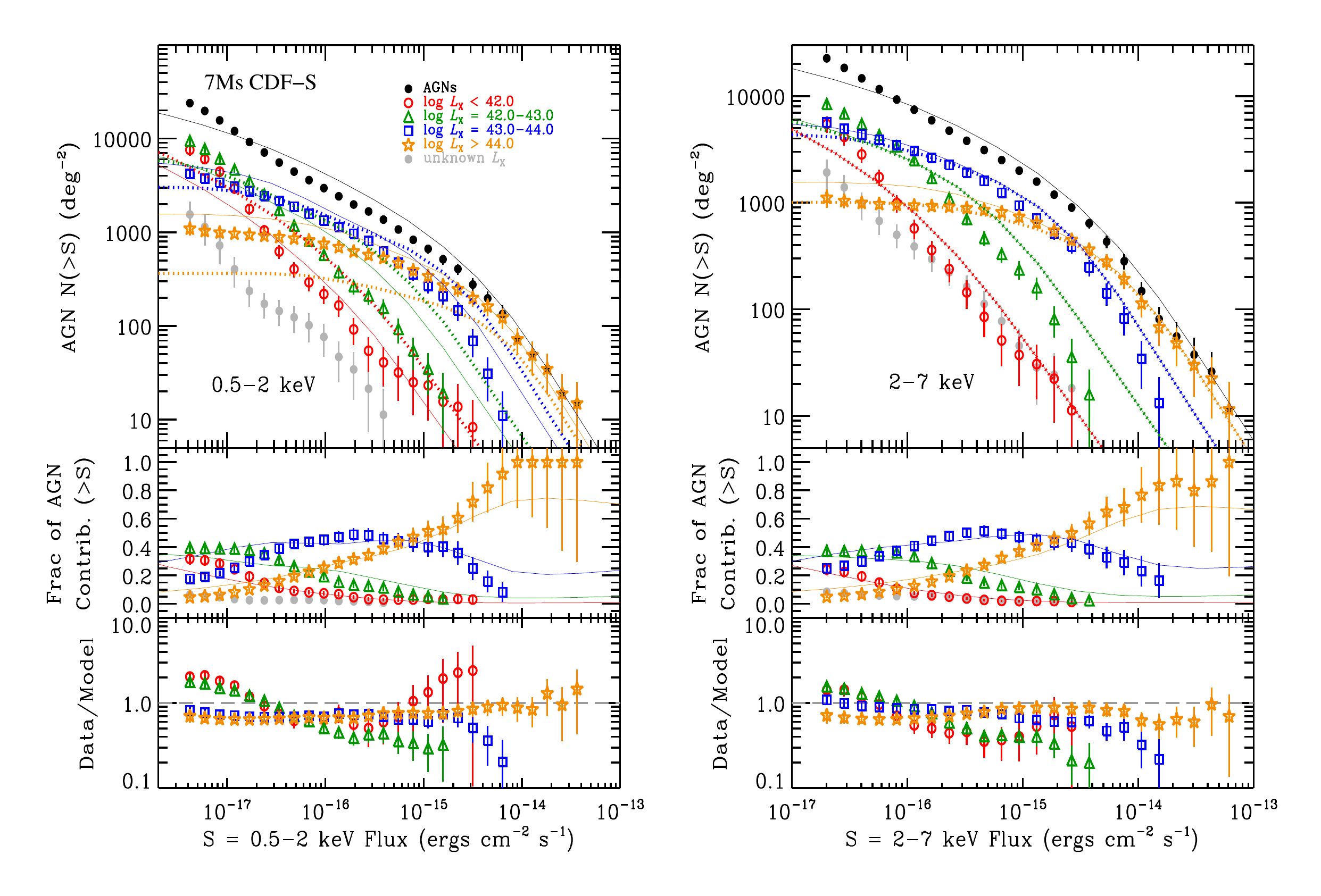}
\caption{Same as Fig.~\ref{fig:ncts-z} (i.e., cumulative AGN number counts for the 7~Ms \cdfs), but apportioned by AGN subsamples with different $L_{\rm X}$ values (courtesy of B.~D.~Lehmer; cf. Fig.~12 of Lehmer et~al. 2012 for the 4~Ms \cdfs\ case).
\label{fig:ncts-lx}}
\end{figure*}

Using the 7~Ms \cdfs\ main catalog and following the Lehmer \etal\ (2012) procedure,
Luo \etal\ (2017; see their Fig.~31) and this paper 
(see Figs.~\ref{fig:ncts-z}, \ref{fig:ncts-nh}, and \ref{fig:ncts-lx}; courtesy of B.~D.~Lehmer) 
present together the unprecedentedly sensitive \xray\ number-counts measurements,
down to $4.2\times10^{-18}$~\flux\ in the 0.5--2~keV band (soft band; SB) and 
$2.0\times10^{-17}$~\flux\ in the 2--7~keV band (hard band; HB;
cf. the sensitivity limits shown in Table~\ref{tab:cdf}), 
for the overall source population and AGNs, respectively, 
confirming and/or extending significantly many previous number-counts results
(e.g., Bauer \etal\ 2004; Lehmer \etal\ 2012): 
(1) AGNs dominate the overall number counts, 
and the number counts of AGNs, starburst and normal galaxies, and Galactic stars
in various \chandra\ passbands
can be adequately described by either double (for AGNs) or single power-law
(for galaxies and stars) functions.
(2) The galaxy power-law slope is steeper than the AGN faint-end power-law slope,
indicating the rapid rise of galaxies toward faint fluxes; and, as predicted,
at the faintest fluxes ($f_{\rm SB}\lsim6.0\times10^{-18}$~\flux) that are 
uniquely accessible to the 7~Ms \cdfs, the galaxy number counts are observed
for the first time to overtake the AGN number counts.
(3) At the SB flux limit, the AGN and galaxy densities reach record highs of
$\approx23900$ and $\approx26600$~deg$^{-2}$ (the former represents the highest AGN sky density measured reliably at any wavelength; cf. the observed
source densities in Table~\ref{tab:cdf}), respectively, which correspond to
$\approx1.0$~billion AGNs and $\approx1.1$~billion \xray\ galaxies in the entire sky.
(4) The measurements of AGN number counts, apportioned by various source properties such as
$z$, \nh, and \lx, are enabled by virtue of the exquisite \cdfs\ \xray\ and multiwavelength data (see Table~\ref{tab:data}), which reveal
that, in both SB and HB, a) AGNs with $z<1.5$ dominate the number counts while
the remainder gradually catch up toward faint fluxes (see Fig.~\ref{fig:ncts-z}), and b)
AGNs with \nh$\ <10^{22}$~cm$^{-2}$ or \lx$\ >10^{43-44}$~\lum\ dominate at the bright end 
while the others dominate or become increasingly important 
toward the faint end (see Figs.~\ref{fig:ncts-nh} and \ref{fig:ncts-lx}).
These results are broadly consistent with the 4~Ms \cdfs\ number-counts results
(Lehmer \etal\ 2012) and the expectations of the Gilli \etal\ (2007) phenomenological
AGN population-synthesis models.
One point worth noting is that 
the fraction of sources with \lx$\ <10^{42}$~\lum\ that are classified as AGNs versus galaxies 
(see Table~\ref{tab:agnid} for the AGN identification criteria)
is higher in the 7~Ms \cdfs\ compared to the 4~Ms \cdfs, and this is reflected in the faint-end number counts with an upturn and also causes the galaxy number counts to not increase as fast 
as expected for the faintest new sources
(also see, e.g., Smol\v{c}i\'{c} et al. 2017a, 2017b for similar faint-end behaviors
of radio AGNs).

With the number-counts estimates in hand, it is relatively straightforward to
resolve the CXRB into individual \xray\ sources by integrating their fluxes.
Luo \etal\ (2017) and Lehmer \etal\ (2012) find that
AGNs dominate the \hbox{0.5--7/8~keV} CXRB, and
the resolved CXRB fractions are $\approx 76\pm 4$\% and $82\pm 13$\% with the 4~Ms \cdfs\
in SB and the \hbox{2--8~keV} band (both being \hbox{$\approx1$--2\%} higher than using the 1--2~Ms CDFs),
rising to $81\pm 4$\% and $93\pm 13$\% with the 7~Ms \cdfs\ in SB and HB, 
when adopting total CXRB intensities of $(8.15\pm 0.58)\times 10^{-12}$
and $(1.49\pm 0.20)\times 10^{-11}$~\flux~deg$^{-2}$ in SB and HB,  
respectively (Hickox \& Markevitch 2006; Kim et~al. 2007).\footnote{These 
resolved CXRB fractions include the bright-end corrections,
which are typically $\approx 23$\% and $\approx 16$\% for the SB and \hbox{2--8~keV} band, 
respectively (e.g., Kim \etal\ 2007; Lehmer \etal\ 2012).}
Although it is still debated whether the resolved CXRB fraction is increasing, constant,
or even decreasing toward higher bandpasses (e.g., Moretti \etal\ 2003; Bauer \etal\ 2004;
Worsley \etal\ 2005; Georgakakis \etal\ 2008; Lehmer \etal\ 2012), 
the above resolved CXRB fractions are in broad agreement with many works
(e.g., Bauer \etal\ 2004; Hickox \& Markevitch 2006; Luo \etal\ 2011; Xue \etal\ 2012),
given the relatively large uncertainties on resolved fractions 
that primarily arise from the use of 
different methodologies and datasets 
(e.g., various AGN selection techniques, data depths, and redshift completeness levels) 
as well as CXRB intensity measurements.
Despite of many efforts, the exact normalization of the CXRB spectrum is still uncertain
by \hbox{$\approx20$--30\%}, due to a combination of several complicated factors that include    
spectral cross-calibrations, instrumental background modeling, foreground contamination by
Galactic collisional thermal plasmas, stray light contamination, 
bright-end corrections, and cosmic variance
(e.g., Bauer et al. 2004; De Luca \& Molendi 2004; 
Hickox \& Markevitch 2006; Kim et al. 2007; Luo et al. 2008;
Moretti \etal\ 2009; Ishida et al. 2011; Tsujimoto et al. 2011; 
Lehmer \etal\ 2012; Cappelluti \etal\ 2017a; Madsen et al. 2017).

There have been some emerging works that study the CXRB from somewhat different
viewpoints, either by pushing the studies of resolving the CXRB to higher energy bands due to
the advent of {\it NuSTAR} (e.g., Ballantyne \etal\ 2011; 
Harrison \etal\ 2016)\footnote{{\it NuSTAR} resolves 
$\approx33$\%--39\% of the CXRB in the \hbox{8--24 keV} band
(depending on the adopted \hbox{20--50 keV} CXRB intensity, and with
an additional $\approx 5$\% statistical uncertainty), 
directly identifying CT AGNs with $N_{\rm H}\lsim 10^{25}$~cm$^{-2}$ (Harrison \etal\ 2016).
Although {\it NuSTAR} can probe much harder (typically a factor of \hbox{$\approx 3$--4}) X-rays than \chandra\ and {\it XMM-Newton},
it is still not properly reaching the CXRB peak.}
or by examining the spectrum and/or origin of the unresolved CXRB 
(e.g., Luo \etal\ 2011; Moretti \etal\ 2012; Salvaterra \etal\ 2012; Xue \etal\ 2012; Cappelluti \etal\ 2017a).  
These studies, in conjunction with other studies, 
help constrain the likely existence, properties, and evolution of 
a missing yet potentially large population of highly obscured (i.e., $N_{\rm H}\gsim 3\times 10^{23}$~cm$^{-2}$) or even CT AGNs 
(CTAGNs hereafter; see Section~\ref{sec:highlyobs}), 
facilitating the refinement of AGN population-synthesis models (e.g., Gilli \etal\ 2007;
Treister \etal\ 2009a), and revealing the various components constituting the unresolved CXRB
that include likely contributions from low-luminosity AGNs (LLAGNs),  
highly obscured AGNs, CTAGNs, non-active galaxies,
and hot gas in groups and clusters (see Section~\ref{sec:group}).

\subsection{Constraining the high-redshift AGN subpopulation and AGN XLF}\label{sec:highz}

AGNs within the first cosmic structures at $z\approx4$--8 and beyond are of
perennial interest because their observations have serious implications for, e.g.,
the primordial BH seeds in the early Universe, 
the overall SMBH growth history (including AGN mass function and luminosity function),
the origin and evolution of correlations between SMBH mass ($M_{\rm BH}$) and galaxy properties,
and the role played by AGNs during intergalactic medium (IGM) heating and reionization
(see, e.g., Reines \& Comastri 2016; Brandt \& Vito 2017 for a review). 
Given that \hbox{X-rays} are much more penetrating and much less affected by galaxy dilution (see Section~\ref{sec:utility}) than other wavelengths, 
\xray\ surveys are therefore critical for assessing such high-redshift SMBH growth.

Indeed, many recent XLF works have been carried out utilizing the CDF data
(note that the 7~Ms \cdfs\ is sufficiently sensitive to reach about 20--3 times below the knee of AGN XLF at $z=0.5$--4) and other \xray\ observations, finding that
there is an exponential decline in the number density of luminous ($L_{\rm X}\gsim 10^{44}$~\lum) AGNs with $z\gsim 3$
(e.g., Barger \etal\ 2003a; Cristiani \etal\ 2004; Silverman \etal\ 2008a; Brusa \etal\ 2009a;
Civano \etal\ 2011; Fiore \etal\ 2012a; Vito \etal\ 2013, 2014a, 2017; Kalfountzou \etal\ 2014;   
Ueda \etal\ 2014; Georgakakis \etal\ 2015; Marchesi \etal\ 2016).
However, the situation is not as clear for lower-luminosity AGNs ($L_{\rm X}\approx10^{43-44}$~\lum), largely owing to 
the faintness, likely heavy obscuration, survey incompleteness, limited sample size, and challenging multiwavelength
follow-up of such sources.
There generally also appears to be a decline in the space density
of $z>3$ moderate-luminosity AGNs, albeit with large uncertainties
(in particular, the constraints on $z>4$ AGNs are even looser due to poorer statistics;
e.g., Fiore \etal\ 2012a; Vito \etal\ 2013, 2014a, 2017;     
Kalfountzou \etal\ 2014; Ueda \etal\ 2014; Georgakakis \etal\ 2015; Marchesi \etal\ 2016).
The space density of such AGNs remains uncertain by more than an order of magnitude
(e.g., Gilli et al. 2011a), effectively having no much discriminating power between
the predictions of the theoretical models that
satisfy other existing observational constraints (e.g., Salvaterra \etal\ 2007; Marulli \etal\ 2008;
Volonteri \& Stark 2011; Habouzit \etal\ 2016; Volonteri \& Reines 2016). 
However, even a small number of such objects, if detected, can remedy this situation and provide critical leverage in modeling early SMBH growth. 
In the long run,
the robust detection of a decent number 
(i.e., a few tens or even more)
of AGNs down to $L_{\rm X}\approx10^{43}$~\lum\ beyond
$z\approx3$--4 would provide invaluable insight on the formation mechanisms responsible for the first quasars
and the early co-evolution history, which has to call for future large \xray\ observatories
that are capable of efficiently carrying out ultra-sensitive surveys over wide sky areas.
Nevertheless, stacking analyses, instead of direct detections, have already proved to be a beneficial attempt along this direction.

\begin{figure}[!t]
\centering
\includegraphics[width=85mm]{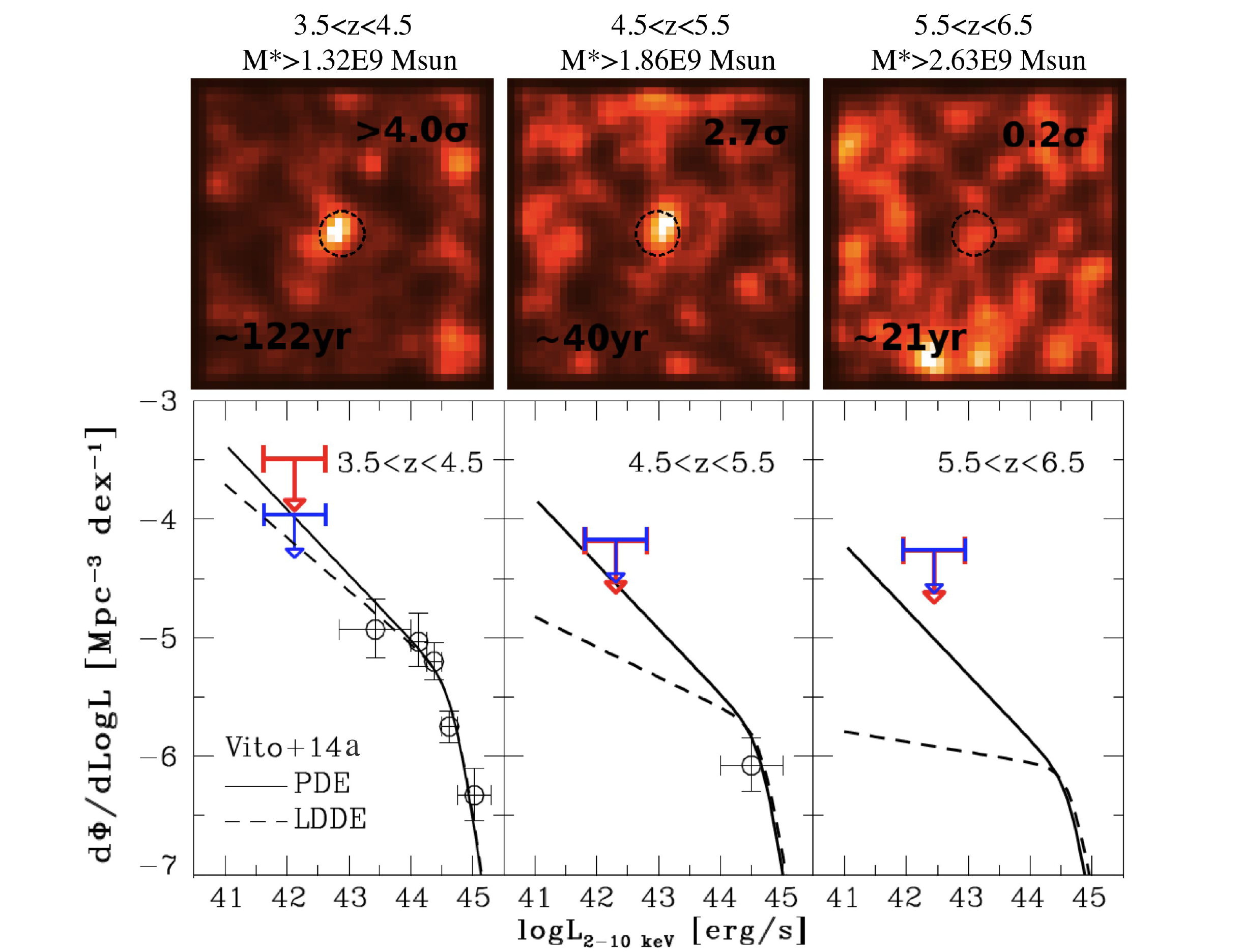}
\caption{(Top) Smoothed stacked $40\times40$ pixel \xray\ images 
(observed SB; each image is normalized to its maximum pixel value) 
of optically selected 
and individually \xray-undetected high-redshift galaxies in the 7~Ms \cdfs, with various redshift
bins, stellar mass ranges, stacked signal significances (e.g., the 2.7~$\sigma$ emission at $4.5<z<5.5$ is a detection
of the highest significance ever in such a redshift range), and stacked effective exposures annotated. Dashed circles indicate the positions of the stacked galaxies.
(Bottom) Upper limits on the faint-end AGN XLF for the cases of maximum AGN contribution 
(i.e., no SF contribution to the stacked signals; red symbols)
and AGN+SF contribution (blue symbols) at high redshifts, indicating a fairly flat faint-end XLF slope at high redshifts.
For comparison, the pure density evolution (PDE) and luminosity-dependent density evolution
(LDDE) models and binned XLF (denoted as black open circles) presented in
Vito \etal\ (2014a) are also shown.
 Adapted from Vito \etal\ (2016).
\label{fig:highz}}
\end{figure}

Recently, Vito \etal\ (2016) utilize both the 7~Ms \cdfs\ and ultradeep CANDELS/GOODS-S data
to sensitively measure the total \xray\ emission from a sample of 2076 $3.5<z<6.5$ optically
selected and individually \xray-undetected galaxies
(see Table~\ref{tab:cdf} for the redshift-range information 
of individually \xray-detected sources),
with a sophisticated stacking procedure
that is validated by simulations.
They detect high-significance ($>3.7\sigma$) stacked \xray\ emission from massive galaxies at
$3.5<z<4.5$, $2.7\sigma$ emission from those at $4.5<z<5.5$ (99.7\% confidence level; highest significance ever in such a redshift 
range), and no
significant signal from those at $5.5<z<6.5$ (see the top panel of Fig.~\ref{fig:highz}; also see, e.g., Willott 2011;  Cowie, Barger \& Hasinger 2012; Fiore \etal\ 2012b; Basu-Zych \etal\ 2013). 
They find that the detected \xray\ emission is likely dominated by
high-redshift XRB populations (also see Cowie \etal\ 2012), by comparing it with the expected
high-redshift XRB emission that is extrapolated from lower-redshift results
assuming a range of XRB model prescriptions (e.g., Ranalli \etal\ 2003; Fragos \etal\
2013a; Lehmer \etal\ 2016; see Section~\ref{sec:gal}), and also by comparing both the total SFRs and SFR densities 
of the stacked galaxies with previous UV-based and SED fitting results of galaxies.
Therefore, they conclude that
(1) the continuous low-rate SMBH growth in individually \xray-undetected galaxies makes negligible
contribution to cosmic SMBH mass assembly at $3.5<z<6.5$, when compared to that in AGNs detected by
deep \xray\ surveys; and
(2) the observational constraints on the faint-end ($L_{\rm X}\approx10^{42}$~\lum) AGN XLF  
at $3.5<z<6.5$ are achieved for the first time,
which indicate a fairly flat faint-end XLF slope (see the bottom panel of Fig.~\ref{fig:highz}),
confirming previous results (e.g., Barger \etal\ 2003a; Ueda \etal\ 2014) 
and extending them down to lower \xray\ luminosities.

Collectively, the AGN XLF studies over a broad stretch of redshift, including those at high redshifts
that are based on direct detections and aided by
the stacking-based AGN XLF constraints such as the aforementioned Vito \etal\ (2016) work,
reveal that
(1) there is an anti-hierachical ``cosmic downsizing'' behavior of AGNs, i.e.,   
the number density of powerful quasars peaks at an earlier cosmic time 
than that of lower-luminosity AGNs (e.g., Cowie \etal\ 2003; Barger \etal\ 2005; 
Hasinger, Miyaji, \& Schmidt 2005; Silverman \etal\ 2008a; 
Yencho \etal\ 2009; Aird \etal\ 2010; Ueda \etal\ 2014; Miyaji \etal\ 2015);
(2) the peak AGN emissivity (i.e., the AGN comoving bolometric luminosity density 
that is in units of \lum~Mpc$^{-3}$)
occurs at $z\approx 1.8$ for AGNs with a broad range of
bolometric luminosity $L_{\rm bol}=10^{43-48}$~\lum\ (see, e.g., Fig.~20 of Ueda \etal\ 2014); 
(3) AGNs are unlikely to dominate reionization at $z\gsim6$,     
with stars playing a leading role instead (e.g., Robertson \etal\ 2010, 2013; Haardt \& Madau 2012; Grissom, Ballantyne \& Wise 2014; Cappelluti \etal\ 2016; Vito \etal\ 2016; Ricci \etal\ 2017; also
see Section~\ref{sec:gal}; 
but see Giallongo \etal\ 2012, 2015); and
(4) the luminosity-dependent density evolution (LDDE) and luminosity and density evolution (LADE) 
XLF models describe the observational data well, 
and the LDDE models appear to be further favored by some high-redshift constraints,
yet with further testing needed due to insufficient statistics at high redshifts 
(e.g., Miyaji \etal\ 2000, 2001, 2015; Aird \etal\ 2010; Ueda \etal\ 2014; 
Vito \etal\ 2014a; Georgakakis \etal\ 2015; Fotopoulou \etal\ 2016).
The basic nature of the above AGN XLF related results (e.g., cosmic downsizing) appears robust, despite of
many relevant details yet to be further worked out.

\subsection{Unveiling the highly obscured and CT AGN subpopulation}\label{sec:highlyobs}

\begin{figure}[!t]
\centering
\includegraphics[width=75mm]{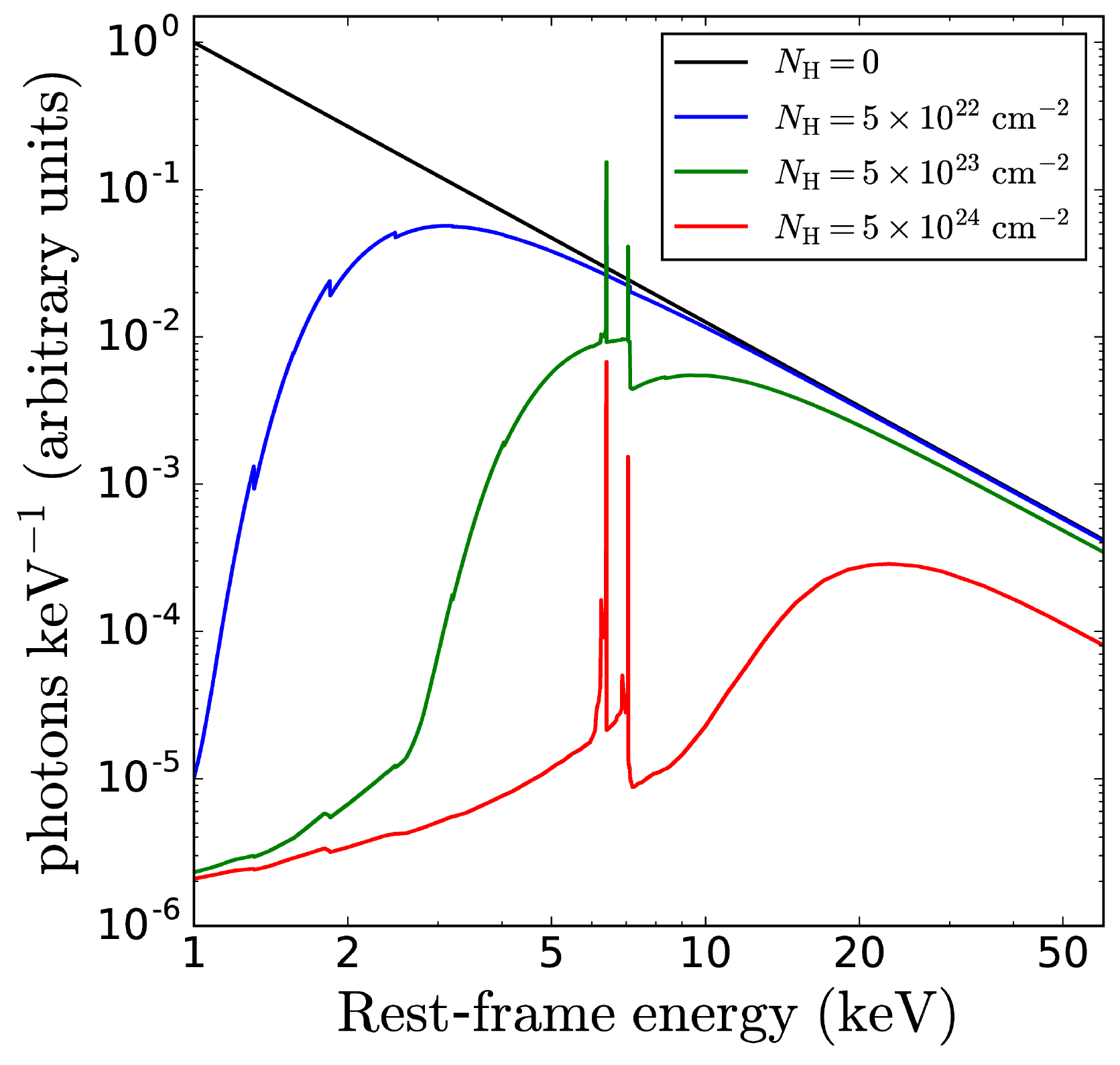}
\caption{Model spectra of AGNs with different obscuration levels. The black, blue, green, and red curves correspond to unobscured (\nh=0), moderately obscured ($N_{\rm H}=5\times 10^{22}$~cm$^{-2}$), highly obscured ($N_{\rm H}=5\times 10^{23}$~cm$^{-2}$), and CT ($N_{\rm H}=5\times 10^{24}$~cm$^{-2}$) AGNs, respectively. The spectra are generated using the MYTorus model with
$\Gamma=1.9$, inclination angle=90~deg, and the same normalization (Murphy \& Yaqoob 2009; courtesy of G.~Yang).
\label{fig:absorption}}
\end{figure}

Hunting for the highly obscured AGNs, CTAGNs in particular, has been a longstanding and challenging quest
for the AGN demography work.
CTAGNs are of special interest given a number of arguments (see, e.g., Comastri 2004; 
Georgantopoulos 2013; Gilli 2013; Vignali 2014 for a review):
(1) There has been growing observational evidence that a substantial fraction of AGNs are
obscured by CT gas both locally (e.g., Maiolino \etal\ 1998; Risaliti, Maiolino, \& Salvati 1999; 
Matt \etal\ 2000) and in the distant universe (e.g., Dwelly \& Page 2006; Tozzi \etal\ 2006; 
Georgantopoulos \etal\ 2009; Merloni \etal\ 2014; Buchner \etal\ 2014, 2015; Lanzuisi \etal\ 2015a; Liu \etal\ 2017; also see
Fig.~\ref{fig:ncts-nh}).
(2) The existence of the CTAGN subpopulation is required by the AGN population-synthesis models for
the CXRB in order to account for the intensity peak of the CXRB spectrum at $\approx20$--30~keV,
and CTAGNs are predicted to be as abundant as moderately obscured AGNs
(e.g., Gilli \etal\ 2007, 2013).
(3) The absorbed energy of CTAGNs at short wavelengths (including optical, UV, and \hbox{X-rays})
is thermally reprocessed by, e.g., the torus, and eventually reemitted in FIR, making them a potential contributor to the cosmic IR background
(e.g., Shi \etal\ 2013a, 2013b).
(4) Finally, distant CTAGNs are believed to represent a crucial phase in SMBH/galaxy co-evolution
models, during which large amounts of gas is funneled to the center and induces both intensive
obscured accretion and powerful star formation as a result of the merger process,
and then feedback processes likely take over by self-regulating SMBH growth and 
quenching star formation (e.g., Page \etal\ 2004; Granato \etal\ 2006; Hopkins \etal\ 2006a; Menci \etal\ 2008; Alexander
\& Hickox 2012).
Despite their importance, the majority of distant CTAGNs escape even from the deep
\xray\ surveys and still remain largely elusive due to their very nature of extreme obscuration
(see Fig.~\ref{fig:absorption} for a demonstration); therefore,
there is no good knowledge even about their space density and cosmological evolution
(e.g., Aird \etal\ 2015a; Buchner \etal\ 2015).
Fortunately, the situation is gradually improving as many CTAGN identification methods have been
developed and many ultradeep \xray\ (e.g., the CDFs) and multiwavelength surveys have been performed
along the quest for CTAGNs.

The signatures of CT emission can be revealed in many ways (see Table~\ref{tab:ctagn} for a brief summary; some of these techniques are closely related but with different emphases), e.g.:
(1) \xray\ spectroscopy is the only unambiguous way to identify bona-fide CTAGNs where deep \xray\ surveys play a key role. With high-quality \xray\ spectra, CT \nh\ values can be determined reliably
through careful spectral modeling,
immediately unveiling the CT nature of sources (e.g., Comastri 2004; Comastri \etal\ 2011).
(2) The presence of a strong Fe~K$\alpha$ line complex (typically with a $\gsim1$~keV 
line equivalent width) around rest-frame 6.4--7~keV in an AGN \xray\ 
spectrum is highly indicative of the source likely being CT (see Fig.~\ref{fig:absorption};
e.g., Comastri 2004; Comastri \etal\ 2011;
Georgantopoulos \etal\ 2013).
(3) An AGN with a characteristic reflection-dominated \xray\ spectrum is almost 
guaranteed to be a CTAGN; such a spectrum is featured by a broad Compton hump peaking 
at \hbox{$\approx 20$--30~keV}, rapid declines toward both low energies (due to absorption) and 
high energies (due to Compton down-scattering), and a power-law shape in the \hbox{$\approx 2$--10~keV} band
with a strong Fe~line complex atop (typically having equivalent width $>1$~keV; e.g., Comastri 2004; Comastri \etal\ 2011).
(4) In case of an \xray\ spectrum having limited counts or not resorting to detailed spectral fitting, 
a (very) large \xray\ hardness ratio or a (very) small $\Gamma_{\rm effective}$ (see Table~\ref{tab:agnid})
can be crudely used to indicate the likely CT nature of the source (e.g., Alexander \etal\ 2011; Gilli \etal\ 2011); furthermore, using \xray\ colors or a combination of \xray\ hardness and \xray-to-MIR flux ratio can select CTAGNs more
efficiently than using a single hardness ratio (e.g., Iwasawa \etal\ 2012; Severgnini \etal\ 2012).
(5) An AGN having excess IR emission (owing to the reemission of the absorbed energy) 
compared to the typical IR/optical emission level of galaxies/non-CT AGNs 
are likely CTAGN candidates; the key to this category of ``IR-excess'' methods is to reduce 
contamination from starburst and normal galaxies (see, e.g., Daddi \etal\ 2007, Alexander \etal\ 2011, 
and Luo \etal\ 2011
for a SFR-excess method; and Sacchi \etal\ 2009 for a high MIR/optical flux-ratio method).
(6) \xray\ stacking takes benefit of \chandra's superb angular resolution,
thus allowing for exploring considerably deeper \xray\ fluxes and revealing the average properties
of the stacked sample of CTAGN candidates (e.g., Luo \etal\ 2011; Xue \etal\ 2012).
(7) Some well-calibrated luminosity ratios appear to have good discriminating power between CT and non-CT AGNs, e.g.: luminosity ratios between \xray\ and high-ionization emission lines from the narrow-line region (thus not suffering from torus extinction; e.g., \hbox{[O III]}~5007$\AA$ and \hbox{[Ne V]}~3426$\AA$;   
e.g., Vignali \etal\ 2006; Ptak \etal\ 2006; Gilli \etal\ 2010; Mignoli \etal\ 2013), and 
luminosity ratios between \xray\ and 6/8/12~$\mu$m band (e.g., Lutz \etal\ 2004; Fiore \etal\ 2009;
Alexander \etal\ 2011; Georgantopoulos \etal\ 2011b; Mateos \etal\ 2015).
(8) An AGN with a very deep Si feature at 9.7~$\mu$m that has an optical depth of $\tau >1$
(i.e., $\tau_{\rm 9.7\ \mu m}>1$) in its IR spectrum is very likely to be a CTAGN (e.g., Georgantopoulos 
\etal\ 2011a).   
(9) A newly-developed metric that utilizes the $>10$~keV spectral curvature to identify CTAGNs with 
{\it Swift}/{\it NuSTAR} data proves to be quite effective (Koss \etal\ 2016).
(10) If the broadband SED decomposition requires a strong AGN torus emission, this can then serve as
a supporting evidence of the source likely being CT (e.g., Lanzuisi \etal\ 2015b; Del~Moro \etal\ 2013, 2016).
Among various selection methods for CTAGNs, \hbox{X-rays} represent a fundamental and powerful probe through direct \xray\ spectroscopy and reliable stacking analysis.
Whenever data allow, it would be beneficial to cross validate the identified CTAGN candidates
using as many methods as possible and meanwhile keep in mind the biases and limitations
(e.g., incompleteness and contamination) associated with each method (see the above relevant references for details).

\begin{table}[!t]
\caption{Techniques to Identify CTAGN Candidates}
\centering
\resizebox{0.48\textwidth}{!}{%
\begin{tabular}{rl}\hline\hline
(1) & CT \nh\ values derived from \xray\ spectral fitting \\
(2) & Strong Fe K$\alpha$ line complex \\
(3) & Characteristic \xray\ reflection spectrum/component \\
(4) & Large \xray\ hardness ratio + \xray\ colors  \\
(5) & IR-excess emission \\
(6) & \xray\ stacking \\
(7) & Some well-calibrated luminosity ratios \\
(8) & Deep 9.7~$\mu$m Si feature with $>1$ optical depth \\
(9) & Spectral curvature above 10 keV \\
(10) & Broadband SED decomposition \\\hline
     & See main text for explanations and references \\
     & therein for caveats.
\end{tabular}}
\begin{flushleft}
\end{flushleft}
\label{tab:ctagn}
\end{table}

\begin{figure}[!t]
\centering
\includegraphics[width=85mm]{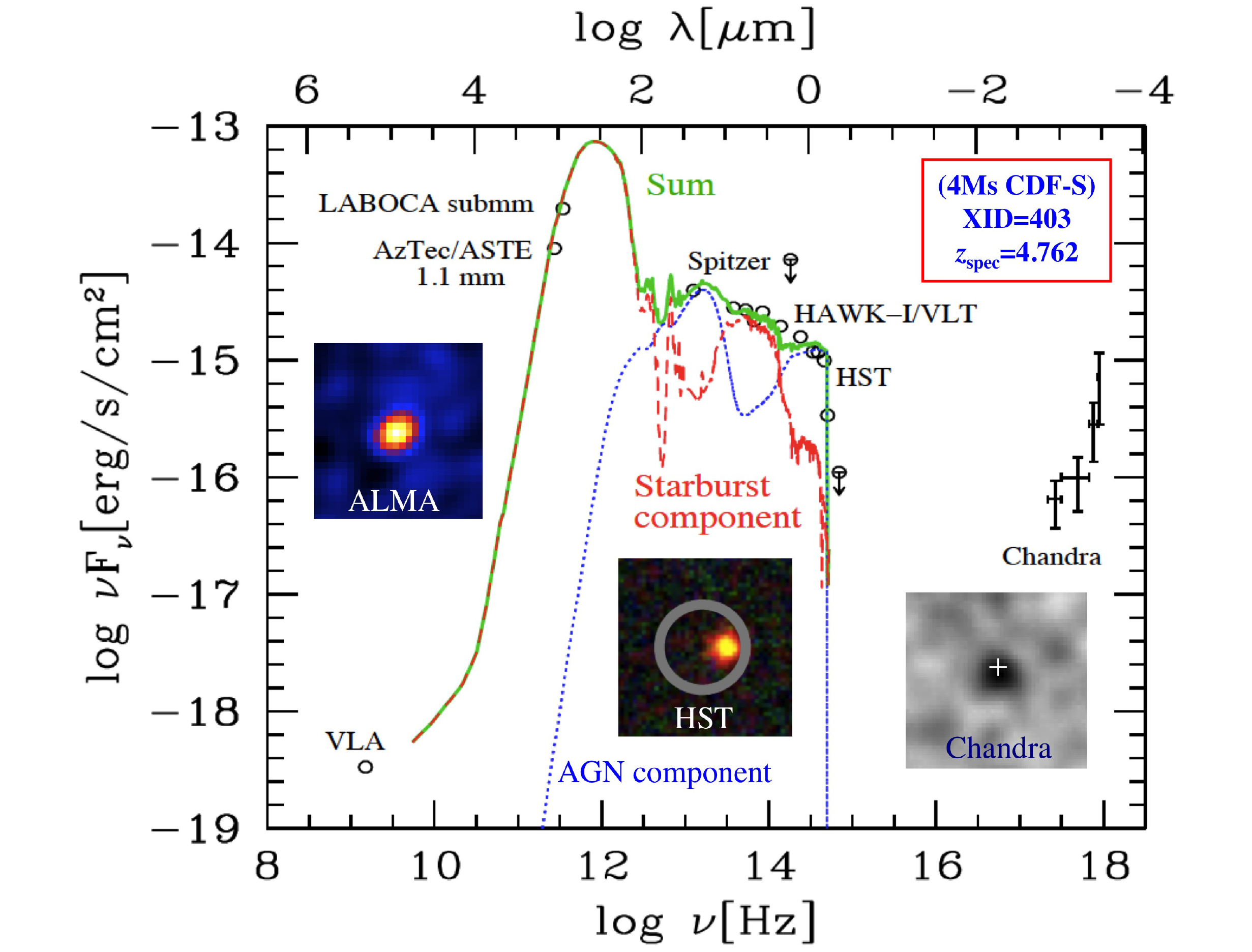}
\caption{Broadband observed SED of a $z_{\rm spec}=4.762$ CTAGN 
(i.e., the most distant bona-fide CTAGN discovered so far that has
XID=403 in the 4~Ms \cdfs) 
and its possible decomposition into a galaxy component (i.e., 
shifted Arp~220 template; red dashed curve) and an AGN component (i.e., sum of a torus and a scattered component; blue dotted curve)
with their sum shown as the green solid curve.
The inset ALMA image is from Gilli \etal\ (2014). 
Adapted from Gilli \etal\ (2011b).
\label{fig:403}}
\end{figure}

\begin{figure*}[!t]
\centering
\includegraphics[width=140mm]{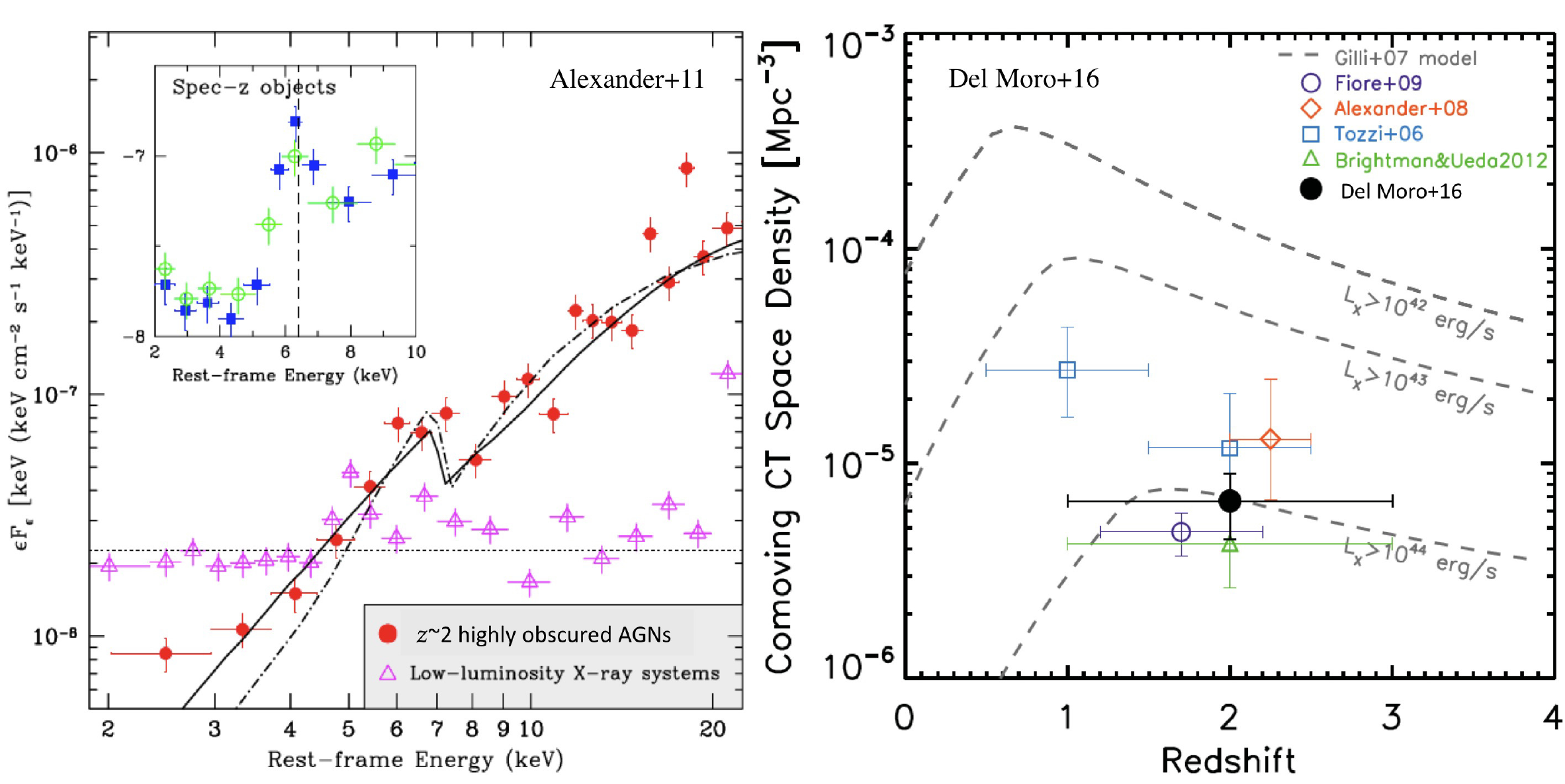}
\caption{(Left) Composite rest-frame 2--20~keV spectra for the $z\approx 2$ highly obscured AGNs 
(filled circles) and low-luminosity \xray\ sources (open triangles) as compared to 
an unabsorbed power-law model (dotted line; intrinsic photon index $\Gamma=2.0$), a pure reflection model 
(solid curve; $\Gamma=1.7$), and a best-fit model to a reflection-dominated local AGN (dot-dashed curve). 
The inset highlights the stacked \xray\ spectra of the highly obscured AGNs with \zs\ 
(filled squares: all sources; open circles: all sources except one with an individually identified Fe K$\alpha$ line).
Several lines of evidence points to the existence of CTAGNs in the studied sample (see Table~\ref{tab:ctagn}). 
(Right) Comoving space densities of CT quasars/AGNs 
from Del~Moro \etal\ (2016; black filled circle),
Brightman \& Ueda (2012; green open triangle),
Fiore \etal\ (2009; purple open circle),
Alexander \etal\ (2008a; orange open diamond),
and Tozzi \etal\ (2006; blue open squares), 
in comparison with the Gilli \etal\ (2007) AGN population-synthesis model predictions
for AGNs with various \xray\ luminosities.
(Left) adapted from Alexander \etal\ (2011) and (Right) adapted from Del~Moro \etal\ (2016).
\label{fig:bzk}}
\end{figure*}

There have been many works that involve the use of one or more of the above methods and prove their utility.
For instance, Gilli \etal\ (2011) discover a \zs=4.762 CTAGN in the 4~Ms \cdfs\ that is confirmed
by \xray\ spectral analysis, thus being the most distant bona-fide CTAGN so far.
This source appears to be caught during a major coeval episode of SMBH accretion 
($L_{\rm 2-10\ keV}\approx 2.5\times 10^{44}$~\lum) and stellar mass assembly 
(being a submillimeter galaxy (SMG) with SFR$\approx 1000M_\odot$~yr$^{-1}$ that is
later confirmed by ALMA observations; Coppin \etal\ 2009; Gilli \etal\ 2014)
at early times (see Fig.\ref{fig:403}).
More interestingly, several lines of evidence point to the existence of outflowing winds from the 
central SMBH, hinting at AGN feedback at work (Gilli \etal\ 2014).
Finding more such AGNs and constraining their number density would be crucial to 
reconstruct the early co-evolution history.
Alexander \etal\ (2011) identify 11 highly obscured AGNs among a sample of $z\approx2$
\hbox{$K<22$~mag} $BzK$-selected galaxies through \xray\ spectral analysis in the 4~Ms \cdfs. 
They find that, among these 11 sources, some display excess IR emission and some prefer a pure
reflection model; furthermore, the stacked \xray\ spectrum of all the sources has a reflection-dominated shape and a strong Fe~line feature, indicating CT absorption in some sources (see the left panel of Fig.~\ref{fig:bzk}); and the estimated lower-limit space-density constraints of highly-obscured AGNs/CTAGNs (i.e., $\gsim f\times 4\times 10^{-5}$~Mpc$^{-3}$
for CTAGNs with $L_{\rm 2-10\ keV}\gsim 10^{43}$~\lum\ and $z\approx 2$, where $f\approx 10$--50\% is
the likely CTAGN fraction within the studied sample) are already comparable to
the results from some other works (e.g., Tozzi \etal\ 2006; Alexander \etal\ 2008;
Fiore \etal\ 2009; Brightman \& Ueda 2012; Del~Moro \etal\ 2016) and the predictions from AGN population-synthesis models, albeit with large uncertainties on both observations
(see the right panel of Fig.~\ref{fig:bzk}) and models 
(e.g., Gilli \etal\ 2007; Treister \etal\ 2009a).  
It appears that the space density of $z\approx 2$ CTAGNs is comparable to that of $z\approx 2$ unobscured AGNs, suggesting a non-negligible contribution of such CTAGNs to the overall SMBH growth at $z\approx 2$.

Most recently, Liu \etal\ (2017) present a systematic spectral analysis for a sample of 276
HB-selected brightest AGNs in the 7~Ms \cdfs. 
After correcting for sample selection biases (e.g., incompleteness and Eddington bias),
they find that: 
(1) The intrinsic \nh\ distribution varies significantly across different redshift ranges
(up to $z\approx 5$)
as a result of strong dependence of \nh\ on both \lx\ and $z$; the overall \nh\ distribution
peaks at $N_{\rm H}\approx 10^{23.5-24}$~cm$^{-2}$, higher than the peak value of 
$N_{\rm H}\approx 10^{23.1}$~cm$^{-2}$ derived with 1~Ms \cdfs\ (Tozzi \etal\ 2006; also see, e.g.,
Burlon \etal\ 2011; Brightman \& Nandra 2011; Buchner \etal\ 2015), mainly 
due to more highly obscured sources being revealed by the 7~Ms \cdfs.
(2) The strong evolution of obscured AGN fraction with $z$ can be formulated as
$f_{\rm obscured}=(0.43\pm 0.07)(1+z)^{0.59\pm 0.12}$; at $z>2$, $f_{\rm obscured}$ 
shows only a weak evolution, likely being saturated (see also, e.g.,
Hasinger 2008; Vito \etal\ 2014a; Ueda \etal\ 2014); and the $f_{\rm obscured}$ measures
are higher than previous ones (e.g., Iwasawa \etal\ 2012; Vito \etal\ 2014a).
(3) After combining the 7~Ms \cdfs\ data with the wider and shallower C-COSMOS data (Lanzuisi \etal\ 2013), it is clear that, at any given (small) redshift bin within the range of 
$0.3<z<4.0$, the average \nh\ (or $f_{\rm obscured}$) decreases with \lx; and
at any given (small) \lx\ bin within the range of  $42<{\rm log}(L_{\rm X})<45$, 
the average \nh\ (or $f_{\rm obscured}$) increases with $z$. 
(4) A total of 22~CTAGNs are identified, corresponding to $\approx 8$\% of the studied sample;
such a CTAGN fraction appears comparable to or slightly lower than CTAGN fractions of 
$\approx 7$--20\% among hard \xray\ selected samples (e.g., Malizia \etal\ 2009; Burlon \etal\ 2011; Vasudevan \etal\ 2013; Akylas \etal\ 2016).
The above dependence of $f_{\rm obscured}$ on \lx\ result presented by Liu \etal\ (2017) confirms many previous works
(Treister and Urry 2006; Hasinger 2008; Brightman \& Nandra 2011; Burlon \etal\ 2011; 
Lusso \etal\ 2013; Brightman \etal\ 2014; 
Merloni \etal\ 2014; Ueda \etal\ 2014) and extends reliably to lower \lx.
Surprisingly, wide-field IR surveys find that $f_{\rm obscured}$ might rise substantially again 
at $L_{\rm bol}\approx 10^{47}$~\lum\ (e.g., Stern et al. 2014; Assef et al. 2015).
Moreover, the dependence of $f_{\rm obscured}$ on $z$ result given by Liu \etal\ (2017) also confirms many previous works
(e.g., La Franca \etal\ 2005; Ballantyne \etal\ 2006; Tozzi \etal\ 2006; Treister and Urry 2006; Hasinger 2008; Treister \etal\ 2009a; Hiroi \etal\ 2012; Iwasawa \etal\ 2012; Vito \etal\ 2013, 2014a; Brightman \etal\ 2014; Merloni \etal\ 2014; Ueda et al. 2014; Buchner \etal\ 2015) and extends reliably to lower \lx.

An improved demography picture for the highly obscured and CT AGNs that better reveals their space density 
and cosmological evolution can be obtained, by combining and reconciling all the bits and pieces of works 
already present in the literature, which are based on direct \xray\ spectroscopy, reliable \xray\ stacking analysis, and/or various multiwavelength diagnostics (e.g., Daddi \etal\ 2007;
Alexander \etal\ 2008a, 2011; Fiore \etal\ 2008, 2009; Treister \etal\ 2009b; 
Gilli \etal\ 2011; Luo \etal\ 2011;
Brightman \& Ueda 2012; Vignali \etal\ 2014; Del~Moro \etal\ 2016),
as well as {\it NuSTAR}'s $>10$~keV constraints (e.g., Alexander \etal\ 2013; Del~Moro \etal\ 2014; Brightman \etal\ 2015; Koss \etal\ 2016).
However, there is still a long way to a complete understanding of the CTAGN demography
given the currently small sample of distant bona-fide CTAGNs; and
the radical change of such a situation would have to rely on new powerful
and preferably hard \xray\ facilities.

Last but not least,  
accumulating evidence shows that the AGN obscuring material may not be necessarily related to
the pc-scale dusty torus as commonly accepted; instead, it may be attributed to a compact starburst
region (e.g., Gilli \etal\ 2014),
the diffuse interstellar medium in the host galaxy (with \nh\ likely up to $10^{22-23.5}$~cm$^{-2}$;
e.g., Simcoe \etal\ 1997; Goulding \etal\ 2012; Buchner \& Bauer 2017; Lanzuisi \etal\ 2017),
or 100~pc-scale dust filaments 
(e.g., Prieto \etal\ 2014).\footnote{Most recently, 
Maiolino \etal\ (2017) find observational evidence for star formation 
(SFR$\approx 15\ M_\odot$~yr$^{-1}$) occurring within
a prominent galactic outflow, which is \hbox{$\approx 7$--9~kpc} away from and perhaps driven 
by the southern obscured AGN in a $z=0.0448$ merging system, likely
playing some role in obscuring the AGN.} 
Such obscuring material existing at relatively large scales likely has multiple functions, e.g.,
fueling AGN accretion, igniting nuclear star formation, and exerting AGN feedback (e.g.,
in the form of AGN-driven outflows) to influence the host
galaxy, which adds an additional complexity to the understanding of the AGN obscuration phenomenon
and makes the quest for highly obscured/CT AGNs even more challenging yet more intriguing.

\subsection{Searching for the low-mass BH/AGN subpopulation}\label{sec:lowmass}

Low-mass BHs (or intermediate-mass BHs; IMBHs;\footnote{A most recent work, which  
discovers an IMBH with $2200^{+1500}_{-800}\ M_\odot$ in the center of the globular
cluster 47~Tucanae, has become a real eye catcher (Kiziltan, Baumgardt, \& Loeb 2017).} 
with $\approx 10^{3-6}$~\msun) are thought to
naturally bridge the evolutionary gap between stellar-mass BHs (with a few tens of solar masses) and SMBHs,
whose studies have important implications for the origin of SMBHs (i.e., the birth and growth
of early BH seeds), the overall SMBH growth history, the common characteristics of BH phenomena 
across all mass scales, and many other aspects
(see, e.g., Miller \& Colbert 2004; Volonteri 2010; Greene 2012; Reines \& Comastri 2016 for a review).
However, such a population of low-mass BHs has been largely elusive, because they are
difficult to find, mainly due to their low masses and thus small gravitational influences
and low luminosities.

Nevertheless, the situation of searching for low-mass BHs has been improving continuously, thanks
to the advent of numerous high-quality multiwavelength observations and various search
techniques.
Within the Local Group, it is possible to utilize stellar or gas dynamics to search for 
low-mass BHs residing in dwarf galaxies (less than a few $10^9$~\msun), which represent a plausible place to host 
such BHs according to,
e.g., the $M_{\rm BH}$--$M_\star$ scaling relation (see, e.g., Kormendy \& Ho 2013 for a review).
A number of such attempts made in some local dwarf galaxies 
either place upper limits of $M_{\rm BH}\lsim 10^4$~\msun\
(e.g., Gebhardt \etal\ 2001; Valluri \etal\ 2005; Lora \etal\ 2009; Jardel \& Gebhardt 2012) 
or find some positive measurements of $M_{\rm BH}\approx 10^{5-6}$~\msun\
(e.g., Seth \etal\ 2010; van den Bosch \& de Zeeuw 2010; den Brok \etal\ 2015).
Going beyond the Local Group, multiwavelength techniques have to be invoked in order to reveal
the signatures of low-mass AGNs typically located in dwarf galaxies. 
For example,
optical spectroscopic features such as high-ionization narrow emission lines,
broad emission lines, and narrow emission-line diagnostics 
reveal many samples of low-mass BHs or candidates with $M_{\rm BH}\approx 10^{5-6}$~\msun\ 
(e.g., Greene \& Ho 2004;
Dong \etal\ 2012; Reines \etal\ 2013; Moran \etal\ 2014; Sartori \etal\ 2015); and
some studies attempt to use MIR color diagnostics to search for low-mass AGNs in
small-bulge and low-mass galaxies (e.g., Satyapal \etal\ 2014; Hainline \etal\ 2016).
It should be noted that the above optical techniques tend to be biased toward AGNs with
relatively high Eddington ratios (\edd=$L_{\rm bol}/L_{\rm Edd}$, where $L_{\rm Edd}$ is Eddington luminosity) 
and MIR approaches are often subject to 
severe contamination because of the confusion between
MIR colors of dwarf starburst galaxies and AGNs.  
In contrast, \xray\ observations do not have such limitations and can
probe AGNs down to very low levels with least bias (see Section~\ref{sec:utility}).

\begin{figure}[!t]
\centering
\includegraphics[width=80mm]{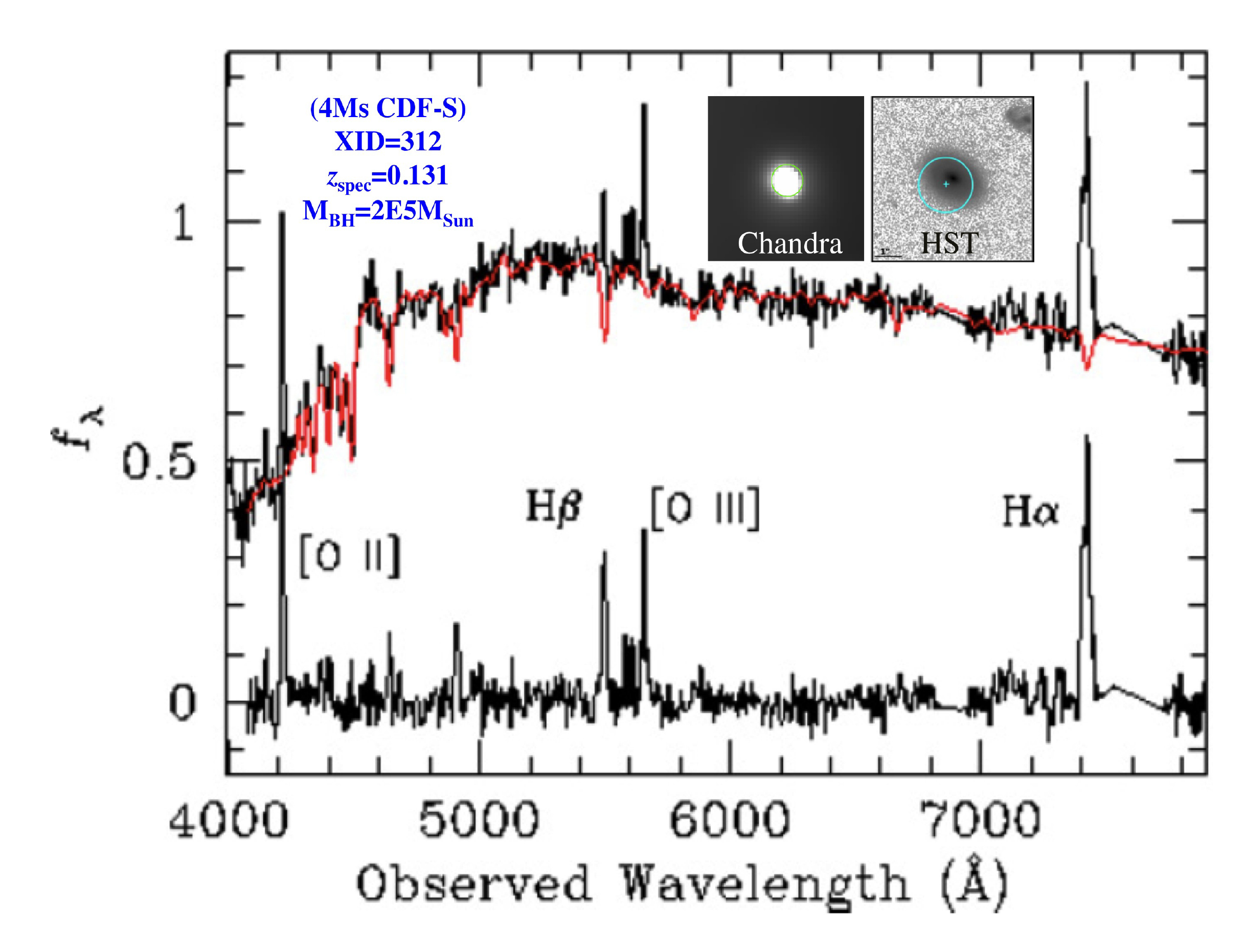}
\caption{An example of low-mass AGNs: 
VLT/FORS2 optical spectra (before and after stellar continuum subtraction; given in relative flux) of a $z_{\rm spec}=0.131$ source (XID=312 in the 4~Ms \cdfs)
that has an estimated $M_{\rm BH}=2\times10^5M_\odot$ based on the virial mass estimate using
the broad ${\rm H}_\alpha$ line. The red curve indicates the best-fit continuum.
Adapted from Schramm \etal\ (2013).
\label{fig:312}}
\end{figure}

Indeed, many \xray\ studies have accumulated growing samples of low-mass AGNs
among a wide variety of low-mass host galaxies mostly in the local universe 
(e.g., Ghosh \etal\ 2008; Jia \etal\ 2011; 
Reines \etal\ 2011, 2014; Kamizasa, Terashima, \& Awaki 2012; Lemons \etal\ 2015; Miller \etal\ 2015;
Baldassare \etal\ 2015, 2017; Plotkin \etal\ 2016).
The search for low-mass AGNs at moderate redshifts calls for deep \xray\ surveys
with direct detections or stacking analyses.
Schramm \etal\ (2013) discover three low-mass AGNs hosted in $M_\star <3\times 10^9$~\msun\
galaxies at $z<0.3$, which are individually detected in the 4~Ms \cdfs\ and have $L_{\rm X}\approx 10^{40}$~\lum\ (with one being variable).
Particularly, one of these sources has its broad H$\alpha$ line detected, providing a virial mass
estimate of $\approx 2\times 10^5$~\msun\ that is consistent with the estimate derived using 
scaling relations between $M_{\rm BH}$ and host galaxy properties (see Fig.~\ref{fig:312}).
Pardo \etal\ (2016) identify 10 low-mass \xray\ AGNs among a sample of $M_\star <3\times 10^9$~\msun\
dwarf galaxies at $z<1$, which are estimated to typically have $\lambda_{\rm Edd}\approx 5$\%.
Most recently, Chen \etal\ (2017) study a sample of 10 low-mass $z<0.3$ AGNs detected in the 
\hbox{3--24 keV} band by the {\it NuSTAR} serendipitous survey, demonstrating the
great potential and advantage of
using hard \xray\ observations to recover low-mass (obscured) AGNs missed by optical spectroscopic surveys and $<10$~keV \xray\ surveys.  
Furthermore, the use of sensitive stacking can significantly push the constraints upon low-mass AGNs out to 
higher redshifts.
By stacking the 4~Ms \cdfs\ data, Xue \etal\ (2012) track down a population of highly obscured AGNs
that are not individually detected in X-rays, 
which make the majority contribution to the unresolved \hbox{6--8 keV} CXRB, and are hosted in faint $1\lsim z\lsim 3$ galaxies that are located on the top of the blue cloud and has $M_\star \approx 2\times 10^{8-9}$~\msun.  
Mezcua \etal\ (2016) stack a large sample of \xray-undetected dwarf starburst and late-type 
galaxies in five redshift bins up to $z=1.5$ using the \chandra\ COSMOS-Legacy survey data, 
and find AGN \xray\ emission (with mean $L_{\rm SB}\approx 10^{39-40}$~\lum) in each redshift bin after removing contributions of XRB and hot gas to the stacked signal.   
It seems promising for subsequent studies to obtain a key advance by finding better
multiwavelength tracers of which galaxies are those likely to host AGNs, 
which can allow much improved \xray\ stacking and diagnosis of physical drivers.

Combining the efforts of searching for low-mass AGNs with optical, MIR, and particularly \xray\ observations,
it appears evident that low-mass AGNs are quite common from the local universe at least out to
moderate redshifts. 
Therefore, it is imperative to further build up statistically meaningful samples of 
low-mass AGNs across a reasonably broad redshift range and with a variety of host properties, 
which will certainly help us piece together a scenario for their evolution,
lend lessons to studies of high-redshift SMBH seeds,
make a critical step forward in understanding the AGN demography in the low-BH mass regime,
and constrain further their behavior in the scaling relations between $M_{\rm BH}$ and host galaxy 
properties (e.g., Greene \etal\ 2010; 
Jiang, Greene, \& Ho 2011; Graham \& Scott 2013, 2015; Reines \& Volonteri 2015; Baldassare \etal\ 2015; Ho \& Kim 2016).

\subsection{Examining the significantly variable AGN subpopulation}\label{sec:var}

Temporal and spectral variability is a defining and ubiquitous feature of AGNs,
and variability studies are valuable for probing AGN physical properties
(e.g., Mushotzky \etal\ 1993; Ulrich \etal\ 1997; Peterson 2001; Vaughan \etal\ 2003).
The versatile utility of AGN variability analyses is rendered in many aspects, e.g.:
(1) A first-order estimation of the physical size of the emission region can be obtained by
searching for the minimum variability timescale according to simple light-crossing time arguments, which is often employed in high-energy (\xray\ and gamma-ray) observations of blazars (e.g., Xue \& Cui 2005; Abdo et al. 2011).
(2) Detailed reverberation-mapping based studies allow for size estimates of various AGN components such as accretion disk, broad-line region, and inner part of dusty torus
(e.g., Koshida \etal\ 2014; Peterson 2014; Fausnaugh \etal\ 2016).
(3) The properties of the absorbing matter such as obscuring wind and gas can be revealed
by examining the changes in absorption (e.g., Filiz Ak \etal\ 2013; Netzer 2015).
(4) The AGN \xray\ power spectral density (PSD) is often modeled with a broken power-law
of red-noise nature 
(e.g., Uttley, McHardy, \& Papadakis 2002; also see Fig.~1 of Zhu \& Xue 2016 for a
demonstration). The break frequency of the PSD,
the amplitude of the high-frequency PSD, and the integral of that part
are found to be closely related to BH properties such as 
$M_{\rm BH}$ and mass accretion rate, thereby enabling measurements of 
such BH properties (e.g., McHardy \etal\ 2004, 2013; Zhou \etal\ 2010; 
Gonz\'{a}lez-Mart\'{i}n \& Vaughan 2012; Kelly \etal\ 2013).
(5) Significant variability has often been used as an effective AGN selection technique
in \xray, UV, and optical bands (e.g., Maoz \etal\ 2005; Trevese \etal\ 2008; 
Boutsia \etal\ 2009; Villforth \etal\ 2010; MacLeod \etal\ 2011; Young \etal\ 2012; 
Falocco \etal\ 2015; also see Table~\ref{tab:agnid}).
(6) AGN variability observations can place constraints on the underlying emission 
processes as well as the models intended to explain the physical origins 
of AGN variability 
(e.g., Xue, Yuan, \& Cui 2006; Kelly \etal\ 2009, 2014; Cai \etal\ 2016;
Liu \etal\ 2016a).

AGN \xray\ variability is of particular interest, because it is generally
more rapid and has larger amplitudes than variability at longer wavelengths; moreover,
\xray\ emission is nearly universal among AGNs (see Section~\ref{sec:utility}) and it originates from the innermost part of the system, thus being
able to probe the immediate vicinity of the SMBH (e.g., Ulrich \etal\ 1997). 
Therefore, many deep \xray\ surveys have been conducted to study AGN variability
down to low and moderate luminosities and up to high redshifts
(e.g., Almaini \etal\ 2000; Paolillo \etal\ 2004, 2017; Mateos \etal\ 2007; 
Papadakis \etal\ 2008; Vagnetti \etal\ 2011, 2016; Lanzuisi \etal\ 2014; 
Shemmer \etal\ 2014; Middei \etal\ 2017; Zheng \etal\ 2017; Li \etal\ in prep.).
The observations of the CDFs, particularly the \cdfs\ and \cdfn, 
cover a timespan up to $\approx 16$~years (see Table~\ref{tab:cdf}), 
resulting in the longest rest-frame timescales
that can possibly be probed for \xray\ variability analyses of a large sample of distant AGNs, and thus enabling a range of exciting variability science.

\begin{figure*}[!t]
\centering
\includegraphics[width=160mm]{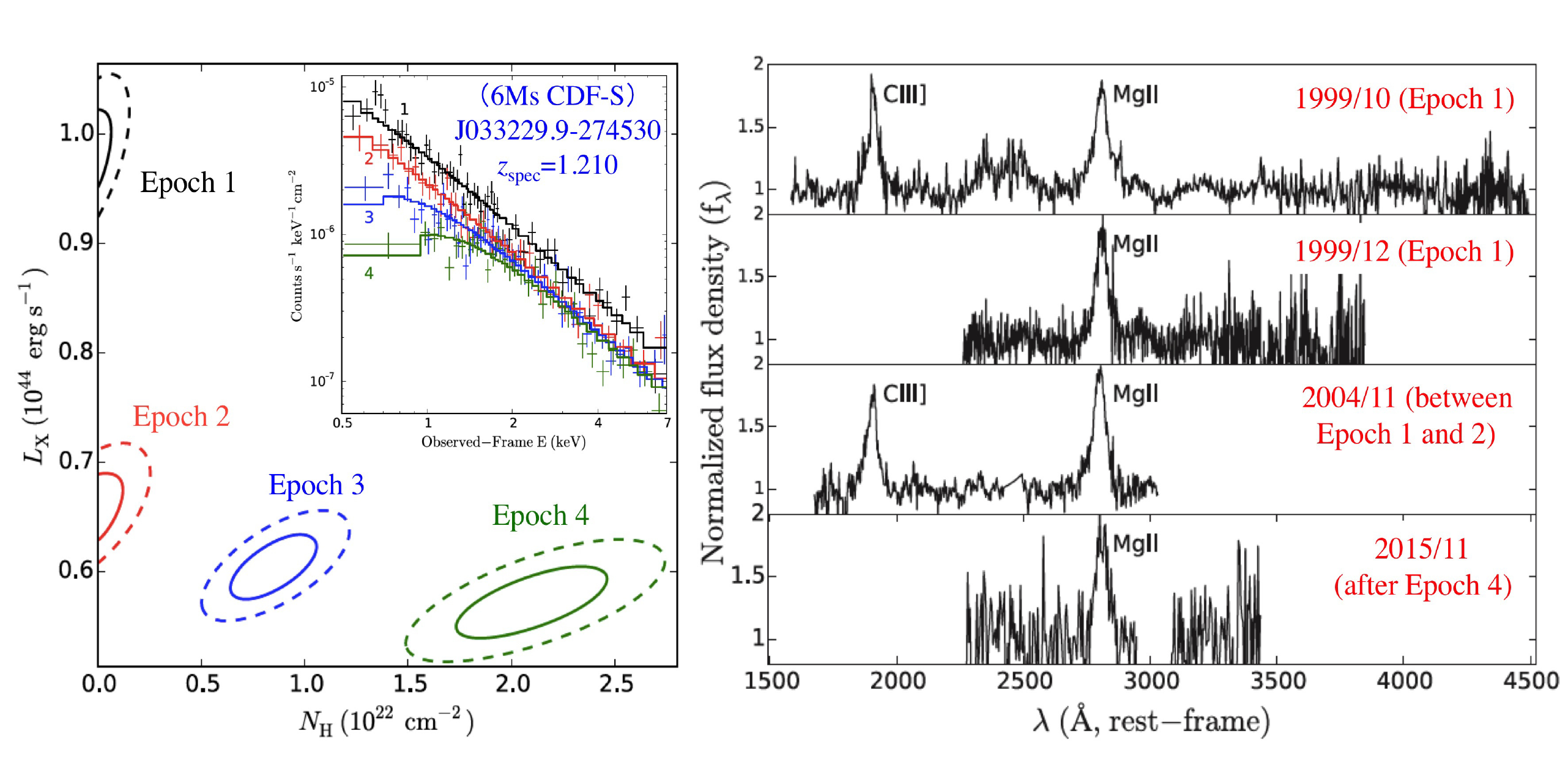}
\caption{(Left) Unfolded \xray\ spectra, best-fit models (data binned for display
purposes only; inset) and \lx--\nh\
confidence contours (1$\sigma$ and 2$\sigma$) for J033229.9$-$274530 in the 6~Ms \cdfs\ during four epochs
(Epoch~1: 1999/10--2000/12; Epoch~2: 2007/09--2007/11; Epoch~3: 2010/03--2010/07;
Epoch~4: 2014/06--2015/01), which displays significant \nh\ variability.
(Right) Four normalized optical spectra of J033229.9$-$274530 taken on different observation dates, 
which always remains as optically type~I. 
Adapted from Yang \etal\ (2016).
\label{fig:varsrc}}
\end{figure*}

Young \etal\ (2012) identify 20~AGNs out of a sample of 92 galaxies
with $z\approx 0.08$--1.02 in the 4~Ms \cdfs, solely based on their significant
long-term \xray\ variability.
These 20 newly-identified AGNs fail all the other non-variability 
AGN identification criteria adopted by the 4~Ms \cdfs\ cataloging
work (Xue \etal\ 2011; also see Table~\ref{tab:agnid}) and were then
regarded as galaxies.
These 20 AGNs have observed variability behavior that cannot be explained
by XRB populations, and appear to be low-luminosity unobscured AGNs given   
their stacked $\Gamma_{\rm effective}=1.93\pm 0.13$, with estimated $M_{\rm BH}$ 
(accretion rates) a factor of 2.4 (22.5) lower than variable
luminous AGNs at the same redshift.
This study underscores the advantage of using \xray\ variability analysis to find 
low-luminosity unobscured AGNs that would otherwise be missed.
Yang \etal\ (2016) conduct systematic long-term \xray\ variability analyses of 
the 68 brightest radio-quiet AGNs at $z\approx 0.6$--3.1 in the 6~Ms \cdfs\ and find:
(1) Among these sources, $\approx 90$\% are variable, indicating widespread photon flux variability;
$\approx 74$\% display \lx\ variability, with quasars having smaller variability amplitudes; and
$\approx 16$\% show \nh\ variability, with variability amplitudes becoming larger for longer time separation.
(2) There are a few sources possessing very interesting variability patterns:
a CTAGN candidate has variable high-energy \xray\ flux, implying the existence of 
$\lsim 0.3$~pc reflecting matter;
a broad absorption line quasar is \lx\ variable; and
a $z_{\rm spec}=1.21$ source appears to be a ``semi-changing-look'' AGN,
transiting from an \xray\ unobscured to obscured state
(also see, e.g., Matt, Guainazzi, \& Maiolino 2003; Ricci \etal\ 2016 for examples of \xray\ changing-look AGNs)
but always remaining as optically type~I (see Fig.~\ref{fig:varsrc};
also see, e.g., LaMassa \etal\ 2015; McElroy \etal\ 2016; Runnoe \etal\ 2016 for examples of optical changing-look AGNs). 
This work showcases the uniqueness of the \cdfs\ dataset in revealing long-term AGN
variability and some unusual variable AGNs.
Paolillo \etal\ (2017) apply various PSD models to study the ensemble \xray\ timing properties of AGNs during a timespan of $>16$ years up to 
$z\approx 4$ in the 7~Ms \cdfs, and find that the average \edd\ appears to be 
largely constant
across this redshift range given the large statistical uncertainties, with only marginal
evidence for a possible increase of \edd\ that peaks at $z\approx 2$--3.
This study shows the potential of using \xray\ variability analysis to trace the overall 
SMBH accretion history, which can be fully developed when future large \xray\
missions enlarge AGN samples and thus reduce statistical uncertainties associated with
such measurements.

\section{Interactions between AGNs and their host galaxies}\label{sec:inter}

\subsection{AGN \xray\ luminosity versus galaxy SFR}\label{sec:lxsfr}

The two relatively independent major astrophysical research fields,
AGNs and galaxies, have begun to be intimately connected with each other since a
series of important observational discoveries of tight relationships between $M_{\rm BH}$
and galaxy bulge properties (e.g., stellar velocity dispersion, luminosity, and mass; e.g.,
Magorrian \etal\ 1998; Ferrarese \& Merritt 2000; Gebhardt \etal\ 2000; 
Tremaine \etal\ 2002; Marconi \& Hunt 2003; Ferrarese \& Ford 2005; 
 G\"{u}ltekin \etal\ 2009; Kormendy \& Ho 2013).
Subsequent results show that
the volume density of SMBH accretion rate (scaled up by a factor of several thousand)
closely tracks the cosmic SFR density out to at least $z\approx 2$
(e.g., Heckman \etal\ 2004; Merloni \etal\ 2004; Silverman \etal\ 2008a; Zheng \etal\ 2009; Aird \etal\ 2010; Mullaney \etal\ 2012b; Kormendy \& Ho 2013; Aird \etal\ 2015b), 
in support of the scenario of a closely related overall SMBH-galaxy growth.
As a result, the co-evolution studies of galaxies and their central SMBHs have 
ever since aroused great interest and 
become a central research theme in modern astrophysics (see, e.g., Cattaneo \etal\ 2009;
Kormendy \& Ho 2013 for a review).

\begin{figure*}[!t]
\centering
\includegraphics[width=160mm]{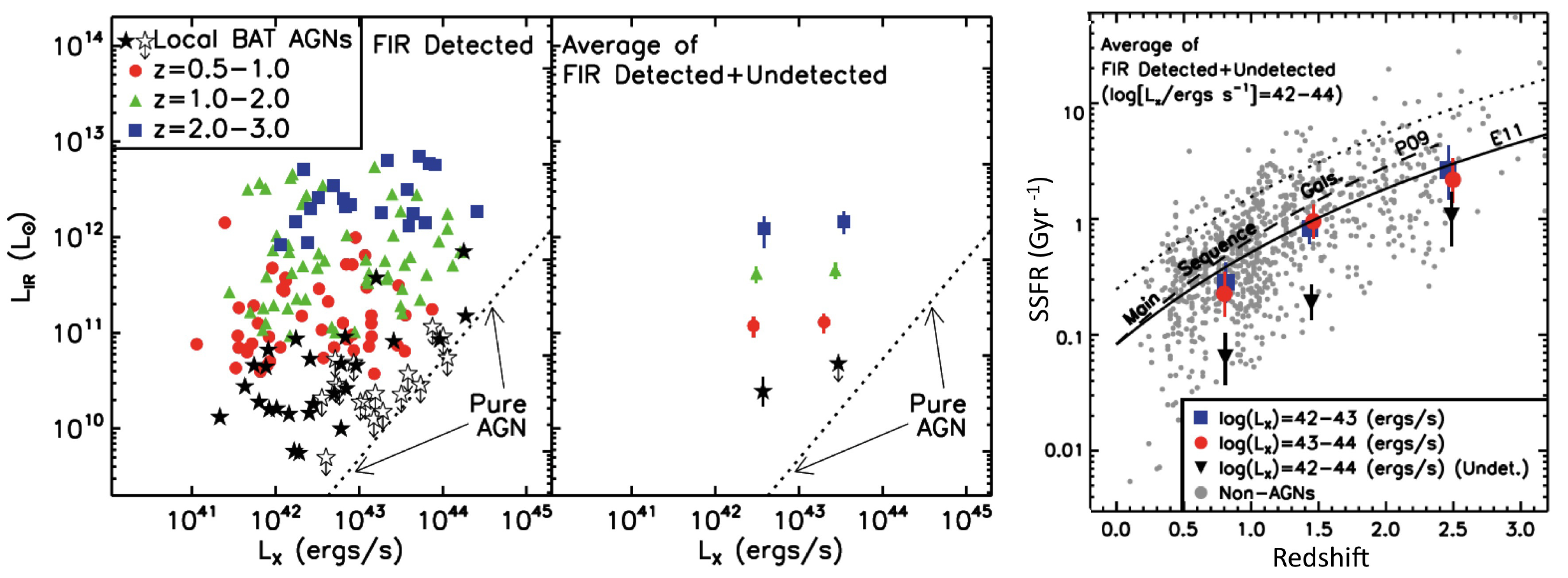}
\caption{(Left) Total IR luminosities ($L_{\rm IR}$) versus $L_{\rm 2-10\ keV}$ for the FIR detected moderate-\xray-luminosity AGNs with $0.5<z<3.0$.
Filled and open stars indicate local ($z<0.1$) BAT/IRAS AGNs for comparison whose 60~$\mu$m flux is likely dominated by the AGN rather than the host galaxy.
The dotted line indicates the $L_{\rm IR}$ expected from a ``pure'' AGN according to Mullaney \etal\ (2011).
(Middle) Same as left, but for the average $L_{\rm IR}$ derived from stacking analyses versus average $L_{\rm 2-10\ keV}$ for each \lx-$z$ bin.
Both the left and middle panels indicate no apparent \lx\ dependence of SFR for $0.5<z<3.0$ moderate-\xray-luminosity AGNs.
(Right) Average specific SFRs (sSFRs) of the \xray\ AGN hosts versus redshift (red and blue symbols), 
being roughly consistent with the star-forming galaxy main-sequence trend.
For comparison, the black and gray symbols indicate the stacks of the FIR-undetected AGNs
and non-AGNs, respectively. 
The solid and dashed curves indicate the main-sequence tracks from Elbaz \etal\ (2011)
and Pannella \etal\ (2009), respectively; the dotted line is a factor of 3 above the Elbaz \etal\ (2011) track (indicating strongly starbursting activities).
Adapted from Mullaney \etal\ (2012a).
\label{fig:lx_lir}}
\end{figure*}

The AGN \xray\ luminosity is related to SMBH accretion and the galaxy SFR is related to
star-formation process; as such, \lx\ and SFR are often taken as the observational 
indicators of respective growth of SMBHs and galaxies, with the relation between
them providing hints for interactions between AGNs and their hosts,
BH accretion modes, and likely effects of AGN feedback
(see, e.g., Fabian 2012 for a review).
Many studies have explored the relations between \lx\ and SFR, providing a wide range of
results, e.g.:
(1) For $z\gsim 1$ moderate-luminosity ($L_{\rm X}\approx 10^{42-44}$~\lum) AGNs,
there is generally no strong evidence for any \lx\ dependence of SFR, 
with SFR being broadly constant over this \lx\ range and AGN host specific SFR (sSFR)
usually being roughly consistent with main-sequence galaxies (e.g., Lutz \etal\ 2010;
Shao \etal\ 2010;
Mullaney \etal\ 2012a; Rosario \etal\ 2012, 2013a; Azadi \etal\ 2015; Stanley \etal\ 2015; 
see Fig.~\ref{fig:lx_lir}; but see Barger \etal\ 2015); 
while at $z\lsim 1$, there appears to be a positive correlation between SFR and \lx\, 
with such a trend being most prominent in the local universe (e.g., 
Shao \etal\ 2010; Rosario \etal\ 2012; Azadi \etal\ 2015; Lanzuisi \etal\ 2017;
also see, e.g., Rowan-Robinson 1995; Netzer \etal\ 2007; Netzer 2009; 
Diamond-Stanic \& Rieke 2012; Matsuoka \& Woo 2015 for similar 
positive SFR-AGN luminosity trends; but see, e.g., Shimizu \etal\ 2017).  
(2) For $z\gsim 1$ high-luminosity ($L_{\rm X}\gsim 10^{44}$~\lum) AGNs,
some studies suggest an increase of SFR or a roughly constant SFR 
toward large \lx\ (e.g., Hatziminaoglou
\etal\ 2010; Lutz \etal\ 2010; Harrison \etal\ 2012; Rovilos \etal\ 2012; Santini \etal\ 2012;
Stanley \etal\ 2015; Lanzuisi \etal\ 2017;
also see, e.g., Lutz \etal\ 2008; Bonfield \etal\ 2011; Dong \& Wu 2016; Harris \etal\ 2016 for similar positive SFR-AGN luminosity trends),
while some suggest a decrease of SFR with increasing \lx, possibly due to AGN feedback 
(e.g., Page \etal\ 2012; Barger \etal\ 2015).

There are several possible causes that can account, at least partially,  
for some of the variations between the above different observational results 
about moderate-luminosity and high-luminosity AGNs.
For example, it is always challenging to obtain robust SFR measurements,
especially for high-luminosity AGNs that can produce significant FIR emission
and thus contaminate that of host galaxies;
limited sample sizes (i.e., poor source statistics) appear to be 
an unavoidable issue most of the time; and, sometimes, cosmic variance may
come into play. 
Additionally, some physical considerations may be the more fundamental causes, e.g.:
(1) The relative timescales of several processes can largely determine whether the observational signatures of suppressed star formation (due to, e.g., AGN feedback) can be
maintained and thus clearly revealed; such processes include, e.g., star-formation activity,
luminous AGN activity (and its variability), 
the delay between the onsets of star formation and AGN activity,
and quenching of star formation (e.g., Di~Matteo \etal\ 2008; Hopkins \& Quataert 2010;
Gabor \& Bournaud 2013; Hickox \etal\ 2014; Neistein \& Netzer 2014).
(2) There are likely two modes in the concomitant AGN activity and star formation that
can transit from each other when conditions allow, 
with one mode primarily associated with low- and moderate-luminosity AGNs
where the evolution is secular, with no close coupling between instantaneous 
AGN luminosity and current galaxy-integrated SFR,
and another mode mainly associated with high-luminosity AGNs
where the evolution is intense and rapid likely through (major) mergers, 
with tight coupling between
AGN growth and host star formation (e.g., Lutz \etal\ 2010; Shao \etal\ 2010; 
Santini \etal\ 2012).
(3) The exact relation between \lx\ and SFR likely depends on whether the star formation is
dominated by a nuclear (i.e., sub-pc scales) or an extended ($\approx$~kpc) component,
with the former case resulting in tighter coupling than the latter
(e.g., Hopkins \& Quataert 2010; Diamond-Stanic \& Rieke 2012).

\begin{figure*}[!t]
\centering
\includegraphics[width=170mm]{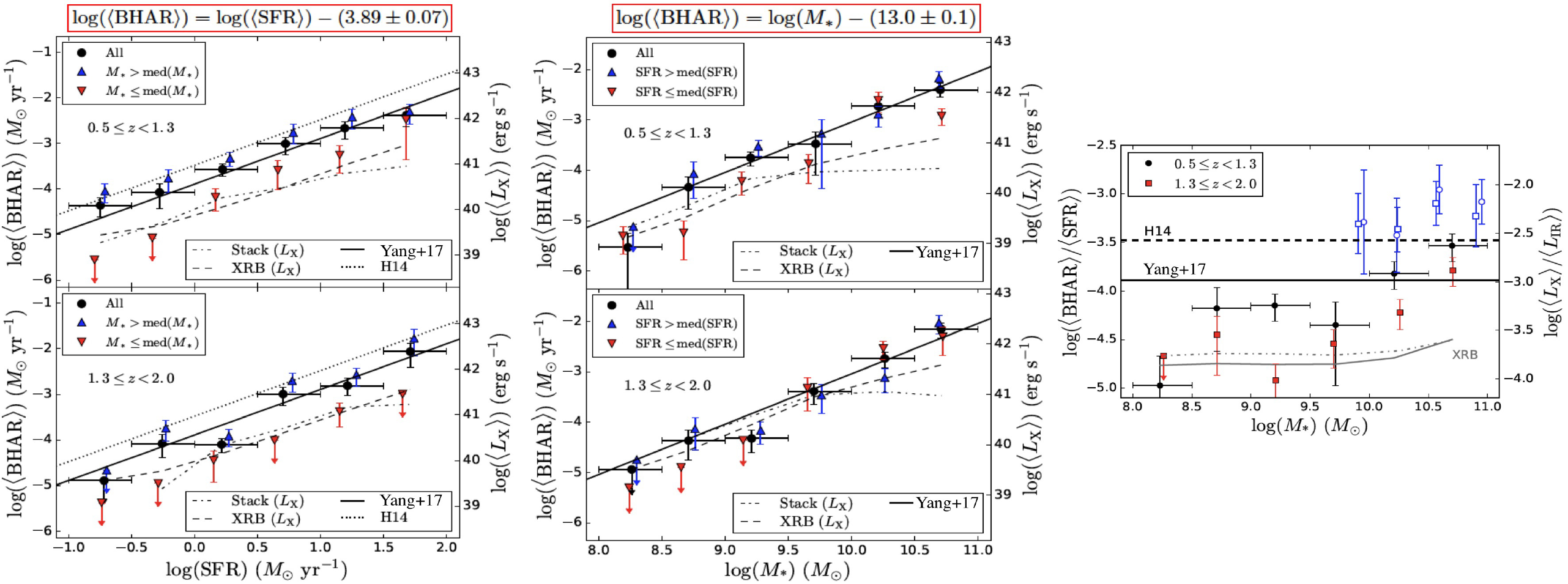}
\caption{(Left) Average black-hole accretion rate ($<$BHAR$>$; converted from $<L_{\rm X}>$ 
that is the sum of both \xray\ detected and undetected sources with galaxy
contamination $L_{\rm X,XRB}$ subtracted) as a function of (median) SFR
for a CANDELS/GOODS-S sample in various redshift and stellar-mass bins.
The solid line indicates the best-fit linear model with a fixed slope of 1, with the equation annotated atop;
the dotted line indicates the Hickox \etal\ (2014) model;
the dashed curves indicate $L_{\rm X,XRB}$ subtracted from each subsample;
and the dash-dotted curves indicate $<L_{\rm X}>$ for stacked sources ($L_{\rm X,XRB}$
not subtracted). 
(Middle) Similar to left, but for $<$BHAR$>$ versus $M_\star$ in various redshift and SFR
bins.  
The comparison between the left and middle panels indicates that
the $<$BHAR$>$-$M_\star$ relation is tighter than the $<$BHAR$>$-SFR relation. 
(Right) $<$BHAR$>$/$<$SFR$>$ as a function of $M_\star$.
The blue open symbols indicate the Mullaney \etal\ (2012b) results (circles and squares: $z\approx1$ and $z\approx2$ samples).
The gray solid and dash-dotted curves indicate $<L_{\rm X,XRB}>$/$<L_{\rm IR}>$ in the low- and high-redshift ranges, respectively. 
It reveals that massive galaxies have larger 
$<$BHAR$>$/$<$SFR$>$ ratios than less-massive galaxies.
Adapted from Yang \etal\ (2017).
\label{fig:bhar}}
\end{figure*}

Alternatively, the connection between AGN activity and star formation can be examined
using a different approach from the above studies,
which finds tight correlations between SFR and AGN luminosity by
averaging AGN and star-formation activities over all galaxies (irrespective
of hosting an AGN or not) in a cosmological volume 
(e.g., Rafferty \etal\ 2011; Symeonidis \etal\ 2011; Mullaney \etal\ 2012b; Chen \etal\ 2013; Delvecchio \etal\ 2015).
Such an approach focuses on investigating long-term average/overall behaviors, thus being 
immune to many detailed factors mentioned above.
As a most recent example, 
Yang \etal\ (2017) study the dependence of sample-mean BH accretion rate ($<$BHAR$>$)
on host SFR and $M_\star$ with a sample of $\approx 18000$ galaxies at $0.5\le z<2.0$,
using the 7~Ms \cdfs\ and CANDELS/GOODS-S observations.
They compare $<$BHAR$>$ for samples with different SFR and/or $M_\star$, down to
levels as low as SFR$\approx 0.1\ M_\odot$~yr$^{-1}$ and $M_\star\approx 10^8\ M_\odot$ 
with reasonable completeness for the first time, finding that (see Fig.~\ref{fig:bhar}):
(1) $<$BHAR$>$ is positively correlated with both SFR and $M_\star$, and both the
$<$BHAR$>$-SFR and $<$BHAR$>$-$M_\star$ relations can be adequately fit using linear
models with slope=1.
(2) The $<$BHAR$>$-$M_\star$ relation is tighter than the $<$BHAR$>$-SFR relation
(confirmed by partial-correlation analyses), 
indicating that BH growth is mainly linked to $M_\star$ rather than SFR, and 
the $<$BHAR$>$-SFR relation is largely secondary and results from the
well-known star-formation main sequence (e.g., Elbaz \etal\ 2011). 
(3) Massive galaxies (i.e., $M_\star \gsim 10^{10}\ M_\odot$) have larger 
$<$BHAR$>$/$<$SFR$>$ ratios than less-massive galaxies, suggesting that
the former have higher AGN fractions and/or higher SMBH fueling efficiencies than the latter
(also see, e.g., Rodighiero \etal\ 2015).
The Yang \etal\ (2017) results are in accordance to the SMBH-galaxy co-evolution scenario,
have important implications for the $M_{\rm BH}$-$M_\star$ scaling (see Section~\ref{sec:coevo}),
and make one worry about likely mass effects in at least some
of the aforementioned \lx-SFR studies that are based on direct detections.

The current research status of the relation between AGN activity and star formation
is somewhat perplexing, as both the observational and theoretical pieces don't all fit together (see, e.g., Harrison 2017 for a review).
Future observational studies can be improved in
many aspects, e.g., 
more reliable SFR measurements for individual sources,
broader parameter spaces (e.g., SFR, $M_\star$, \lx, $z$) to be probed,
better source statistics,
and higher sample completeness.
These improved studies will provide more sensitive tests of how AGN activity impacts star formation and make a step forward in understanding AGN triggering mechanisms.

\subsection{Conducive host galaxy properties for AGN activity}\label{sec:factor}

\begin{figure}[!t]
\centering
\includegraphics[width=80mm]{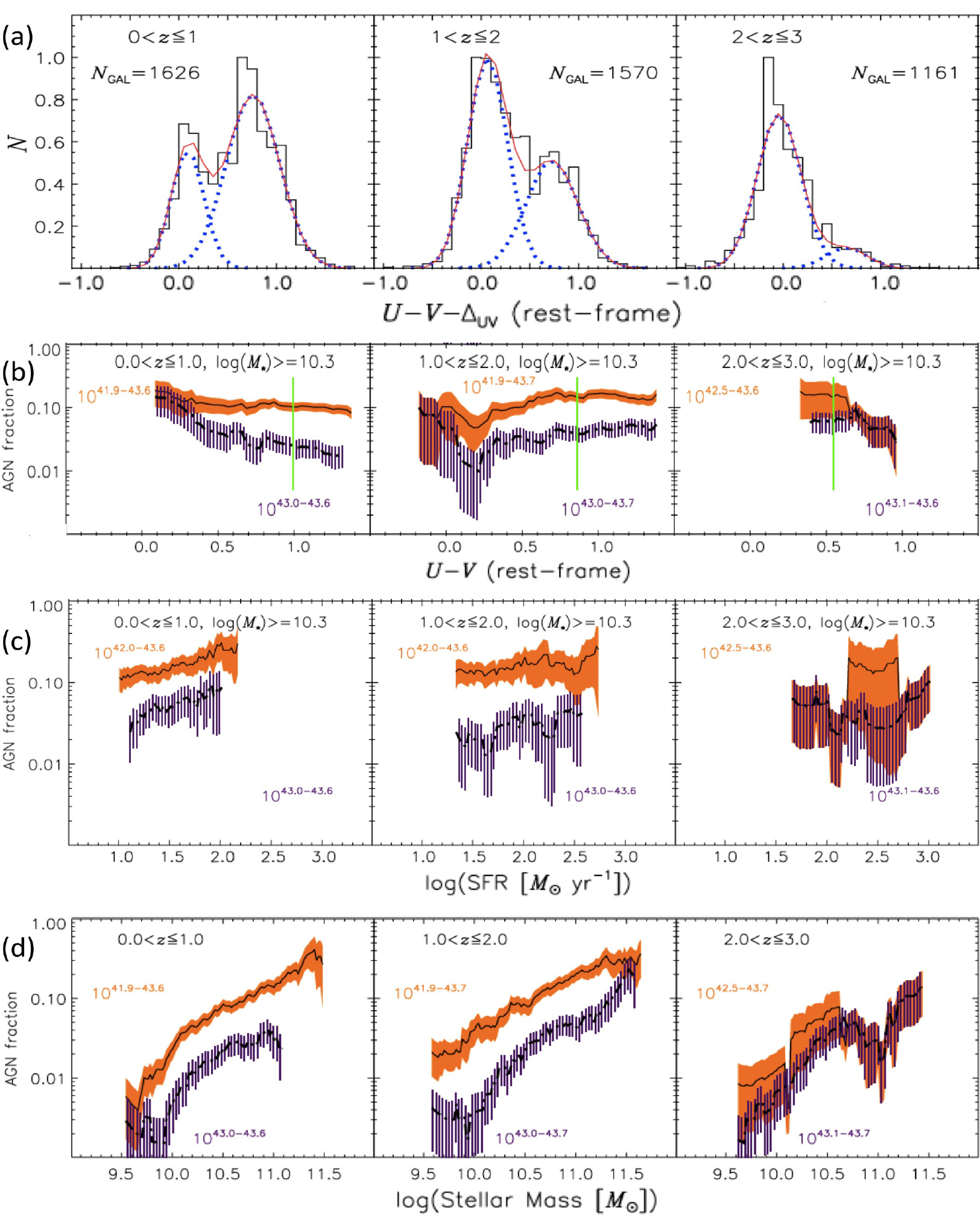}
\caption{[Row (a)] Normalized (peak=1) histograms of rest-frame dust-extinction corrected $U-V$ color (bin size=0.1)
for non-AGN galaxies (black histograms; total numbers annotated) in various redshift ranges,
which all display a color bimodality.
The two best-fit Gaussian components are indicated with blue dotted curves, with their sum shown as red solid curves.
[Rows (b--d)] AGN fraction as a function of rest-frame $U-V$ host color (in bins of $\Delta(U-V)=0.4$), SFR (in bins of $\Delta {\rm logSFR}=0.5$), and $M_\star$ (in bins of $\Delta {\rm log}(M_\star)=0.5$) in different $z$ and \lx\ bins, respectively.
The green vertical lines in Row (b) roughly separate blue and red galaxies according to
the Bell \etal\ (2004) separation scheme.
The log$(M_\star)\ge 10.3$ cut in Rows (b) and (c) effectively ensures mass-matched
samples of AGN hosts and non-AGN galaxies for reliable comparison (see main text for details). 
Adapted from Xue \etal\ (2010).
\label{fig:cmd}}
\end{figure}

\begin{figure}[!t]
\centering
\includegraphics[width=70mm]{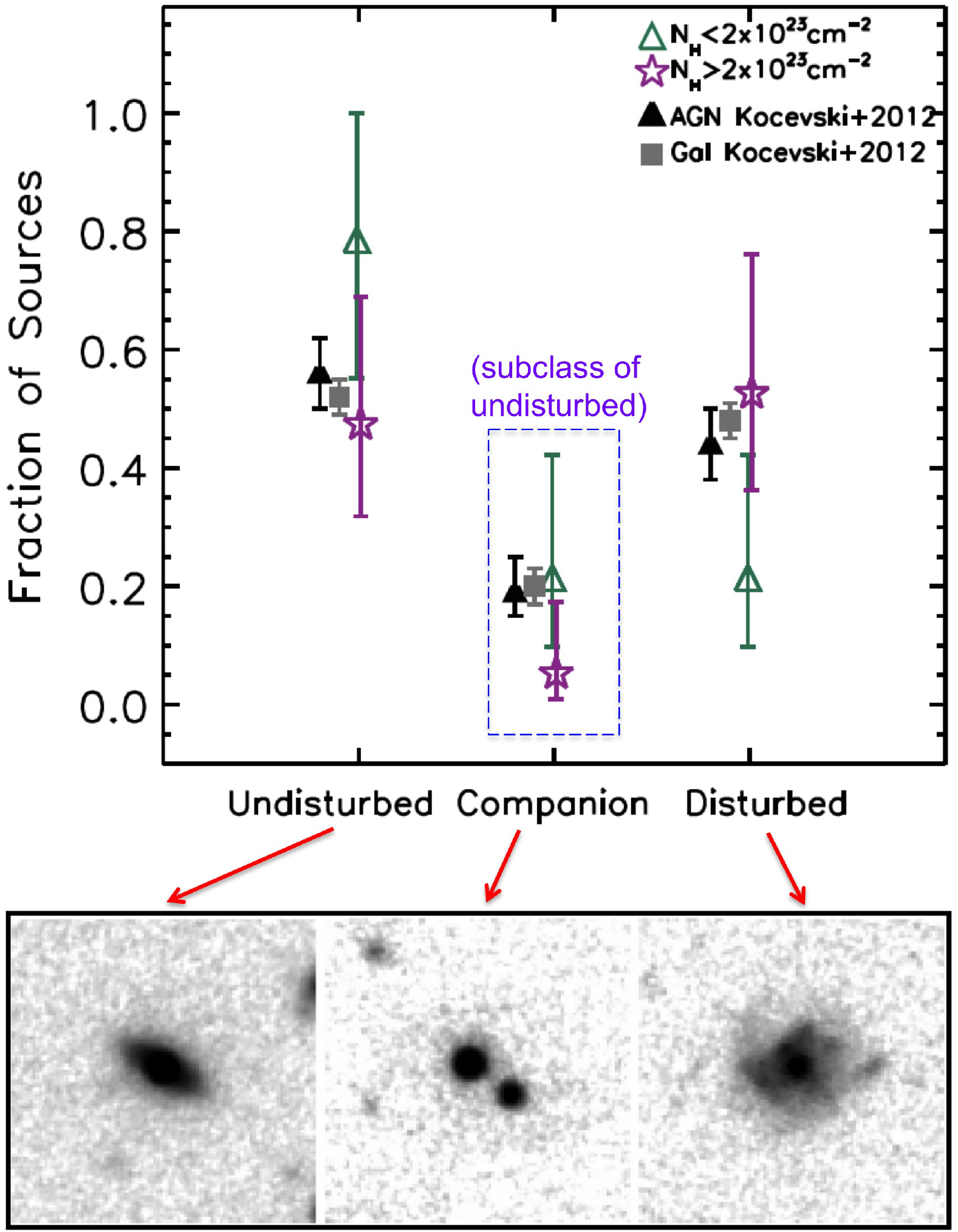}
\caption{(Top) Fraction of $z\approx2$ unobscured/moderately obscured (green symbols)
and highly obscured (purple symbols) 
MIR-luminous quasar hosts with various disturbance classes that are classified using the {\it HST} F160W images (see Bottom for example 6~arcsec$\times$6~arcsec cutouts). 
The Kocevski \etal\ (2012) $z\approx2$ AGN (black symbols) and non-AGN (gray symbols) 
samples are shown for comparison.
Adapted from Del~Moro \etal\ (2016).
\label{fig:morph}}
\end{figure}

What kinds of host galaxy environments are most AGN-friendly? 
This is a question that invites numerous observational efforts, which are made
over a broad redshift range (up to $z\approx 3$--4) and obtain many interesting 
results, e.g.:  
(1) Distant \xray\ AGN host galaxies 
are typically more massive than non-AGN galaxies, with the AGN fraction
(above a given AGN cutoff luminosity) 
increasing strongly toward larger stellar masses
(e.g., Akiyama 2005; Papovich \etal\ 2006; Alonso-Herrero \etal\ 2008; 
Brusa \etal\ 2009b; Xue \etal\ 2010; Aird \etal\ 2012; Mullaney \etal\ 2012a; 
Wang \etal\ 2017; see Row (d) of Fig.~\ref{fig:cmd}).
(2) There is a color bimodality among galaxy populations, with the majority being
red(-sequence) and blue(-cloud) galaxies and the minority green(-valley) galaxies
lying in between (e.g., Strateva \etal\ 2001; Bell \etal\ 2004; Brammer \etal\ 2009;
Xue \etal\ 2010; see Row~(a) of Fig.~\ref{fig:cmd}).
\xray\ AGN host galaxies are generally found to be optically more luminous and redder than
non-AGN galaxies (e.g., Barger \etal\ 2003b; Nandra \etal\ 2007; Silverman \etal\ 2008b;
Xue \etal\ 2010; Bongiorno \etal\ 2012).
However, when stellar mass-matched samples are considered,
both AGN hosts and non-AGN galaxies are comparably luminous with similar colors, and 
the AGN fraction is largely constant or even slightly decreasing toward red host colors
(e.g., Silverman \etal\ 2009; Xue \etal\ 2010; Rosario \etal\ 2013a; 
Hern\'{a}n-Caballero \etal\ 2014; see Row~(b) of Fig.~\ref{fig:cmd}).
(3) Most \xray\ AGNs reside in star-forming and starburst hosts, with the AGN fraction
increasing significantly toward higher SFRs
(e.g., Silverman \etal\ 2009; Xue \etal\ 2010; Rafferty \etal\ 2011; Rosario \etal\ 2013a; Wang \etal\ 2017).
Nevertheless, when mass-matched samples are examined, once again,
the above trend of rising AGN fraction upon larger SFRs 
becomes less prominent or even non-detectable, and 
there is no apparent difference in terms of star-formation properties between AGN hosts and
star-forming main-sequence galaxies
(e.g., Xue \etal\ 2010; Mullaney \etal\ 2012a; see Row~(c) of Fig.~\ref{fig:cmd}).
(4) The first morphological studies of \xray\ AGN hosts indicate that 
these AGNs preferentially reside in bulge-dominated systems (e.g.,
Grogin \etal\ 2005; Pierce \etal\ 2007).
However, the mass-matched technique does the magic once more:
clear differences between morphological types of AGN hosts and non-AGN galaxies
mostly disappear when the AGN and galaxy samples are matched in mass, i.e.,
\xray\ AGNs reside in a broad range of host-galaxy types that include, e.g., 
disk-dominated, bulge-dominated, irregular, and point-like morphology classes
(or simply categorized as undisturbed and disturbed classes; 
e.g., Kocevski \etal\ 2012; Fan \etal\ 2014; Villforth \etal\ 2014;
see Fig.~\ref{fig:morph}).

Based on the above \xray\ AGN results, 
it appears clear that, among a variety of host galaxy properties (e.g.,
stellar mass, optical luminosity and colors, SFR, and morphology) likely 
conducive for AGN activity,
stellar mass plays the fundamental driving role in AGN triggering,
while the other properties play a secondary role;
this is hardly surprising, given the
mass-light/luminosity (e.g., Zibetti, Charlot, \& Rix 2009), 
mass-color (i.e., massive galaxies generally tend to be redder; e.g., Xue \etal\ 2010),   
mass-SFR (i.e., the star-formation main sequence; e.g., Elbaz \etal\ 2011),
and mass-morphology (i.e., E/S0 galaxies generally dominate the higher-mass population;
e.g., Bundy, Ellis, \& Conselice 2005) correlations.
There are several physically plausible causes for the dominant role of stellar mass
in triggering AGN activity, e.g.,
massive galaxies (1) have stronger gravitational pull to make gas fall into galaxy 
centers and fuel SMBHs eventually, 
(2) are more likely to have nuclear bars to induce gas inflow efficiently, and 
(3) tend to have larger SMBHs that are more capable of accreting gas from their vicinity
(see, e.g., Section~4.2 of Yang \etal\ 2017 for details).  
Therefore, the conditions that are most conducive for AGN activity appear to be 
a massive host galaxy and a large reservoir of gas (e.g., Silverman \etal\ 2009;
Vito \etal\ 2014b).
In combination with the fact that galaxy stellar masses can be estimated more reliably 
than colors and luminosities (e.g., van~Dokkum \etal\ 2006),
consequently, evolutionary studies of galaxies and SMBHs are best probed using stellar-mass selected samples (e.g., van~Dokkum \etal\ 2006; Kriek \etal\ 2008)
and/or mass-matched samples (e.g., Silverman \etal\ 2009; Xue \etal\ 2010).
As such, the digesting of the above results derived with mass-matched samples
plainly indicates at least two points: 
(1) there is essentially no substantial difference between
many physical properties (e.g., optical luminosity and colors, SFR, and morphology)
of \xray\ AGN hosts and coeval non-AGN galaxies over a broad range of redshift
(but see, e.g., Mullaney \etal\ 2015b; Wang \etal\ 2017);
and (2) currently there appears no strong direct observational evidence for feedback effects
from moderate-luminosity \xray\ AGNs (e.g., Xue \etal\ 2010; Bongiorno \etal\ 2012;
Rosario \etal\ 2013b; Azadi \etal\ 2017; but see, e.g., 
Wang \etal\ 2017).\footnote{Interestingly, 
theoretical studies predict diverse roles played by moderate-luminosity AGNs,
whose feedback may be either important (e.g., Fabian \etal\ 2008; Raimundo \etal\ 2010) 
or not (e.g., Hopkins \& Hernquist 2006).}

Gas-rich major mergers and triggering of distant AGNs are often closely related
in theoretical considerations (e.g., Sanders \etal\ 1988;
Di~Matteo \etal\ 2005; Hopkins \etal\ 2008).
Observationally, however, only $\lsim 20$\% of the AGN population at $z\approx 0$--2.5 
have clear signatures of major mergers (e.g., Koss \etal\ 2010; 
Silverman \etal\ 2011; Kocevski \etal\ 2012; Cotini \etal\ 2013;
Villforth \etal\ 2014), which is much below theoretical expectations.
Interestingly, recent morphological studies find that
major mergers only trigger the most luminous AGNs (e.g., Treister \etal\ 2012;
Rumbaugh \etal\ 2017; but see, e.g., Villforth \etal\ 2017); and
highly obscured AGN hosts
tend to undergo dynamical compaction (e.g., Chang \etal\ 2017), interactions,
or mergers (e.g., Kocevski \etal\ 2015; Lanzuisi \etal\ 2015a; Del~Moro \etal\ 2016).
In particular,
Del~Moro \etal\ (2016) study a sample of $z\approx 2$ MIR-luminous quasars, and find that  
the highly obscured quasars tend to reside in galaxies with  
disturbed morphologies while
the unobscured/moderately obscured quasars preferentially lie in undisturbed hosts, 
but the disturbed quasar hosts only constitute the minority ($\approx 40$\%) of  
the entire sample (see Fig.~\ref{fig:morph}), which is consistent with other
findings (e.g., Kocevski \etal\ 2015; Lanzuisi \etal\ 2015a).
The above results indicate that secular processes such as galaxy bars and disk
instabilities, as well as minor mergers, may be the main fueling mechanisms for majority AGN populations.
Therefore, it remains appealing for future morphological studies to quantify reliably and accurately 
the respective contribution of each likely AGN fueling mechanism
as a function of, e.g., \lx\ and redshift.

\subsection{SMBH growth behavior revealed by $\Gamma$-\edd\ relation and \edd\ distribution}\label{sec:edd}

\edd\ is an important parameter in accretion disk theories, which can be used as
a primary indicator of SMBH growth rate and accretion mode.\footnote{In practice, the observable specific accretion rate (i.e., \lx/$M_\star$) is often taken as a proxy for \edd; thus, for convenience and succinctness,
the mixed use of \lx/$M_\star$ and \edd\ is adopted here, i.e.,
\edd\ actually refers to \edd\ itself or \lx/$M_\star$.}
Recent observational results about the AGN $\Gamma$-\edd\ relation 
reveal a V-shape correlation: as \edd\ decreases continuously, the $\Gamma$-\edd\ correlation 
changes from being positive (e.g., Wang \etal\ 2004;
Shemmer \etal\ 2006, 2008; Risaliti \etal\ 2009;
Zhou \& Zhao 2010; Brightman \etal\ 2013; Fanali \etal\ 2013; Yang \etal\ 2015) into being negative (e.g., Constantin \etal\ 2009; Gu \& Cao 2009; 
Younes \etal\ 2011; Jang \etal\ 2014; Yang \etal\ 2015; Kawamuro \etal\ 2016),
with the transition occurring at \edd$\approx 1$\%.\footnote{See, e.g., 
Fig.~1 of Gu \& Cao (2009) and Fig.~2 of Yang \etal\ (2015) for a demonstration of the
V-shape $\Gamma$-\edd\ correlation, which is also seen in some well-observed individual AGNs 
(e.g., Sobolewska \& Papadakis 2009; Emmanoulopoulos \etal\ 2012; Connolly \etal\ 2016)
as well as in XRB populations (e.g., Yang \etal\ 2015).}
It is generally believed that these observational results reflect that, as the accretion rate decreases, the SMBH accretion mode changes accordingly, from the original standard thin accretion disk into radiatively inefficient accretion flow
(e.g., Gu \& Cao 2009; Brightman \etal\ 2013; Jang \etal\ 2014; Yuan \& Narayan 2014).
Recently, Yang \etal\ (2015) propose a coupled hot accretion flow-jet model,
which can well explain the overall $\Gamma$-\edd\ correlation and even predicts
that, toward the very low \edd\ regime (e.g., \edd$\lsim 10^{-5}$), 
this correlation will display a small bump and then level off eventually.
Interestingly, Liu \etal\ (2016b) find that, toward the very high \edd\ regime
(e.g., \edd$\gsim 0.1$), the $\Gamma$-\edd\ correlation appears to be largely constant
or even negative, which is different from previous findings.
Therefore, it would be useful to build a large, uniform, and (relatively) complete
AGN sample with, e.g., the CDF observations in conjunction with other data, 
to study in detail the $\Gamma$-\edd\ correlation and its likely dependence
on redshift and host galaxy properties (Sun \etal\ in prep.),
in order to thoroughly compare observations with various SMBH accretion models.
If reliably calibrated (see, e.g., Section~4.3 of Brandt \& Alexander 2015
for the challenges in obtaining reliable estimates of $\Gamma$ and \edd), 
the $\Gamma$-\edd\ correlation may be used to
conveniently estimate some important AGN parameters (e.g., \edd, $M_{\rm BH}$, and 
mass accretion rate), which will facilitate better understanding accretion processes and evolution of AGNs.

Observational \edd\ estimates for $z\approx 0$--4 \xray\ AGNs span a broad range of
\edd$\approx 10^{-5}$--1, with the majority having 10$^{-4}$--0.1 
(e.g., Babi\'{c} \etal\ 2007; Ballo \etal\ 2007; Alonso-Herrero \etal\ 2008;
Brusa \etal\ 2009b;
Hickox \etal\ 2009; Raimundo \etal\ 2010; Trump \etal\ 2011; Lusso \etal\ 2012;
Matsuoka \etal\ 2013; Azadi \etal\ 2015; Suh \etal\ 2015; Bernhard \etal\ 2016).
On average, \edd\ seems to increase with redshift at any given $M_{\rm BH}$, but
has no clear evolution with redshift at any given $L_{\rm bol}$ (e.g., Lusso \etal\ 2012).
Further careful analyses reveal that
the intrinsic \edd\ distribution for \xray\ AGNs appears to follow   
a ``universal'' power-law with a slope independent of both $M_\star$ and redshift
out to $z\approx 2.5$ (Aird \etal\ 2012, 2013; Bongiorno \etal\ 2012; Azadi \etal\ 2015;
Wang \etal\ 2017),
implying that the basic physical processes responsible for triggering and fueling
the overall AGN population might be essentially the same;
such a power-law distribution is in contrast to a lognormal distribution,
with the former also seen for, e.g., AGNs in quiescent galaxies 
(e.g., Kauffmann \& Heckman 2009; but see, e.g., Jones \etal\ 2016)
and the latter seen for, e.g., the observed \edd\ distribution for \xray\ AGNs
(e.g., Aird \etal\ 2012) as well as the \edd\ distribution for optically luminous AGNs
(e.g., Kollmeier \etal\ 2006) and AGNs in star-forming galaxies 
(e.g., Kauffmann \& Heckman 2009).

\begin{figure*}[!t]
\centering
\includegraphics[width=140mm]{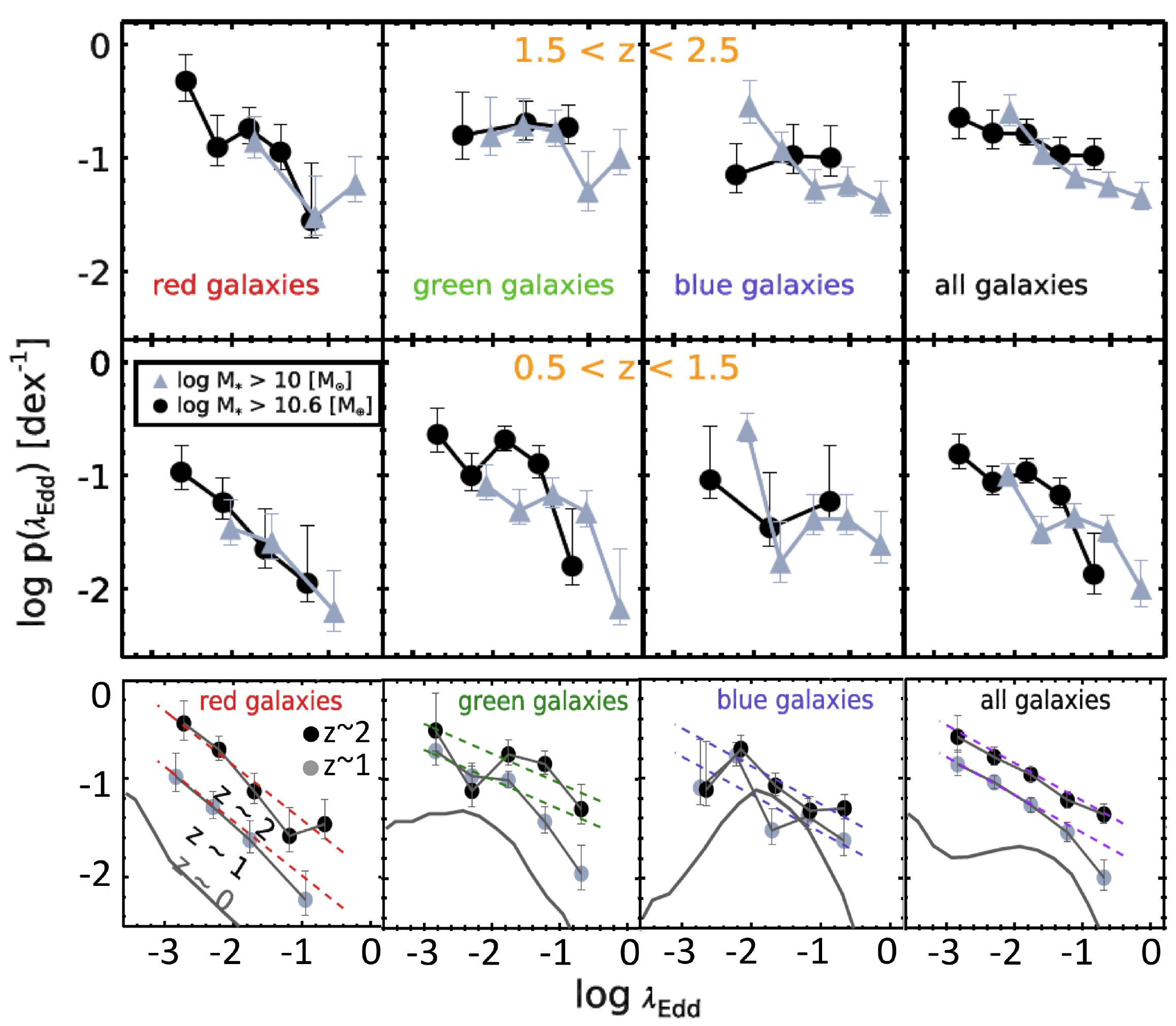}
\caption{Binned Eddington ratio distributions (${\rm p}(\lambda_{\rm Edd})$, i.e.,
the probability of galaxies hosting an AGN with \edd\ in a given $M_\star$ and $z$ bin,
with $1/V_{\rm max}$ correction applied)
for a sample of moderate-luminosity ($L_{\rm 0.5-8\ keV}=10^{41.9-43.7}$~\lum) AGNs
in various bins of redshifts, $M_\star$, 
and intrinsic (i.e., dust-extinction corrected) host colors. 
In the bottom panel, no $M_\star$ subdivision is applied;
the best-fit ${\rm p}(\lambda_{\rm Edd})$ models at two redshift ranges ($z\approx1$ and $z\approx2$) are shown as dashed lines based on the Extended Maximum Likelihood fitting of the unbinned data; and the gray curves indicate the observed binned ${\rm p}(\lambda_{\rm Edd})$ of local Type~II AGN hosts from Kauffmann \& Heckman (2009).
It seems clear that there is a strong dependence of SMBH accretion (${\rm p}(\lambda_{\rm Edd})$ 
as a proxy here)
on the host colors (see main text for details). Adapted from Wang \etal\ (2017).
\label{fig:edd}}
\end{figure*}

Specifically, Wang \etal\ (2017) study the \edd\ distribution (${\rm p}(\lambda_{\rm Edd})$) 
for a mass-complete sample of $0.5<z<2.5$ 
moderate-luminosity \xray\ AGNs in the two GOODS fields, and find some interesting results
(see Fig.~\ref{fig:edd}):
(1) ${\rm p}(\lambda_{\rm Edd})$ for the overall galaxy 
population appears to be a power-law with a slope of $-0.4$,
being consistent with the slope obtained by Jones \etal\ (2016) but 
slightly shallower than the slope of $-0.6$ measured by, e.g., Aird \etal\ (2012).
(2) However, in terms of redshift evolution, ${\rm p}(\lambda_{\rm Edd})$ is different 
for galaxies with different intrinsic colors, such that
red galaxies having more rapid redshift evolution (i.e., 
${\rm p}(\lambda_{\rm Edd})\propto (1+z)^{3.7}$, 
agreeing with previous results for the overall galaxy population; 
e.g., Aird \etal\ 2012; Bongiorno \etal\ 2012)
than both blue and green galaxies (note that ${\rm p}(\lambda_{\rm Edd}) \propto (1+z)^{1.8}$ for the overall galaxy population, which includes red, green, and blue galaxies).
(3) There is marginal evidence for red galaxies having a steeper power-law slope ($-0.6$) of
${\rm p}(\lambda_{\rm Edd})$ than both blue and green galaxies, with the former slope
in agreement with previous measurements (e.g., Kauffmann \& Heckman 2009; Aird \etal\ 2012).
The Wang \etal\ (2017) results reveal the strong dependence of SMBH accretion
on their host colors, and thus caution that
ambiguous conclusions may be drawn without taking such a color dependence into account.

All the above works on \xray\ AGN \edd\ distribution
suffer, to various degrees, from issues such as limited sample sizes, in particular toward
higher redshifts and lower \xray\ luminosities, thus resulting in conclusions
associated with large uncertainties.
Therefore, future studies using larger and relatively complete samples, 
more uniform AGN identification criteria,
and more reliable \edd\ estimates would be critical to nail down the 
detailed behavior of intrinsic \xray\ AGN \edd\ distribution as a function of, e.g., 
redshift, host stellar mass, and intrinsic colors, thereby providing
sharp insights into SMBH growth processes across cosmic time.

\subsection{Coeval growth of SMBHs and their hosts}\label{sec:coevo}

The coeval growth of SMBHs and their host galaxies is vividly reflected (in part) in the 
observed tight $M_{\rm BH}$-$M_{\rm bulge}$ correlation
(e.g., Kormendy \& Gebhardt 2001; 
McLure \& Dunlop 2002; Marconi \& Hunt 2003; 
Ferrarese \& Ford 2005; Lauer \etal\ 2007; 
G\"{u}ltekin \etal\ 2009; Sani \etal\ 2011; Vika \etal\ 2012),
which is largely in the form of $M_{\rm BH}\sim M_{\rm bulge}^{1.2}$
(e.g., Wandel 1999; H\"{a}ring \& Rix 2004; Kormendy \& Ho 2013; 
McConnell \& Ma 2013) 
with an intrinsic scatter of $\approx 0.3$~dex (e.g., Kormendy \& Ho 2013).
The tight $M_{\rm BH}$-$M_{\rm bulge}$ correlation can be explained either by
AGN feedback where the coupling between AGN triggering and feedback-regulated
star formation plays a key role (e.g., Di~Matteo \etal\ 2005; Hopkins \etal\ 2006b;
Sijacki \etal\ 2007; Booth \& Schaye 2009; Fabian~2012), 
or by galaxy mergers based on the central limit theorem
(e.g., Peng~2007; Hirschmann \etal\ 2010; Jahnke \& Macci\`{o} 2011).
In addition to the $M_{\rm BH}$-$M_{\rm bulge}$ relation, 
many works study the $M_{\rm BH}$-$M_{\star}$ relation instead and find similar tight
correlations (e.g., Kormendy \& Richstone 1995; Magorrian \etal\ 1998; 
G\"{u}ltekin \etal\ 2009; Merloni \etal\ 2010; Schulze \& Gebhardt 2011; McConnell \& Ma 2013; Reines \& Volonteri 2015; Sun \etal\ 2015),
given the facts that 
$M_{\star}$ is an important physical property (see Section~\ref{sec:factor}) being relatively easy to measure, 
$M_{\rm bulge}$ is approximately equal to $M_{\star}$ for bulge-dominated systems, and 
it is challenging to separating $M_{\rm bulge}$ from $M_{\star}$ at $z\gsim 1$.   
Together, scaling relations between SMBHs and their hosts (including, e.g., 
$M_{\rm BH}$-$M_{\rm bulge}$ and 
$M_{\rm BH}$-$M_{\star}$) are fundamentally important to understanding SMBH and galaxy evolution across cosmic time.

Obviously, observational studies of the evolution of the $M_{\rm BH}$-$M_{\star}$ 
($M_{\rm bulge}$) relation are of great interest yet with diverse results.
For example, many works find that the $M_{\rm BH}/M_{\star}$ ($M_{\rm BH}/M_{\rm bulge}$)
ratio evolves positively with redshift,
with high-redshift SMBHs being a factor of several overmassive relative to their host 
galaxies compared to their local counterpart systems 
(e.g., Peng \etal\ 2006a, 2006b; Ho 2007; Merloni \etal\ 2010;
Bennert \etal\ 2011; Bongiorno \etal\ 2014; Shankar \etal\ 2016).
However, some other works find no difference in the $M_{\rm BH}/M_{\star}$ ratio
between high-redshift and local measurements
(e.g., Jahnke \etal\ 2009; Schramm \& Silverman 2013; Sun \etal\ 2015).
Additionally, there are works finding high-redshift undermassive SMBHs that
are located below the local scaling relation (e.g., Borys \etal\ 2005; 
Alexander \etal\ 2008b; Urrutia \etal\ 2012). 
The likely causes for the above inconsistent results are manifold, e.g., 
poor statistics due to limited sample sizes,
use of unrepresentative subsamples,
lack of corrections for observational biases 
(e.g., Eddington bias, biased $M_{\rm BH}$ estimates for high-redshift sources,
flux-limited samples being biased toward larger $M_{\rm BH}/M_{\star}$ ratios;
e.g., Lauer \etal\ 2007; Shen \& Kelly 2010; Schulze \& Wisotzki 2011, 2014; 
Bongiorno \etal\ 2014; Shankar \etal\ 2016),
as well as large uncertainties associated with $M_{\rm BH}$ measurements
(sometimes also $M_{\star}$ estimates, e.g., for luminous quasars; e.g., Bongiorno \etal\ 2012; Shen~2013).

\begin{figure}[!t]
\centering
\includegraphics[width=80mm]{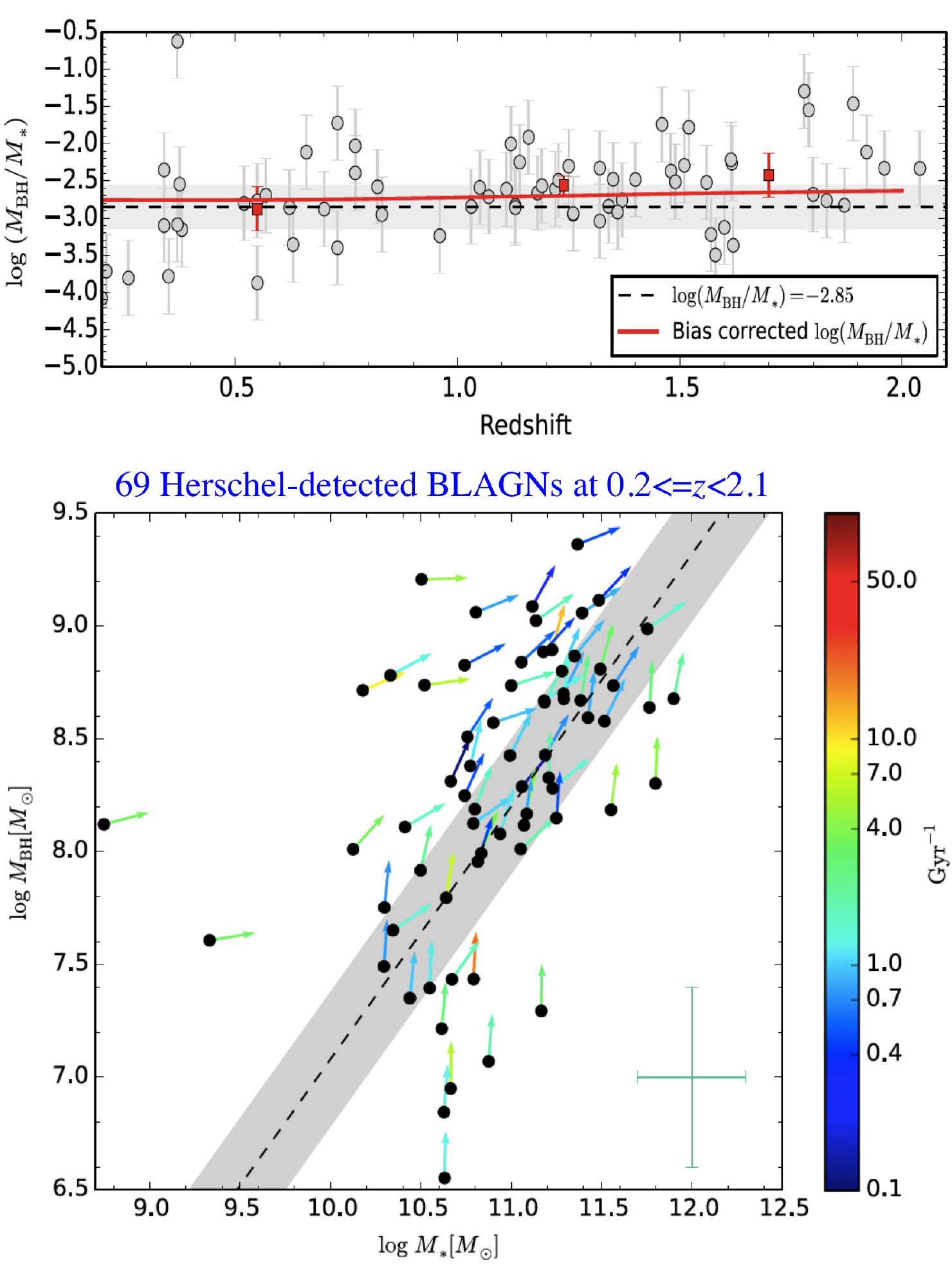}
\caption{(Top) $M_{\rm BH}/M_\star$ as a function of $z$.
The three red squares indicate average $M_{\rm BH}/M_\star$ for sources with
$z<1.0$, $1.0\le z<1.5$, and $z\ge 1.5$.
The red solid line indicates the local H\"{a}ring \& Rix (2004) $M_{\rm BH}-M_{\rm bulge}$  relation ($M_{\rm bulge}\approx M_{\star}$ here as these local host galaxies are
bulge-dominated; dashed line; the shaded region indicates its uncertainty) as it would be observed under the biases of the sample presented here. 
No evolution of the $M_{\rm BH}$-$M_\star$ relation is observed at $0.2\le z<2.1$.
(Bottom) ``Flow patterns'' of SMBHs and their host galaxies in the $M_{\rm BH}-M_\star$ plane,
with arrows showing the evolutionary directions
and colors indicating the absolute value of the total specific growth rate (i.e., $\sqrt{{\rm s}\dot{M}^2+{\rm sSFR}^2}$,
where s$\dot{M}$ is the specific BH mass accretion rate).
The large cross indicates $1\sigma$ uncertainties of $M_{\rm BH}$ and $M_\star$. 
The self-maintenance behavior of the $M_{\rm BH}$-$M_\star$ relation is evident.
Adapted from Sun \etal\ (2015).
\label{fig:smy}}
\end{figure}

Recently, Sun \etal\ (2015) carry out a unique study that simultaneously
measures high-redshift $M_{\rm BH}$-$M_{\star}$ relation and its instantaneous 
evolution (i.e., SMBH and galaxy growth rates), using a sample of 69 
{\it Herschel}-detected broad-line AGNs at $0.2\le z<2.1$ from the COSMOS
and \cdfs\ fields and accounting for biases due to
both sample selection limits and the steep slope of $M_{\star}$ distribution
at the high-mass end (see Fig.~\ref{fig:smy}).  
They find:
(1) There is no evolution in the $M_{\rm BH}$-$M_{\star}$ relation from
$z\approx 2$ to the present time.
(2) The $M_{\rm BH}$-$M_{\star}$ relation has a ``self-maintenance'' function, i.e.,
AGNs and their hosts lying off the $M_{\rm BH}$-$M_{\star}$ relation 
(i.e., ``outliers'') tend to
have evolutionary vectors anti-correlated with their $M_{\rm BH}/M_{\star}$ ratios,
such that they will evolve back toward the canonical $M_{\rm BH}$-$M_{\star}$ track given
sufficient amounts of evolutionary time.
(3) The preferred AGN duty cycle that best maintains the non-evolving 
$M_{\rm BH}$-$M_{\star}$ relation appears to be a non-evolving 
(i.e., within a factor of $\approx 4$) value of $\approx 0.1$. 
These results are in line with the scenario that the same gas reservoir
fuels both SMBH accretion and galaxy star formation, and have  
important implications for mass assembly histories of SMBHs and their host galaxies.

Future studies on the $M_{\rm BH}$-$M_{\star}$ scaling relation 
can gain further insights with higher confidence levels on the formation and co-evolution
history of SMBHs and their hosts, by enlarging samples of representative AGNs,
controlling for various observational biases, and
reducing uncertainties associated with $M_{\rm BH}$ estimates.
Furthermore, additional distinct constraints upon the connection between SMBHs and 
their host galaxies can be placed, by improving measurements of host galaxy properties (e.g., SFR, morphology, gas content) and then examining likely dependence of 
the $M_{\rm BH}$-$M_{\star}$ scaling relation on them, thereby further unraveling 
their co-evolutionary details.

\section{Evolution of starburst and normal galaxy X-ray emission}\label{sec:gal}

%
%

Remarkable insights into the formation and evolution of populations of XRBs have been gained through \chandra\ studies of local galaxies
(see, e.g., Fabbiano 2006 for a review); only ultradeep \chandra\ observations have 
enough sensitivity to study the \xray\ properties of starburst and normal galaxies at large
cosmological distances either by individually detecting them or by sensitively stacking their signal,
thus providing a direct view of the cosmic history of XRB production and evolution.
Indeed, the CDFs have detected a significant \xray\ galaxy population, e.g., 28.3\% of the 1008 7~Ms \cdfs\
main-catalogs sources are galaxies with $0.038\le z\le 2.636$ (see Table~\ref{tab:cdf}).
As predicted by previous \xray\ number-counts studies, the contribution of galaxies to
the cumulative \xray\ number counts is rapidly rising toward low fluxes, and galaxies outnumber AGNs
in the SB for the first time around the 7~Ms \cdfs\ flux limits (see Fig.~31 of Luo \etal\ 2017 and  Section~\ref{sec:numcts}).
These detections will enable substantial improvement of galaxy XLF characterization and detailed study of XLF dependence upon redshift, galaxy morphology, type, SFR, and stellar age (e.g., Georgakakis \etal\ 2006; Ptak \etal\ 2007;
Tremmel \etal\ 2013; Lehmer \etal\ 2014).
Furthermore, \chandra\ stacking analyses have revealed evolution of the 
high-mass \xray\ binary (HMXB) populations in late-type galaxies as a response to 
the rapidly increasing cosmic SFR with redshift (e.g., Reddy \etal\ 2005; Laird \etal\ 2006; 
Lehmer \etal\ 2008; Cowie \etal\ 2012):
$z\approx 0.5$--1.4 late-type galaxies are 5--15 times more \xray\ luminous (per unit stellar mass) 
than their local counterparts, showing evidence for ``downsizing'' of XRB populations. 
In contrast, stacking results of luminous early-type galaxies indicate little evolution of their hot interstellar gas up to $z\approx 1.2$, providing insights into how AGN activity 
keeps the long-term balance between gas heating and cooling (e.g., Danielson \etal\ 2012).

A number of studies based on \chandra\ observations have constrained
scaling relations between \lx\ and SFR for HMXB populations (i.e., $L_{\rm X}{\rm (HMXB)/SFR}$)
and between \lx\ and $M_\star$ for low-mass \xray\ binary (LMXB) populations 
(i.e., $L_{\rm X}{\rm (LMXB)}/M_\star$; e.g., Bauer \etal\ 2002a; Ranalli \etal\ 2003; Colbert \etal\ 2004; Hornschemeier \etal\ 2005; Lehmer \etal\ 2010; Boroson \etal\ 2011; Mineo \etal\ 2012, 2014;
Vattakunnel \etal\ 2012; Zinn \etal\ 2012; Basu-Zych \etal\ 2013;
Symeonidis \etal\ 2014).
However, the large scatters in these  \xray\ scaling relations
cannot be accounted for by measurement errors and/or statistical fluctuations 
(e.g., Hornschemeier \etal\ 2005; Mineo \etal\ 2012), and hence point to some real physical variations 
of, e.g., stellar ages, metallicities, and star formation histories that likely influence XRB formation and evolution significantly (see, e.g., Madau \& Dickinson 2014 for a review).
Fragos \etal\ (2013a) present a much improved framework of theoretical XRB population-synthesis 
models that is supported by many subsequent observational tests, 
being able to track XRB population evolution throughout cosmic history,
make predictions for redshift evolution of the scaling relations, and 
identify a ``best-fit'' theoretical model for the local scaling relations.
With the CDF and new deep multiwavelength data available, it is now plausible to
use stacking techniques to isolate large populations of galaxies, obtain their global physical
properties, investigate their population-averaged \xray\ emission, and compare with
XRB population-synthesis models in great detail
(e.g., Laird \etal\ 2006; Lehmer \etal\ 2007, 2008; Cowie \etal\ 2012; Basu-Zych \etal\ 2013).

\begin{figure*}[!t]
\centering
\includegraphics[width=160mm]{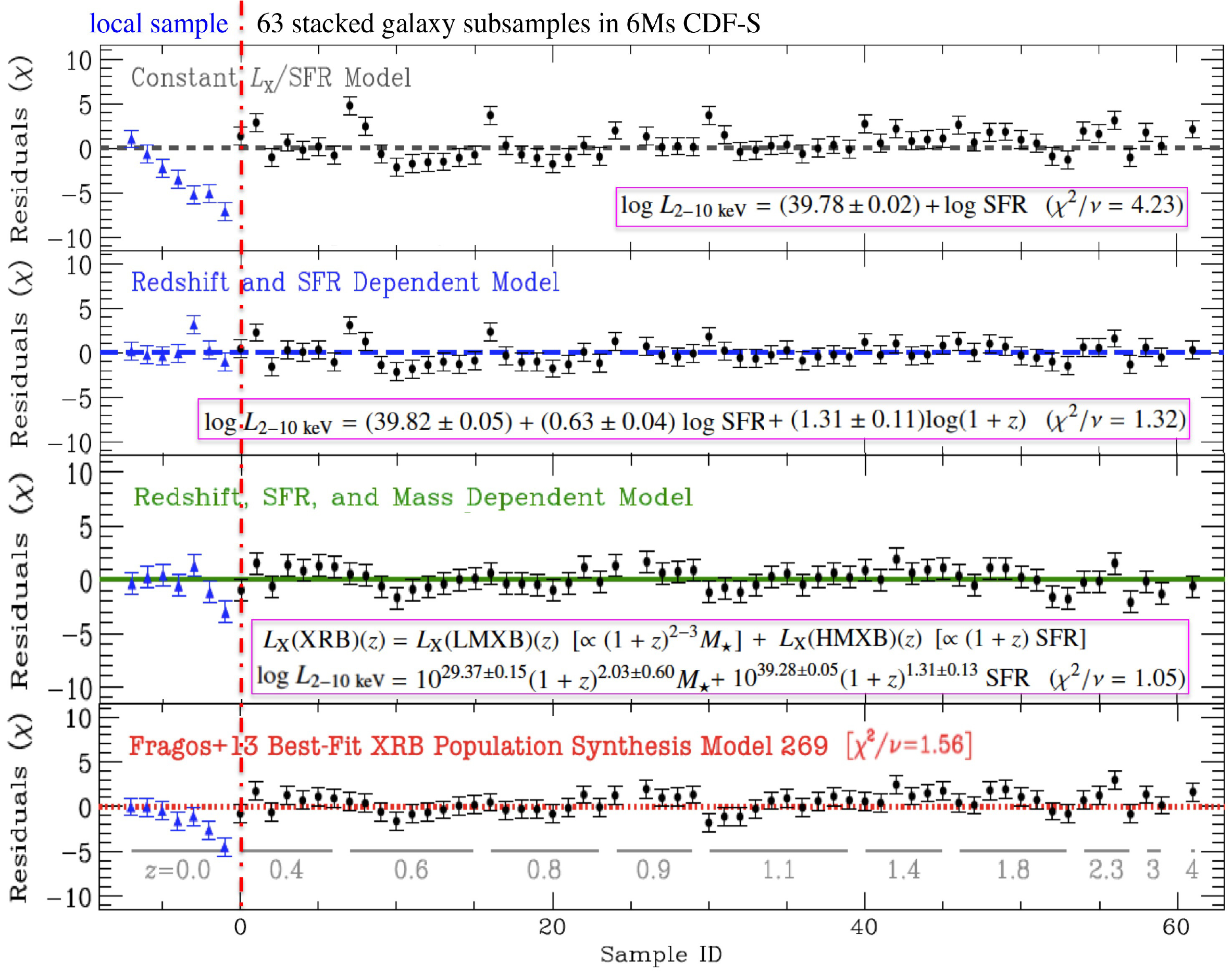}
\caption{Fitting global non-AGN \xray\ emission: residuals to model fits for a combination of 
the Lehmer \etal\ (2010) local sample (blue triangles) and 63 stacked galaxy subsamples in the 6~Ms \cdfs\ (black circles) using three different empirical models (with the third
one providing the best characterization of $L_{\rm X}{\rm (XRB)}$) and the Fragos \etal\ (2013a) best-fit
XRB population-synthesis model (No. 269), 
with model names, best-fit formula, and reduced $\chi^2$ values annotated. 
Adapted from Lehmer \etal\ (2016).
\label{fig:xgal}}
\end{figure*}

Recently, Lehmer \etal\ (2016) make use of the 6~Ms \cdfs\ data through reliable \xray\ stacking analyses to examine the dependence of
galaxy XRB emission on SFR, $M_\star$, and redshift (in the range of $z\approx 0$--7), thereby conducting the most powerful and robust tests to date 
of the Fragos \etal\ (2013) model predictions.
Their findings are as follows:
(1) Scaling relations involving SFR, $M_\star$, and redshift simultaneously best characterize global
galaxy \xray\ emission (\lx), in stark contrast to a widely assumed ``universal'' \lx/SFR relation 
(see Fig.~\ref{fig:xgal}).
(2) HMXB and LMXB populations appear to evolve as  
$L_{\rm 2-10\ keV}{\rm (HMXB)/SFR}\propto (1+z)$ and
$L_{\rm 2-10\ keV}{\rm (LMXB)}/M_\star \propto (1+z)^{2-3}$, respectively, at least up to $z\approx 2.5$,
which is consistent with basic XRB population-synthesis model predictions that
attribute the increase in the HMXB/LMXB relation with redshift primarily to 
effects related to the decrease in metallicities/stellar ages (Fragos \etal\ 2013a; also see Aird,
Coil, \& Georgakakis 2017).
However, the marginal agreement between the observational data and the Fragos \etal\ (2013)
best-fit model necessitates minor revisions of such XRB population-synthesis models 
(see Fig.~\ref{fig:xgal}).
(3) LMXBs likely dominate galaxy \xray\ emissivity (i.e., \lx\ per volume) out to $z\approx 1$--2, while HMXBs take over at 
higher redshifts. 
(4) The overall galaxy \xray\ emissivity peaks around $z\approx 1.5$--3, 
mimicking the cosmic SFR density, but declines more slowly at $z\gsim 3$ than the latter owing to
the rising \lx/SFR scaling with redshift; extrapolation of these results indicates that
galaxies provide a larger \xray\ emissivity than AGNs at $z\gsim 6$--8, thereby dominating the
reionization process, as expected by XRB population-synthesis models (see, e.g., Fragos \etal\ 2013a, 2013b; also see, e.g., Vito \etal\ 2016 mentioned in Section~\ref{sec:highz}).

This Lehmer \etal\ (2016) work, together with many previous relevant studies,
can be significantly improved, e.g., by utilizing
statistically significant samples of local (through direct detections)
and distant galaxies (mainly relying on stacking techniques) with
a variety of physical characteristics (e.g., $M_\star$, SFR, stellar age, metallicity,
and morphology), from an optimal combination of current and scheduled, wide and deep \chandra\ surveys.   
Without question, these works will be greatly boosted even further, thanks to the employment of future
powerful \xray\ observatories (see Section~\ref{sec:sum}).

\section{Census of X-ray galaxy groups and clusters}\label{sec:group}

Theoretically, there is almost no doubt that the growth and evolution of 
SMBHs and galaxies are environmentally dependent (e.g., Kauffmann 1996; De~Lucia \etal\ 2006). 
However, in terms of detailed observational manifestations,
it is still not so clear about the exact role that large-scale structures (LSSs; including
galaxy groups, clusters, and superclusters) play regarding AGN triggering and star-formation activities (see, e.g., McNamara \& Nulsen 2012 for a review on AGN feedback in LSSs), 
which is often complicated by additional likely dependencies 
such as host stellar mass, AGN luminosity, and redshift. 
Galaxy groups and clusters, lying at the high end of the cosmic density spectrum,
have been intensively observed by \hbox{X-ray} observations that
are subject to the least observational biases (e.g., Brandt \etal\ 2001; 
Bauer \etal\ 2002b; Giacconi \etal\ 2002; 
Finoguenov \etal\ 2006, 2007, 2010, 2015; Silverman \etal\ 2010; Allen, Evrard, \& Mantz 2011; Tanaka \etal\ 2012, 2013; Ehlert \etal\ 2013, 2014, 2015; Patel \etal\ 2015; Koulouridis \etal\ 2016; Miniati \etal\ 2016; Rumbaugh \etal\ 2017), given the utility of extragalactic \xray\ surveys (see Section~\ref{sec:utility});
but studies based on such observations sometimes obtain inconsistent results.
For example,
significant suppression of \xray\ AGNs in clusters (e.g., lower AGN fractions than in field populations) has been observed by some authors (e.g., Koulouridis \& Plionis 2010; Haines \etal\ 2012; Ehlert \etal\ 2013, 2014), but not seen by other authors
(e.g., Martini \etal\ 2013; Melnyk \etal\ 2013; Koulouridis \etal\ 2014).
Therefore, more observational efforts are needed to better understand the
interplay between AGN and star-formation activities and their surrounding structures, 
and hopefully reconcile the inconsistent results.

Compared to \xray\ clusters, \xray\ galaxy groups prove to be more challenging objects
to study in \hbox{X-rays}, which essentially rely on \xray\ observations with 
larger depths. 
Fortunately, both \chandra\ and {\it XMM-Newton} observations have elevated the research field of
galaxy groups to a new level, with many surveys producing large catalogs of \xray\
selected groups over a wide redshift range of $0<z<1.5$ (e.g., Brandt \etal\ 2001; Bauer \etal\ 2002b; Giacconi \etal\ 2002; Finoguenov \etal\ 2007, 2009, 2010; Bielby \etal\ 2010).
One of the current focuses is to accumulate a statistical sample of high-redshift (i.e., $z\gsim 1.5$) groups, as they are likely progenitors of the local clusters in a statistical sense, thus being key
to better understand the origin of the environmental dependence seen locally.
In this regard, a number of such systems up to $z\approx 2$ have been discovered,
although some of these systems need further spectroscopic observations for confirmation, e.g.:  
Papovich \etal\ (2010) and Tanaka \etal\ (2010) independently confirm a $z=1.62$
group in the Subaru/{\it XMM-Newton} Deep Field;
Henry \etal\ (2010) report a possible $z=1.75$ \xray\ group;
Gobat \etal\ (2011) find a $z=2.07$ group by color selection;
and Andreon \etal\ (2009) and Spitler \etal\ (2012) present a few $z\approx 2$
groups based on \zp\ selection.
These high-redshift groups, combined with those at lower redshifts, 
provide an ideal resource to understand the evolution of groups and examine
galaxy formation and evolution across the environments and cosmic time, and is also a
powerful probe of cosmology (e.g., Finoguenov \etal\ 2010, 2015).

\begin{figure*}[!t]
\centering
\includegraphics[width=170mm]{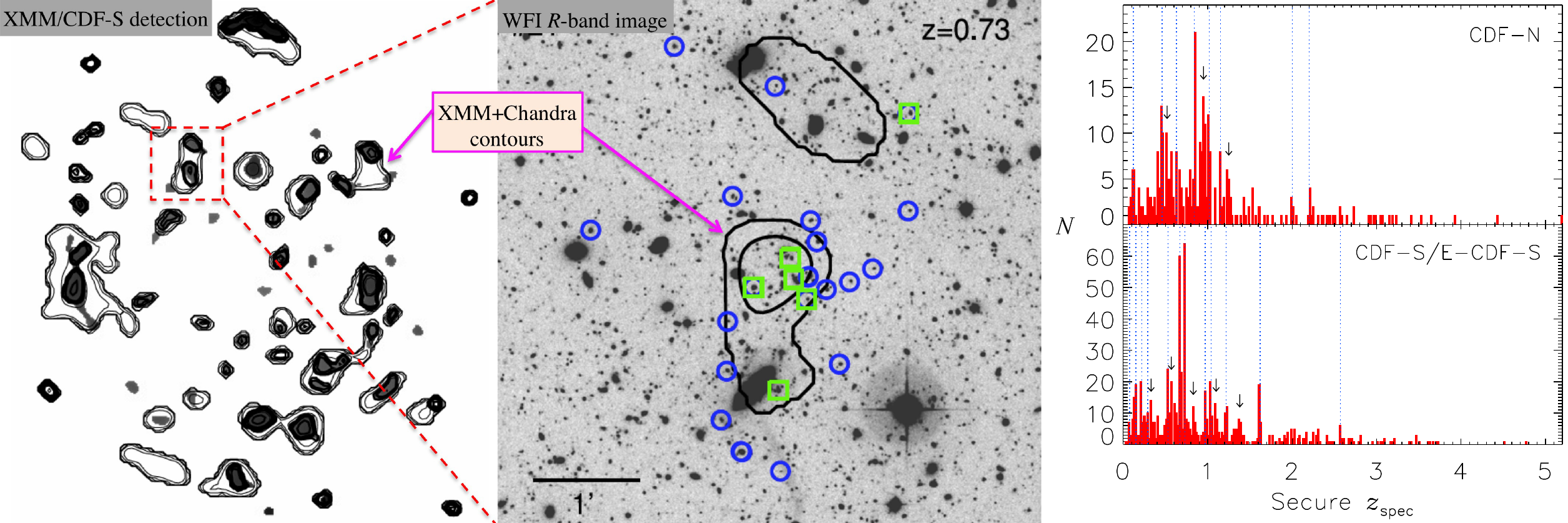}
\caption{(Left) {\it XMM-Newton} detection of the extended emission on a 32~arcsec scale in the full \ecdfs\ field,
overlaid with the contours that show the extended emission detected in the combined \chandra\ 
and {\it XMM-Newton} images on the 32 and 64~arcsec scales.
(Middle) The most prominent spectroscopically identified $z\approx0.73$ group/cluster
in the \ecdfs\ (i.e., the highest redshift spike in the Right-bottom panel),
with the contours overlaid onto the WFI $R$-band image showing 
the extended \xray\ emission detected by {\it XMM-Newton} and \chandra. \xray\ sources
with $|z_{\rm spec}-0.73|<0.01$ are labeled as circles (GOODS spectroscopy) and squares (VLT/VIMOS and Keck/DEIMOS spectroscopy).
(Right) Secure \zs\ distributions of the CDF \xray\ sources in bins of $\Delta$\zs=0.02
(the \cdfs\ \zs's are compiled by Luo \etal\ 2017, and the \cdfn\ and \ecdfs\ \zs's by Xue \etal\ 2016).
The vertical dotted lines indicate the previously known redshift spikes (i.e., associated with LSSs) identified by \xray\ sources, and the downward arrows indicate additional likely redshift spikes
identified in this paper.
(Left) adapted from Finoguenov \etal\ (2015) and (Middle) adapted from Silverman \etal\ (2010).
\label{fig:ext}}
\end{figure*}

Recently, Finoguenov \etal\ (2015) combine the ultradeep \chandra\ and {\it XMM-Newton}
observations in the \cdfs/\ecdfs\ to carry out a systematic search for \xray\ groups
down to an unprecedented flux level of $2\times 10^{-16}$~\flux, with extensive 
spectroscopic observations providing identifications of galaxy groups out to high redshifts.
They produce an \xray\ group catalog that consists of a total of 46 spectroscopically
identified systems up to $z=1.61$ and down to low masses (see the left panel of Fig.~\ref{fig:ext}), and investigate their properties, finding that:
(1) The number counts and XLF of the \ecdfs\ \xray\ groups are in broad agreement
with expectations of the $\Lambda$CDM cosmological model (also see, e.g., Liu \etal\ 2015).
(2) The low-luminosity \xray\ groups are confirmed to be low-mass systems through
one- and two-point statistics as well as weak-lensing analysis.
(3) The scaling relations between the \xray\ luminosity and total group mass
is verified and extended to regimes of lower masses and higher redshifts, based on
stacked weak lensing and clustering analyses.
These groups, with broad ranges of mass, luminosity, and redshift,
make a good and representative case of groups constituting 
the most common environment for galaxy evolution.
Moreover, there are a few notable individual structures in the \ecdfs.
For instance, the two most prominent and populated LSSs in the \ecdfs\ are located at
$z=0.67$ and $z=0.73$, each traced by over 60
\xray\ AGNs and many more galaxies (see the middle panel
and Right-bottom panel of Fig.~\ref{fig:ext}); 
and interestingly, the former LSS is in a shape
of thick sheet with a radial size of 67.7~Mpc extending over the full \ecdfs\ field,
and in contrast, the latter LSS is thin (18.8~Mpc) and filamentary 
(e.g., Silverman \etal\ 2010).
Another eye-catching system is the $z=1.61$ group that is the most distant \xray\ group 
identified in the \ecdfs\ and examined in detail by Tanaka \etal\ (2013).
They find that 
this group is actually the lowest-mass one (with $(3.2\pm 0.8)\times 10^{13}M_\odot$) 
ever confirmed at $z>1.5$, and
exhibits a surprisingly prominent red sequence of quiescent early-type galaxies
whose star formation is likely shut down by the bright AGN group members.

The above example works demonstrate that the CDFs are good sky patches for the census of \xray\
groups and clusters, as further elucidated in the right panel of Fig.~\ref{fig:ext} that 
shows a number of CDF \zs\ spikes indicating (likely) LSSs, where the most up-to-date secure \zs\ compilations from Xue \etal\ (2016) and Luo \etal\ (2017) are adopted.
Indeed, the CDFs are rich in LSSs:
in the \cdfn, the previously known \xray\-source-traced LSSs 
located at $z=0.12$, 0.46, 0.63, 0.843, 1.02, 1.15, 2.0, and 2.2 are all recovered by this paper
(indicated by dotted lines; e.g., Barger \etal\ 2002, 2003b; Gilli \etal\ 2005; Reddy \etal\ 2006);
and in the \ecdfs, those at $z=0.077$, 0.15, 0.218, 0.29, 0.53, 0.67, 0.73, 0.97, 1.04, 1.22, 1.62,
and 2.57 are recovered by this work (e.g., Gilli \etal\ 2003; Treister \etal\ 2009c; Silverman \etal\
2010; Dehghan \& Johnston-Hollitt 2014).
In addition to the known CDF LSSs, a total of 8 potential LSSs are identified using \xray\ sources as tracers for the first time (indicated by downward arrows), which are worth 
further verification and quantification.
Needless to say, a census of \xray\ galaxy groups and clusters as complete as possible in the 
fields like the CDFs would be most beneficial to the improved understanding of the evolution of these
systems themselves and the environmental dependence of SMBH accretion and galaxy star formation.

\section{Additional Science}\label{sec:addsci}

In addition to the aforementioned exciting science that can be done with the CDFs,
there are furthermore a wide variety of interesting topics that can be tackled using the CDFs, e.g.:
(1) Bauer \etal\ (2017) report a remarkable new fast \xray\ 
transient discovered by Luo \etal\ (2014b) 
in course of the extension of \cdfs\ observations from 4~Ms to 7~Ms,
whose intriguing \xray\ and multiwavelength properties effectively rule out the vast
majority of previously known high-energy transients, leaving out only 
a few exotic theoretical possibilities
that still cannot completely explain all observed properties.
The inferred rate of such events is crudely comparable to that of gamma-ray bursts,   
which indicates the discovery of an untapped regime for a known transient type or a new class of transient phenomenon with its nature to be determined. 
(2) Cappelluti \etal\ (2017b) report a 3$\sigma$ detection of an $\approx 3.5$~keV emission line
in the CXRB spectrum derived with a combined $\approx 10$~Ms \chandra\ exposure of the \cdfs\
and COSMOS Legacy surveys, and discuss the likely origins of this observed line
that include the iron line background, S~XVI charge exchange, and sterile neutrino decay.
(3) Using the 4~Ms \cdfs, {\it HST}, and {\it Spitzer} data, 
Mitchell-Wynne \etal\ (2016) perform and compare the cross-correlation analyses between \xray\ and
optical/NIR cosmic background intensity fluctuations and find that the sources responsible for the cosmic IR background at 3.6 and 4.5~$\mu$m are at least partly dissimilar to 
those at $\lsim 1.6\ \mu$m (also see, e.g., Cappelluti \etal\ 2013; Helgason \etal\ 2014).
(4) There are a significant number of SMGs identified in the CDFs, which are of great interest given their
important role in the SMBH/galaxy co-evolution picture (e.g., Alexander \& Hickox 2012; 
also see Section~\ref{sec:inter}).
Using the CDF and multiwavelength data, many topics can be studied in detail, including, e.g., 
the incidence of AGNs in SMGs, the location of SMGs in color-magnitude/mass diagrams, and 
coeval growth and evolutionary track of SMBHs and SMGs (e.g., Alexander \etal\ 2005b; Xue \etal\ 2010;
Wardlow \etal\ 2011; Wang \etal\ 2013).
(5) There is an extended \xray\ emission identified in the \cdfn\ that is best explained as
an inverse Compton ghost of a giant radio source (e.g., Fabian \etal\ 2009), calling for a 
systematic search for such sources in the CDFs.
(6) There are a few tens of individual off-nuclear XRBs at $z\approx 0.05$--0.3
discovered in the CDFs, allowing for investigating their properties and redshift evolution 
as a population (e.g., Hornschemeier \etal\ 2004; Lehmer \etal\ 2006; also see
Section~\ref{sec:gal}).
(7) There is a minority population of ``interloping'' Galactic stars in the CDFs, whose long-term evolution of magnetic activity can be studied by examining their \xray\ emission (e.g., Feigelson \etal\ 2004).

\section{Summary and prospects}\label{sec:sum}

The CDFs represent the amazing outcome of the multi-decade efforts of numerous people
(counting from the proposal of \chandra\ submitted to NASA by R.~Giacconi and H.~Tananbaum in 1976 to
the last \cdfs\ observation taken in March 2016),
act as a major thrust among extragalactic \xray\ surveys that are complemented effectively by deep multiwavelength observations, 
contribute critically
to our dramatically improved characterization of the \hbox{0.5--8 keV} CXRB sources,
enable a wide range of scientific topics (including AGNs, starburst and normal galaxies,
groups and clusters of galaxies, LSSs, etc.),
and launch literally hundreds of exciting research papers.

In this paper, I have highlighted some recent key observational results that are mostly from the CDFs
and enabled by the revolutionary and versatile scientific capabilities of \chandra, 
including the AGN demography (see Section~\ref{sec:demo}), 
the interactions between AGNs and their host galaxies (see Section~\ref{sec:inter}), 
the evolution of starburst and normal galaxy X-ray emission (see Section~\ref{sec:gal}), 
and the census of X-ray galaxy groups (see Section~\ref{sec:group}) through cosmic time. 
The beauty and power of the CDF treasure trove will surely be further augmented given
the consistent and full exploitation of the latest CDF catalogs and products (Xue \etal\ 2016; Luo \etal\ 2017) in conjunction with the ever improving multiwavelength observations.

Despite the great advances in the above research areas,
it has almost always been challenging to tell a consistent and credible story in each of these areas or
regarding some of the areas as a whole,
which requires reconciling/distinguishing inconsistent results (owing to, e.g.,
differences in sample selection criteria, analysis techniques, data qualities and depths, 
incompleteness levels, and/or sample sizes, as well as likely effects of cosmic 
variance)\footnote{Perhaps these different and even inconsistent observational 
manifestations highlight, to some degree, the vast complexities and personalities of galaxies
and AGNs that essentially preclude any simple ``universal'' scenarios regarding many
questions about them. }
and then stringing up all the reasonable pieces.
Furthermore, there are many significant open questions that should persist for a foreseeable future, e.g.:
(1) How to ultimately resolve the CXRB?
(2) What processes dictate the formation and growth of the first SMBHs?
(3) What drives the AGN downsizing behavior?
(4) What are the origins of AGN variability?
(5) How to obtain a census of highly obscured AGNs, CTAGNs, LLAGNs, and low-mass BHs as completely as possible? 
(6) What is the exact role of AGN feedback and what is the link between SMBH accretion and star formation?
(7) How do accreting XRB populations evolve over most of cosmic time?
(8) How do LSSs affect AGN activity?
To address these questions, there are at least three aspects to move forward 
with each having both short-term and long-term goals associated with current and future facilities, respectively.

\begin{figure*}[!t]
\centering
\includegraphics[width=170mm]{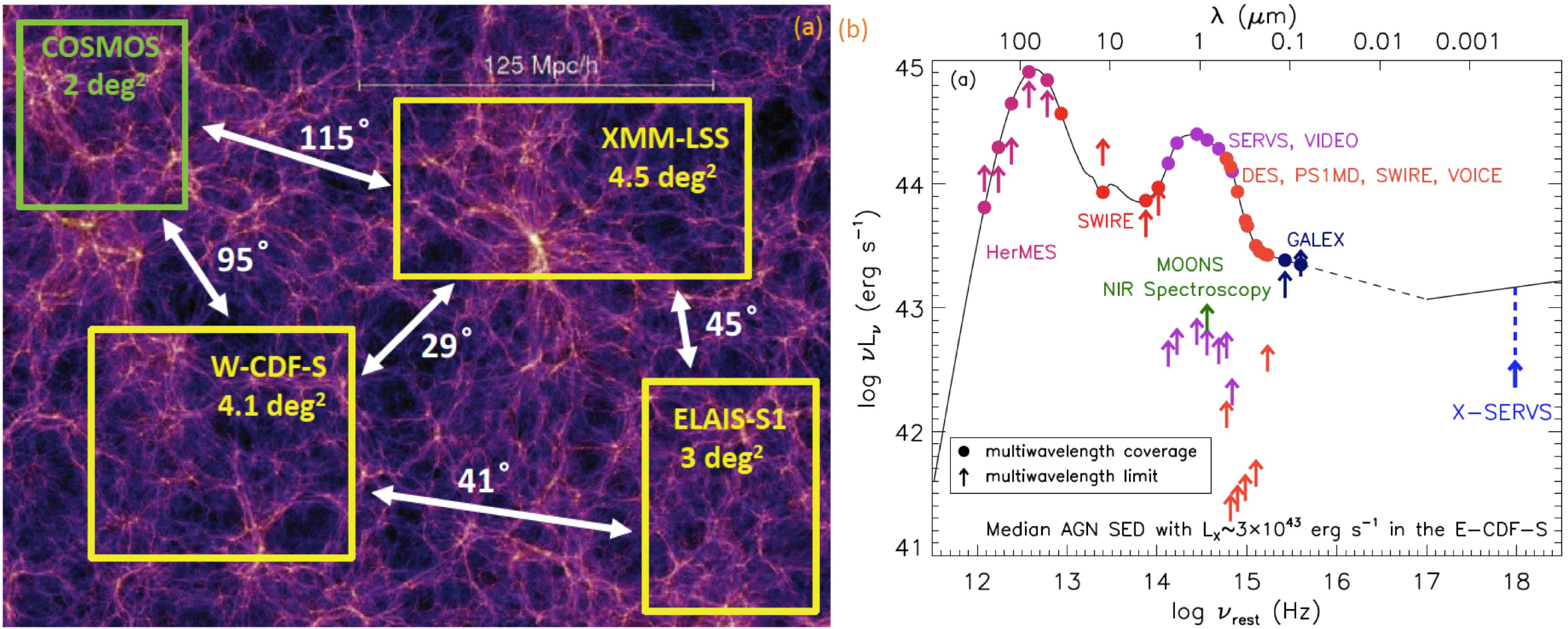}
\caption{(Left) Projected density map at $z=1.4$ with comoving size $350\times450$~Mpc$^2$ from the Millennium Simulation (Springel \etal\ 2005), with the sizes of COSMOS and the porposed X-SERVS fields as well as their respective separations annotated.
(Right) Median SED of AGNs with $L_{\rm X}\approx 3\times 10^{43}$~\lum\ in the \ecdfs\
combined with a composite FIR starburst galaxy SED (Sajina et al. 2012) at $z=1$. 
The data points and upward arrows indicate the current/scheduled X-SERVS multiwavelength coverage and their $z=1$ limiting luminosities; and the rightmost blue arrow indicates
the proposed critical \xray\ coverage that is currently missing 
(courtesy of B.~Luo).
\label{fig:prop}}
\end{figure*}

$\bullet$ {\it Going wider.}
Ultradeep pencil-beam surveys such as the CDFs (see Table~\ref{tab:cdf}) are inevitably subject to the effects of cosmic variance.
To remedy this situation and facilitate the science that requires larger solid-angle coverages, a straightforward solution is to widen the spatial coverages of the small-area surveys
at moderate depths (thus being relatively easily achievable).
Indeed, the CDF team has been in the process of proposing a $\approx 12$~deg$^2$ \chandra/{\it XMM-Newton} survey at 30/50~ks depth of the SERVS areas of Wide-\cdfs\ (W-\cdfs, centered at \cdfs), 
ELAIS-S1, and XMM-LSS, with 1.3~Ms of {\it XMM-Newton} observing time awarded already (PI: W.~N.~Brandt; see
Fig.~\ref{fig:prop}).
These three fields all have multiple intensive radio-to-UV observations
performed or scheduled but lack the critical \xray\ coverage. 
Therefore, the addition of the proposed \xray\ observations will powerfully leverage
those multiwavelength surveys by detecting thousands of new \xray\ AGNs and
hundreds of new \xray\ groups/clusters, thereby
dramatically advancing studies
of SMBH growth across the full range of cosmic environments
(from voids, groups, clusters, to the largest structures found in cold dark matter simulations), links
between SMBH accretion and star formation, exceptional AGNs and protoclusters at high
redshifts, and other topics. 
In the long run, the funded ESA-led mission {\it Athena} (band: \hbox{0.3--12~keV}; large collecting area: 
$\approx 2.0$--2.5~m$^2$; large field of view: $\approx 40\times 40$~arcmin; good angular
resolution: $\approx 3$--5~arcsec; scheduled launch: 2028; Nandra \etal\ 2013)
will be $\approx 100$ times more efficient in carrying out deep and wide surveys than \chandra\ and {\it XMM-Newton}
for a given combination of solid-angle coverage and flux limit 
(above $f_{\rm SB}\gsim (1-2)\times 10^{-17}$~\flux, which corresponds to {\it Athena}'s source
confusion limit and sensitivity limits of $\approx 2$~Ms \chandra\ exposures; see Table~\ref{tab:cdf}),
thus substantially revolutionizing extragalactic \xray\ studies.  
A large strategic mission concept, {\it Lynx} (called {\it \xray\ Surveyor} previously; e.g., Weisskopf
\etal\ 2015), 
is essentially to build a super \chandra\ with significantly enlarged collecting area
(thus tremendously increased sensitivities) and still sub-arcsecond angular resolution;
if selected and funded, it will take the sharpest \xray\ vision even fainter and farther, thereby prodigiously transforming the field as {\it Athena} would do.

$\bullet$ {\it Going harder.}
Sensitive hard \xray\ ($\gsim 10$~keV) surveys can open up a large volume of discovery space 
uncharted by \chandra\ and {\it XMM-Newton}, deciphering the even more energetic universe.
Deep hard \xray\ observations that fully encompass the CXRB peak
can provide additional critical insights into the origin of the CXRB and unveil the highly obscured and CT AGNs.
In this regard, {\it NuSTAR}, as the first focusing hard \xray\ observatory in orbit, has been a pathbreaker that is significantly influencing the field.
{\it NuSTAR} has already observed the CDFs with 200~ks depth (see Footnotes~\ref{ft:cdfs}~and~\ref{ft:cdfn}), being able to detect Seyfert-like sources ($L_{\rm X}<10^{44}$~\lum) up to $z\approx 0.5$.
Future more sensitive hard \xray\ observations would be expected from {\it HEX-P} 
(PI: F.~Harrison) if funded, which has been proposed to be a natural successor to {\it NuSTAR}.

$\bullet$ {\it Going more.}
Additional sky coverages at the CDF flux levels are critically important as they substantially improve the statistical sample sizes of the faintest \xray\ sources and also allow for a basic assessment of the effects of cosmic variance.
It is well expected that the CDF-like surveys will become commonplace 
once the next-generation large \xray\ observatories such as {\it Athena} and {\it Lynx}
are put into operation.  
Specifically, {\it Lynx} will have the capability of efficiently performing even deeper surveys
than the CDFs, thus effectively going deeper and wider simultaneously (see the above 
{\it Going wider} point).

\vspace{0.2in}
\noindent {\bf Acknowledgements:}
\vspace{0.2in}

I thank the referee for careful reading and constructive comments that helped
improve this review.
I thank W.N.~Brandt ``who led the effort to obtain the deepest \chandra\ fields'' (excerpted from the citation of his 2016 Bruno Rossi Prize), 
thank B.D.~Lehmer for producing Figures~\ref{fig:ncts-z}, \ref{fig:ncts-nh}, and \ref{fig:ncts-lx} exclusively for this paper, thank B.~Luo for providing Figure~\ref{fig:prop},
and thank G.~Yang for making Figure~\ref{fig:absorption}.
I thank D.M.~Alexander, F.E.~Bauer, W.N.~Brandt, B.D.~Lehmer, B.~Luo, M.Y.~Sun,
F.~Vito, and G.~Yang for helpful feedback, discussions, and information sharing, as well as consistent collaboration.
I also thank all my other collaborators for working on various projects together.
The following funding support is gratefully acknowledged:
the 973 Program (2015CB857004), 
the CAS Frontier Science Key Research Program (QYZDJ-SSW-SLH006), 
the National Natural Science Foundation of China (NSFC-11473026, 11421303), 
the Fundamental Research Funds for the Central Universities, 
the Thousand Young Talents Plan of China, and
the CAS Strategic Priority Research Program ``The Emergence of
Cosmological Structures'' (XDB09000000).

\end{document}